\documentclass[12pt]{article}

\usepackage{amssymb}
\usepackage{amsmath}
\usepackage{amscd}
\usepackage{latexsym}
\usepackage{graphicx}

\numberwithin{equation}{section}
\usepackage{cite}
\usepackage{enumitem}

\newcommand{\be}{\begin{equation}}
\newcommand{\ee}{\end{equation}}
\newcommand{\Dlt}{\Delta}
\newcommand{\dlt}{\delta}
\newcommand{\prt}{\partial}
\newcommand{\br}{{\bf r}}
\newcommand{\bk}{{\bf k}}

\newcommand{\bp}{{\bf p}}

\newcommand{\bt}{\beta}
\newcommand{\vp}{\varphi}
\newcommand{\ep}{\varepsilon}
\newcommand{\al}{\alpha}
\newcommand{\ra}{\rightarrow}
\newcommand{\sgm}{\sigma}

\newcommand{\gm}{\gamma}
\newcommand{\om}{\omega}
\newcommand{\Om}{\Omega}
\newcommand{\Gm}{\Gamma}

\newcommand{\lbd}{\lambda}
\newcommand{\Lbd}{\Lambda}

\newcommand{\rgl}{\rangle}
\newcommand{\lgl}{\langle}

\begin{document}

\begin{center}

{\Large{\bf Interplay Between Approximation Theory and Renormalization Group} \\ [5mm]

V.I. Yukalov } \\ [3mm]

{\it Bogolubov Laboratory of Theoretical Physics, \\
Joint Institute for Nuclear Research, Dubna 141980, Russia \\ [2mm]
and \\ [2mm]
Instituto de Fisica de S\~ao Carlos, Universidade de S\~ao Paulo, \\
CP 369,  S\~ao Carlos 13560-970, S\~ao Paulo, Brazil  \\ [5mm]
E-mail: yukalov@theor.jinr.ru} 
\end{center}

\vskip 5cm

\begin{abstract}

The review presents general methods for treating complicated problems that cannot
be solved exactly and whose solution encounters two major difficulties. First, there
are no small parameters allowing for the safe use of perturbation theory in powers
of these parameters, and even when small parameters exist, the related perturbative
series are strongly divergent. Second, such perturbative series in powers of these
parameters are rather short, so that the standard resummation techniques either yield
bad approximations or are not applicable at all. The emphasis in the review is on the
methods advanced and developed by the author. One of the general methods is {\it
Optimized Perturbation Theory} now widely employed in various branches of physics,
chemistry, and applied mathematics. The other powerful method is {\it Self-Similar
Approximation Theory} allowing for quite simple and accurate summation of divergent
series. These theories share many common features with the method of renormalization
group, which is briefly sketched in order to stress the similarities in their ideas
and their mutual
interconnection.

\end{abstract}

\vskip 2mm


\newpage

{\bf Contents}

{\bf 1. Introduction}

\vskip 5mm
{\bf 2. Optimized perturbation theory}

\vskip 2mm
   2.1. Control functions

   2.2. Fastest convergence

   2.3. Optimization conditions

   2.4. Asymptotic conditions

   2.5. Partition functions

   2.6. Thermodynamic potentials

   2.7. Eigenvalue problem

   2.8. Exact solutions

   2.9. Sequence convergence

   2.10. Reexpansion trick: induced form

   2.11. Reexpansion trick: arbitrary form

   2.12. Critical temperature: initial conditions

   2.13. Critical temperature: reexpansion trick

   2.14. Critical temperature: trapped bosons

   2.15. Fluid string

   2.16. Fluid membrane

   2.17. Linear Hamiltonians

   2.18. Nonlinear Hamiltonians

   2.19. Hamiltonian envelopes

   2.20. Magnetic systems

   2.21. Field-theory models

\vskip 5mm
{\bf 3. Self-similar approximation theory}

\vskip 3mm
   3.1. Approximation cascade

   3.2. Approximation flow

   3.3. Stability analysis

   3.4. Free energy

   3.5. Eigenvalue problem

   3.6. Choice of initial approximation

   3.7. Fractal transform

   3.8. Self-similar root approximants

   3.9. Self-similar nested approximants

   3.10. Self-similar exponential approximants

   3.11. Self-similar additive approximants

   3.12. Self-similar factor approximants

   3.13. Self-similar combined approximants

   3.14. Acceleration of convergence

   3.15. Electron gas

   3.16. Harmonium atom

   3.17. Schwinger model

   3.18. Large-variable prediction: partition function

   3.19. Large-variable prediction: eigenvalue problem

   3.20. Exact solutions: initial-value problem

   3.21. Exact solutions: boundary-value problem

   3.22. Vortex-line equation

   3.23. Critical phenomena: spin systems

   3.24. Critical phenomena: Bose gas

   3.25. Critical exponents

   3.26. Time series

   3.27. Probabilistic scenarios

   3.28. Discrete scaling

   3.29. Complex renormalization

   3.30. Quark-gluon plasma

   3.31. Gell-Mann-Low function: Quantum Chromodynamics

   3.32. Gell-Mann-Low function: Quantum Electrodynamics

   3.33. Gell-Mann-Low function: Quartic Field Theory

   3.34. Other applications

\vskip 5mm
{\bf 4. Basics of renormalization group}

\vskip 3mm
   4.1. Renormalization-group classes

   4.2. Critical phenomena

   4.3. Universality classes

   4.4. Kadanoff transformations

   4.5. Momentum scaling

   4.6. Fixed points

   4.7. Wilson method

   4.8. Fractional dimension

   4.9. Differential formulation

   4.10. Decimation procedure

   4.11. Exact semigroup

   4.12. Field-theory group

   4.13. Invariant charge

\vskip 5mm
{\bf 5. Discussion}

\newpage

\section{Introduction}

The overwhelming majority of problems describing realistic phenomena and systems
cannot be solved exactly. Sometimes it is possible to resort to cumbersome computer
calculations. However, such a numerical treatment is not always realizable and,
moreover, is not sufficient for understanding the physics of the considered
phenomenon. The most widespread way of analyzing complicated problems is by invoking
some kind of perturbation theory with respect to parameters characterizing the system.
But it is a rare occasion when there exist small parameters making perturbation theory
sensible. Usually, such small parameters are absent. Even when there happens a small
parameter, the resulting perturbative series, nevertheless, are usually divergent.

There exist methods allowing for the extrapolation of series, derived for asymptotically
small parameters, to finite values of the latter. The most popular are the method of
Pad\'e approximants \cite{Baker_1} and Borel summation (see, e.g. \cite{Kleinert_2}).
However, these methods not always are applicable. For instance, Borel summation requires
the knowledge of large-order behaviour of perturbative terms, which in the majority of
cases is not known.

The present article reviews an original approach based on the methods developed by the
author, which are {\it Optimized Perturbation Theory} (Sec. 2) and {\it Self-Similar
Approximation Theory} (Sec. 3). These methods are essentially more general than that
of Pad\'{e} approximants, including the latter as just a particular case. They provide
accurate approximations having in hands only a few perturbative terms. It becomes possible
to extrapolate asymptotic series in powers of small parameters to finite and even to
infinite values of the latter. In those cases, when a number of perturbative terms are
known, the approach results in approximants being in very good agreement with numerical
calculations, when these are available. In those cases, when the considered problem
enjoys an exact solution, the described theory usually also yields this exact solution.
The approach shares many ideas with the method of renormalization group. And vice versa,
in the method of renormalization group it is possible to use the ideas of optimal control
theory. This is why the basics of the renormalization group approach are sketched in
Sec. 4. The parallel exposition of approximation theory and renormalization group not
only explains the underlying grounds of the former, but may also give hints on the
possible development and mutual enrichment of both approaches. Section 5 contains a
brief summary of the main ideas of the described approaches and emphasizes the common
points in approximation theory and renormalization group techniques.

Throughout the paper, the system of units is used, where the Planck and Boltzmann
constants are set to unity.

\section{Optimized Perturbation Theory}

In literature, one often manipulates by the term "perturbation theory", as opposed
to "nonperturbative methods", keeping in mind that perturbation theory gives only
asymptotic series with respect to some parameter, while nonperturbative methods result
in a sequence of approximants that do not form asymptotic series, possessing a more
complicated structure. However, such a counterposition is basically senseless. From
the mathematical point of view, there already exists a distinction between asymptotic
and nonasymptotic series. So, there is no need to involve such an ugly term as
"nonperturbative methods". In the general mathematical definition, perturbation theory
is a systematic method of defining successive approximations \cite{Nayfeh_297}. In the
present section, we describe a method called Optimized Perturbation Theory.

{\it Optimized Perturbation Theory is a systematic method defining a sequence of
successive approximants, whose convergence is governed by control functions}.

\subsection{Control Functions}

Optimized perturbation theory was advanced in Refs. \cite{Yukalov_3,Yukalov_4}.
Then it has been applied to treating anharmonic quantum crystals
\cite{Yukalov_3,Yukalov_4,Yukalov_5,Yukalov_6,Yukalov_7,Yukalov_8}, the theory of
melting \cite{Yukalov_5,Yukalov_10}, and many other systems and phenomena. The basic
idea of this theory is the introduction of control functions making divergent series
convergent. This theory sometimes appears in literature under different names, such
as modified perturbation theory, renormalized perturbation theory, variational
perturbation theory, controlled perturbation theory, self-consistent perturbation
theory, oscillator-representation method, delta expansion, optimized expansion,
nonperturbative expansion, and so on. However all these guises imply the same idea
of introducing control functions optimizing the convergence of a perturbative sequence.

Suppose we are looking for a real function $f(x)$ of a real variable $x$. For
simplicity, we consider here real-valued functions and variables, while the extension
to complex-valued quantities can be straightforwardly done by considering several
functions and variables. When the sought function is defined by a complicated problem
that cannot be solved exactly one invokes a kind of perturbation theory yielding a
sequence of approximations $f_k(x)$, with $k=0,1,2,\ldots$. As a rule, the sequence
$\{f_k(x)\}$ is divergent. The main idea of optimized perturbation theory is to
rearrange the sequence $\{f_k(x)\}$ to another form $\{F_k(x,u_k)\}$ by introducing
control parameters $u_k$ that become control functions $u_k=u_k(x)$ making the
rearranged sequence convergent. Such a rearrangement, that can be denoted as
\be
\label{2.1}
 \hat R[u_k] f_k(x) = F_k(x,u_k) \;  ,
\ee
can be realized is three ways.

First of all, we need to incorporate control parameters $u_k$ into the considered
sequence. For a while, we assume that control functions can be defined by some
optimization conditions to be discussed later. And here we only describe the ways
how control functions can be incorporated into a perturbative sequence.

\vskip 2mm

{\bf A}. {\it Introducing control parameters through initial conditions}

\vskip 2mm

It is possible to start the solution procedure with an initial approximation containing
trial parameters $u_k$. For example, let the sought function be defined by an equation
\be
\label{2.2}
\hat E[f(x)] = 0 \;  .
\ee
We can start solving the equation with an initial approximation $F_0(x,u)$ following
an iterative procedure
\be
\label{2.3}
 F_{k+1}(x,u_{k+1}) = F_k(x,u_k) + \hat E[F_k(x,u_k) ] \;  .
\ee
Then, defining the control functions $u_k = u_k(x)$ so that to make the sequence
$\{F_k(x,u_k)\}$ convergent, we get the optimized approximants
\be
\label{2.4}
 \tilde f_k(x) = F_k(x,u_k(x)) \;  .
\ee

Or, suppose we consider a system characterized by a Hamiltonian $H$. If the system
is complicated and cannot be solved exactly, we can take as a zero approximation a
simpler Hamiltonian $H_0(u)$ containing a parameter $u$. Then the system Hamiltonian
can be represented as
\be
\label{2.5}
  H = H_0(u) + \ep [ H - H_0(u) ] \qquad ( \ep \ra 1 ) \; ,
\ee
where $\varepsilon$ is a dummy parameter to be set to one at the end, after using
perturbation theory in powers of this parameter. Calculating in the $k$-th order of
the perturbation theory an observable quantity
\be
\label{2.6}
 F_k(x,u_k) = \lgl \hat A(x) \rgl_k \;  ,
\ee
corresponding to an operator of local observable $\hat{A}(x)$, and defining control
functions $u_k = u_k(x)$, we get the optimized approximant (\ref{2.4}).

As is clear, it is straightforward to deal in the same way with several types of
control functions and with several observable quantities. Instead of a Hamiltonian
one can consider a Lagrangian.

\vskip 2mm

{\bf B}. {\it Introducing control parameters by means of functional transformation}

\vskip 2mm

The other way of introducing control parameters is by transforming each member of the
sequence $\{f_k(x)\}$ by means of a transformation containing a parameter $u_k$,
which can be denoted as
\be
\label{2.7}
 \hat T [u_k] f_k(x) = F_k(x,u_k) \;  .
\ee
Then, defining control functions $u_k = u_k(x)$, we make the inverse transformation
and obtain the optimized approximants
\be
\label{2.8}
  \tilde f_k(x) = \hat T^{-1} [u_k(x)] F_k(x,u_k(x) ) \; .
\ee

One of the simplest examples is the fractal transform
$$
\hat T [u] f_k(x) = x^u f_k(x)
$$
to be considered in Sec. 3.

\vskip 2mm

{\bf C}. {\it Introducing control parameters by reexpansion trick}

\vskip 2mm

The third method of introducing control parameters is by making the change of variables
and invoking a reexpansion procedure that can be done is several ways.

\vskip 2mm

(i) It is possible to make the change of the variable
$$
x = \al(x,z,u) \; , \qquad z = z(x,u)
$$
containing a parameter $u$. Then one reexpands $f_k(\al(x,z,u))$ in powers of $x$
keeping $z$ and $u$ fixed, thus getting
$$
f_k(\al(x,z,u)) \simeq \sum_{n=0}^k b_n(z,u) x^n \qquad (x \ra 0 ) \;   .
$$
After this, one substitutes here $z = z(x,u)$, which gives
$$
F_k(x,u_k) =  \sum_{n=0}^k b_n(z(x,u_k),u_k) x^n \;  .
$$
Defining control functions $u_k = u_k(x)$, one comes to the optimized approximant
(\ref{2.4}).

\vskip 2mm
(ii) The other possible way is to make the change of the variable
$$
 x = x(z,u) \; , \qquad z = z(x,u)
$$
and then reexpand $f_k(x(z,u))$ in powers of $z$ getting
$$
 f_k(x(z,u)) \simeq \sum_{n=0}^k b_n(u) z^n \qquad (z \ra 0 )\;  .
$$
Substituting here $z=z(x,u)$ yields
$$
 F_k(x,u) \equiv  \sum_{n=0}^k b_n(u) z^n(x,u) \;  .
$$
Defining control functions $u_k = u_k(x)$ results in the optimized approximant
(\ref{2.4}).

\vskip 2mm

(iii) One more method is as follows. Let $f_k(x,m)$ depend on a parameter $m$. One
makes the substitution
$$
m \ra u + \ep(m - u) \; , \qquad x \ra \ep x
$$
and expands $f_k$ in powers of $\varepsilon$, obtaining
$$
 f_k(\ep x, u + \ep(m - u) ) \simeq \sum_{n=0}^k b_n(x,m,u) \ep^n \; .
$$
Then one sets $\varepsilon = 1$, which gives
$$
 F_k(x,m,u) \equiv \sum_{n=0}^k b_n(x,m,u) \; .
$$
Defining control functions $u_k = u_k(x,m)$, one obtains the optimized approximant
\be
\label{2.9}
 \tilde f_k(x,m) = F_k(x,m,u_k(x,m)) \;  .
\ee

\vskip 2mm
(iv) One sometimes needs to find the value $f(x)$ at large $x$ including $x\ra\infty$.
Then it is possible to change the variable $x=x(z,u)$ so that $z \ra 1$ when $x\ra\infty$.
Expanding $f_k$ in powers of $z$, one gets
$$
 f_k(x(z,u)) \simeq  \sum_{n=0}^k b_n(u) z^n  \qquad ( z \ra 0) \; .
$$
After defining control functions $u_k = u_k(z)$, the optimized approximant for
$f(\infty)$ reads as
\be
\label{2.10}
 \tilde f_k(\infty) = \lim_{z\ra 1} \sum_{n=0}^k b_n(u_k(z)) z^n \; .
\ee

As examples of the corresponding changes of the variable, we can mention
$$
 x = \frac{cz}{(1-z)^u} \;  ,
$$
where $c$ is a parameter, and the conformal mapping
$$
x = \frac{4u^2z}{(1-z)^2} \;  ,  \qquad
 z = \frac{\sqrt{x+u^2}-u}{\sqrt{x+u^2}+u} \; .
$$
The latter is a variant of the Euler transformation.

\subsection{Fastest Convergence}

Let the control parameters $u_k$ be incorporated into the rearranged sequence by means
of one of the ways described above. But we need to formulate conditions for defining
control functions $u_k=u_k(x)$, such that the sequence $\{F_k(x,u_k)\}$ be convergent.

By the Cauchy criterion the sequence $\{F_k(x,u_k)$ is convergent if and only if, for
any $\varepsilon > 0$ there exists a number $n_\varepsilon$ such that
\be
\label{2.11}
| F_{k+p}(x,u_{k+p}) - F_k(x,u_k) | < \ep
\ee
for all $k \geq n_\varepsilon$ and $p \geq 0$.

The Cauchy convergence criterion and the methods of optimal control theory motivate
us to define the optimal control functions serving as minimizers of the fastest
convergence cost functional \cite{Yukalov_11}
$$
  C[u] = \sum_{k=0}^\infty |F_{k+p}(x,u_{k+p}) - F_k(x,u_k) | \;  .
$$
This is equivalent to looking for control functions satisfying the {\it fastest
convergence criterion} that defines the control functions as the minimizers of the
difference of the terms in the Cauchy criterion:
\be
\label{2.12}
 \min_u |F_{k+p}(x,u_{k+p}) - F_k(x,u_k) | \qquad ( k \geq 0 , ~ p \geq 0 ) \;  .
\ee

The value of $F_k(x,u_k)$ varies under the variation of the control function $u_k$
and the variation of the approximation order index $k$. Therefore, approximately, we
can write
\be
\label{2.13}
 F_{k+p}(x,u_{k+p}) \cong F_{k+p}(x,u_k) + \frac{\prt F_k(x,u_k)}{\prt u_k} \;
(u_{k+p} - u_k ) \; .
\ee
Hence criterion (\ref{2.12}) takes the form
\be
\label{2.14}
 \min_u \left | F_{k+p}(x,u_k) - F_k(x,u_k) + \frac{\prt F_k(x,u_k)}{\prt u_k} \;
(u_{k+p} - u_k )  \right | \; .
\ee

If either the value of $F_k(x,u_k)$ weakly depends on $u_k$ or the control functions
slowly vary with varying the order index $k$, then the main term in expression
(\ref{2.14}) is the difference of the first two terms. Therefore we have to minimize
this difference by means of the {\it minimal difference condition}
\be
\label{2.15}
  \min_u  | F_{k+p}(x,u_k) - F_k(x,u_k) | \; .
\ee
The absolute minimum is given by the equation
\be
\label{2.16}
  F_{k+p}(x,u_k) - F_k(x,u_k) = 0 \;  .
\ee

On the other hand, if the value of $F_k(x,u_k)$ weakly varies with the change of the
the approximation order index $k$, then the main term in criterion (\ref{2.14}) is that
containing the derivative. Thence this term is to be minimized by the {\it minimal
derivative condition}
\be
\label{2.17}
 \min_u \left |  \frac{\prt F_k(x,u_k)}{\prt u_k} \; (u_{k+p} - u_k )  \right | \; .
\ee
The absolute minimum of the latter condition implies that
\be
\label{2.18}
 (u_{k+p} - u_k )\; \frac{\prt F_k(x,u_k)}{\prt u_k} = 0 \; .
\ee
This has to be understood in the following way. If the control functions of different
orders are not equal to each other, then the derivative has to be zero,
\be
\label{2.19}
  \frac{\prt F_k(x,u_k)}{\prt u_k} = 0 \qquad ( u_{k+p} \neq u_k ) \; .
\ee
But if the latter equation has no solutions for the control function $u_k = u_k(x)$,
then we can assume the equality of the control functions,
\be
\label{2.20}
 u_{k+p} = u_k \; , \qquad   \frac{\prt F_k(x,u_k)}{\prt u_k} \neq 0 \; .
\ee
When the derivative equation (\ref{2.19}) has no solution, the control functions can
be found from the minimization of this derivative,
\be
\label{2.21}
 \min_u \left | \frac{\prt F_k(x,u_k)}{\prt u_k} \right | \; , \qquad
u_k = u_k(x) \; .
\ee

As is easy to see, the minimal difference and minimal derivative conditions are
equivalent to each other. Really, if the minimal difference condition (\ref{2.16})
is valid, then the fastest convergence criterion (\ref{2.14}) yields the minimal
derivative condition (\ref{2.18}). While, if the latter condition holds true, then
the minimal difference condition (\ref{2.16}) follows. In practice, one can employ
that condition which is more convenient. These conditions are analogous to fixed-point
conditions in the renormalization group approach (see Sec. 4).

\subsection{Optimization Conditions}

The previous section shows how the fastest convergence criterion generates the
optimization conditions defining control functions. Two such optimization conditions,
being equivalent to each other, are obtained: the minimal difference and minimal
derivative conditions. These conditions follow from the fastest convergence criterion
(\ref{2.12}), with the use of representation (\ref{2.13}). This representation has the
meaning of the low-order expansion of $F_k(x,u_k)$ as a function of two variables,
the approximation order $k$ and the variable $u_k$, while $x$ being fixed.

Generally, a function $f(t,z)$ of two continuous variables can be represented as a
Taylor expansion with respect to points $t_0$ and $z_0$,
$$
 f(t,z) = \sum_{m,n=0}^\infty \frac{1}{m!n!} \;
\left [ \frac{\prt^{m+n}f(t,z)}{\prt t^m\prt z^n} \right ]_{t_0,z_0}
(t - t_0)^m ( z - z_0)^n \;.
$$
In our case, one of the variables, that is $k$, is discrete. Therefore for this
variable, instead of the derivative, we need to employ the finite difference
$$
\Dlt_p F_k \equiv F_{k+p} - F_k \;   .
$$
In the second order, the finite difference is
$$
 \Dlt^2_p F_k = F_{k+2p} - 2F_{k+p} + F_k \;  ,
$$
and  so on. Then the general expansion of $F_{k+p}(x,u_{k+p})$ reads as
\be
\label{2.22}
 F_{k+p}(x,u_{k+p}) = \sum_{m,n=0}^\infty \frac{\Dlt^m_p}{m!n!} \;
\frac{\prt^n F_k(x,u_k)}{\prt u^n_k} \; ( u_{k+p} - u_k)^n \; .
\ee
In the first order, we return back to representation (\ref{2.13}). In the second order, we
have
$$
F_{k+p}(x,u_{k+p}) \cong F_{k+p}(x,u_k) +
\frac{\prt F_k(x,u_k)}{\prt u_k} \; ( u_{k+p} - u_k) +
\frac{1}{2} \frac{\prt^2 F_k(x,u_k)}{\prt u_k^2} \; ( u_{k+p} - u_k)^2 +
$$
\be
\label{2.23}
 +
\frac{1}{2} [ F_{k+2p}(x,u_k) - 2F_{k+p}(x,u_k) + F_k(x,u_k) ] +
 ( u_{k+p} - u_k) \; \frac{\prt}{\prt u_k} \; [ F_{k+p}(x,u_k)  - F_k(x,u_k) ]  \;.
\ee

With the general expansion (\ref{2.22}), the fastest convergence criterion (\ref{2.12})
leads to the minimization
\be
\label{2.24}
 \min_u | F_{k+p}(x,u_{k+p}) - F_k(x,u_k) | \leq \min_u
\sum_{m,n=0}^\infty \frac{|u_{k+p}-u_k|^n}{m!n!} \; \left |
\frac{\prt}{\prt u_k^n}\; \Dlt_p^m F_k(x,u_k) \right | \; .
\ee
To second order, this gives
$$
\min_u | F_{k+p}(x,u_{k+p}) - F_k(x,u_k) | \leq
\min_u \left\{ | F_{k+p}(x,u_k) - F_k(x,u_k) | +
\left | ( u_{k+p} - u_k )\; \frac{\prt F_k(x,u_k)}{\prt u_k}  \right | +
\right.
$$
$$
+
\frac{1}{2} \; | F_{k+2p}(x,u_k) - 2F_{k+p}(x,u_k) + F_k(x,u_k) | +
\frac{1}{2}\; \left | ( u_{k+p} - u_k )^2\; \frac{\prt^2 F_k(x,u_k)}{\prt u^2_k}\right |
+
$$
\be
\label{2.25}
+  \left.
\left | ( u_{k+p} - u_k ) \frac{\prt}{\prt u_k}
[ F_{k+p} (x,u_k) -  F_k(x,u_k)] \right | \right \} \;  .
\ee
Minimizing here the first difference term, we get the minimal difference condition
(\ref{2.16}). The minimization of the second term yields the minimal derivative
condition (\ref{2.18}) or (\ref{2.19}). When these conditions do not have solutions
or suffer from the multiplicity of solutions, one can resort to the minimization of
other terms in criterion (\ref{2.24}). In that way, we can employ the higher-order
minimal difference conditions, as
\be
\label{2.26}
 F_{k+2p}(x,u_k) - 2F_{k+p}(x,u_k) + F_k(x,u_k) = 0 \;  ,
\ee
the higher-order minimal derivative conditions
\be
\label{2.27}
 \frac{\prt^n F_k(x,u_k)}{\prt^n u_k} = 0 \qquad ( n = 1,2,\ldots ) \; ,
\ee
and mixed minimal derivative-difference conditions, such as
\be
\label{2.28}
 \frac{\prt}{\prt u_k} \; [ F_{k+p}(x,u_k) - F_k(x,u_k) ] = 0 \;  .
\ee

Note that when the considered problem can be described by several quantities of
interest, say $f(x)$ and $h(x)$, it is admissible to introduce control functions for
one of these sought quantities, also using it for the other. That is, if we have the
sequences $F(x,u_k)$ and $H(x,v_k)$, corresponding to $f(x)$ and $h(x)$, but associated
with the same problem, we may define control functions $u_k = u_k(x)$ and then set
$u_k(x) = v_k(x)$. This is allowed because both these sought quantities follow from
the same problem and in that sense are interrelated. For instance, studying various
characteristics of an anharmonic system, we can define the control functions through
the optimization conditions related to the moments $\lgl r^n\rgl $ characterizing the
properties of wave functions
\cite{Yukalov_4,Yukalov_5,Yukalov_6,Yukalov_7,Yukalov_8,Yukalov_9,Yukalov_10}.

\subsection{Asymptotic Conditions}

When the behaviour of the sought function $f(x)$ in the vicinity of a point $x_0$ is
known, so that
\be
\label{2.29}
 f(x) \simeq g(x) \qquad ( x \ra x_0 ) \;  ,
\ee
then control functions can be found from the asymptotic condition
\be
\label{2.30}
  F_k(x,u_k) \simeq g(x) \qquad ( x \ra x_0 ) \; .
\ee

If the behaviour of the sought function is known near several points $x_i$, with
$i = 1,2,\ldots$, where
\be
\label{2.31}
  f(x) \simeq g_i(x) \qquad ( x \ra x_i ) \;  ,
\ee
then it is possible to introduce several control functions satisfying the given asymptotic
conditions
\be
\label{2.32}
   F_k(x,u_{1k},u_{2k},\ldots) \simeq g_i(x) \qquad ( x \ra x_i )
\ee
defining $u_{ik} = u_{ik}(x)$. In that case, the optimized approximant is
\be
\label{2.33}
  \tilde f_k(x) =  F_k(x,u_{1k}(x),u_{2k}(x),\ldots) \; .
\ee

\subsection{Partition Functions}

Optimized perturbation theory has been used for a variety of problems. To show
how the theory works, it is instructive to illustrate it by several simple examples,
whose mathematical structure is typical of many problems of statistical physics and
field theory.

One such a case illustrates the calculation of partition functions in statistical
physics or effective potentials in field theory. Let us consider the so-called
zero-dimensional  $\vp^4$ theory represented by the partition function
\be
\label{2.34}
  Z(g) = \frac{1}{\sqrt{\pi}} \int_{-\infty}^\infty e^{-H[\vp]}\; d\vp \; ,
\ee
with the effective Hamiltonian
\be
\label{2.35}
  H[\vp] = \vp^2 + g \vp^4 \; ,
\ee
where $g \geq 0$ is a coupling parameter. This example has been treated in
several publications, e.g. in \cite{Kleinert_2,Yukalov_12}. Below we follow
Ref. \cite{Yukalov_12}.

Perturbation theory for this problem can be compared with the known exact expression
\be
\label{2.36}
 Z(g) = \frac{1}{\sqrt{4\pi g}} \; \exp\left ( \frac{1}{8g} \right )
K_{1/4}\left ( \frac{1}{8g} \right ) =
\frac{1}{(2g)^{1/4}} \exp\left ( \frac{1}{8g} \right )
D_{-1/2} \left ( \frac{1}{\sqrt{2g} } \right ) \; ,
\ee
in which $K_\nu$ is a MacDonald function and  $D_\nu$ is the function of parabolic
cylinder. For a large variable $x \ra \infty$, we have
$$
 K_{\nu}(x) \simeq  \left ( \frac{\pi}{2x } \right )^{1/2} e^{-x} \left [
1 + \frac{4\nu^2-1}{8x} + \frac{(4\nu^2-1)(4\nu^2-3^2}{2(8x)^2} \right ]
\qquad ( x \ra \infty ) \;  ,
$$
from where the weak-coupling expansion of $Z(g)$ becomes
\be
\label{2.37}
 Z(g) \simeq 1 - \; \frac{3}{4}\; g + \frac{105}{32} \; g^2 \qquad ( g\ra 0 ) \; .
\ee
For the small variable $x\ra 0$, the MacDonald function is
$$
 K_{\nu}(x) \simeq \frac{\pi}{2\sin(\pi\nu)} \left [  \frac{(x/2)^{-\nu}}{\Gm(1-\nu)}
\; - \; \frac{(x/2)^{\nu}}{\Gm(1+\nu)}\; + \; \frac{(x/2)^{2-\nu}}{\Gm(2-\nu)}\; -
\; \frac{(x/2)^{2+\nu}}{\Gm(2+\nu)} \right ] \qquad (x \ra 0 ) \; ,
$$
where
$$
 \Gm(x) \Gm(1-x) = \frac{\pi}{\sin(\pi x)} \;  .
$$
Then the strong-coupling limit of $Z(g)$ reads as
\be
\label{2.38}
 Z(g) \simeq \frac{1}{2\sqrt{\pi}} \; \left [ \Gm \left ( \frac{1}{4} \right ) g^{-1/4}
- \Gm \left ( \frac{3}{4} \right ) g^{-3/4} \right ] \qquad ( g \ra \infty) \; .
\ee
With the values of the gamma function
$$
 \Gm \left ( \frac{1}{4} \right ) = 3.625609 \; , \qquad
\Gm \left ( \frac{3}{4} \right ) = 1.225417  \; ,
$$
we get
\be
\label{2.39}
 Z(g) \simeq 1.022765 g^{-1/4} - 0.345684 g^{-3/4} \qquad (g \ra \infty )\;  .
\ee

If we employ the expansion in powers of $g$, then, using the integral
$$
  \int_{-\infty}^\infty x^{2n} \exp(-x^2) \; dx =
\frac{\sqrt{\pi}}{2^n} \; (2n-1)!! = \Gm\left ( n+ \frac{1}{2} \right ) \; ,
$$
we come to the sum
\be
\label{2.40}
 Z(g) = \sum_{n=0}^\infty \frac{(-1)^n}{\sqrt{\pi}\; n!} \;
\Gm\left ( 2n+ \frac{1}{2} \right ) g^n \;   .
\ee
This sum is strongly divergent at any nonzero $g$ because of the factorial growth
of the coefficients
$$
 \frac{1}{n!}\; \Gm\left ( 2n+ \frac{1}{2} \right )\simeq
\left ( \frac{4n}{e} \right )^n \ra \infty \qquad ( n \ra \infty)  \;   .
$$
Such a divergence is typical of many problems in statistical physics and field theory.

In the optimized perturbation theory, we start with the initial Hamiltonian
\be
\label{2.41}
  H_0[\vp] = u + \om^2\vp^2 \; ,
\ee
containing the trial parameters $u$ and $\omega$. Hamiltonian (\ref{2.35}) can be
rewritten as
\be
\label{2.42}
 H[\vp] = H_0[\vp] + \ep \Dlt H \qquad ( \ep \ra 1 ) \;  ,
\ee
with the perturbative term
\be
\label{2.43}
  \Dlt H =  H[\vp] - H_0[\vp] = u + ( \om^2 - 1 ) \vp^2 - g \vp^4 \; .
\ee

Expanding the partition function (\ref{2.34}) in powers of $\varepsilon$ and setting
the latter to one, we find the successive terms $Z_k(g,u,\omega)$ of zero order
\be
\label{2.44}
 Z_0(g,u,\om) = \frac{e^{-u}}{\om} \; ,
\ee
first order
\be
\label{2.45}
  Z_1(g,u,\om) = Z_0(g,u,\om) + e^{-u} \left ( \frac{u}{\om} +
\frac{\om^2-1}{2\om^3} \; - \; \frac{3g}{4\om^5} \right ) \; ,
\ee
second order
$$
 Z_2(g,u,\om) = Z_1(g,u,\om) + e^{-u} \left [ \frac{u^2}{2\om} +
\frac{u(\om^2-1)}{2\om^3}  \right. +
$$
\be
\label{2.46}
   \left.   +
\frac{3(\om^2-1)^2-6gu}{8\om^5} \; - \;
\frac{15g(\om^2-1)}{8\om^7} + \frac{105g^2}{32\om^9}   \right ] \;  ,
\ee
and so on.

We can define the control functions $u_k = u_k(g)$ and $\omega_k = \omega_k(g)$ from
the optimization conditions
\be
\label{2.47}
 \frac{\prt}{\prt u_k} \; Z_k(g,u_k,\om_k) = 0 \; , \qquad
\frac{\prt}{\prt \om_k} \; Z_k(g,u_k,\om_k) = 0 \; .
\ee
To first order $k =1$, these conditions yield the equations
\be
\label{2.48}
 2 ( 1 + 2u) \om^4 - 2\om^2 - 3g = 0\; , \qquad
2 ( 3 + 2u) \om^4 - 6\om^2 - 15g = 0\; ,
\ee
in which $u = u_1$ and $\omega = \omega_1$. From these equations, one has the
relations
\be
\label{2.49}
\om^2 = \frac{1}{1+4u} \; , \qquad u = - \; \frac{\om^2-1}{4\om^2} = -\;
\frac{3g}{4\om^4} \; ,
\ee
using which, we come to the equation
\be
\label{2.50}
  \om^4 - \om^2 - 3g = 0 \; .
\ee
The real solution of the latter equation gives the control function
\be
\label{2.51}
 \om(g) = \frac{1}{\sqrt{2}} \; ( 1 + \sqrt{1+12g} )^{1/2} \;  ,
\ee
from where the other control function is
\be
\label{2.52}
 u(g) = -\;\frac{3g}{(1 + \sqrt{1+12g})^2 } \; .
\ee

In the weak-coupling limit, the control functions behave as
$$
\om(g) \simeq 1 + \frac{3}{2}\; g \; - \; \frac{45}{8} \; g^2 \; ,
$$
\be
\label{2.53}
 u(g) \simeq -\; \frac{3}{4}\; g + \frac{9}{2} \; g^2 - 27 g^3 \qquad
( g \ra 0) \;  ,
\ee
and in the strong-coupling limit, as
$$
\om(g) \simeq (3g)^{1/4} + \frac{1}{4}\; (3g)^{-1/4}  +
\frac{1}{32} \; (3g)^{-3/4} \; ,
$$
\be
\label{2.54}
u(g) \simeq -\; \frac{1}{4} + \frac{1}{4}\; (3g)^{-1/2} +
\frac{9}{16} \; (3g)^{-1} \qquad (g \ra \infty )\;   .
\ee

In the second order $k=2$, conditions (\ref{2.47}) do not provide real solutions, because
of which we can set $u_2(g)=u_1(g)\equiv u(g)$ and $\om_2(g)=\om_1(g)\equiv\om(g)$. Then
the optimized approximants for the first and second orders are
\be
\label{2.55}
\tilde Z_k(g) \equiv Z_k(g,u(g),\om(g)) \qquad ( k = 1,2) \;   .
\ee
In the weak-coupling limit, these approximants behave as
\be
\label{2.56}
\tilde Z_1(g) \simeq 1 - \; \frac{3}{4}\; g + \frac{81}{32}\; g^2 \; ,  \quad
\tilde Z_2(g) \simeq 1 - \; \frac{3}{4}\; g + \frac{105}{32}\; g^2 \qquad ( g \ra 0)
\ee
and in the strong-coupling limit, as
$$
 \tilde Z_1(g) \simeq e^{1/4} \left [ (3g)^{-1/4} -\;
\frac{1}{4}\; (3g)^{-3/4}\right ] \; ,
$$
\be
\label{2.57}
 \quad
\tilde Z_2(g) \simeq e^{1/4} \left [ \frac{13}{12}\;(3g)^{-1/4} -\;
\frac{7}{6}\; (3g)^{-3/4}\right ] \qquad ( g \ra \infty ) \; .
\ee
Simplifying the latter, we get
$$
\tilde Z_1(g) \simeq 0.975648 g^{-1/4} - 0.140823 g^{-3/4} \; ,
$$
\be
\label{2.58}
 \tilde Z_2(g) \simeq 1.056952 g^{-1/4} - 0.246440 g^{-3/4} \qquad
( g \ra \infty ) \; .
\ee

The optimized approximants provide good approximations in the whole range of the
coupling parameter $ g \in [0, \infty)$. The overall accuracy of the approximants
can be characterized by the maximal error
\be
\label{2.59}
 \ep_k \equiv \sup_g \left | \frac{\tilde Z_k(g)}{Z(g)} \; - \; 1
\right | \times 100\% \; .
\ee
For the approximants (\ref{2.55}), we find $\ep_1=5\%$ and $\ep_2=3\%$.

Let us stress that here two control functions are used. One describing an effective
uniform potential $u_k$ and the other, an effective frequency (or mass) $\om_k$. If
we use only one control function, say $\omega_k$, setting $u_k \equiv 0$, the accuracy
of the approximants becomes slightly worse \cite{Yukalov_12}, giving $\ep_1=7\%$ and
$\ep_2=5\%$.

\subsection{Thermodynamic Potentials}

A dimensionless thermodynamic potential can be expressed through the partition
function by the definition
\be
\label{2.60}
 f(g) = - \ln Z(g) \;  .
\ee
In field theory, this is called a generating functional.

Let us accept the partition function (\ref{2.34}) from the previous section.
Following Ref. \cite{Yukalov_13}, we take for the zero approximation the Hamiltonian
\be
\label{2.61}
 H_0[\vp ] = \om^2 \vp^2 \;  ,
\ee
containing one trial parameter $\omega$. The total Hamiltonian takes the form
\be
\label{2.62}
 H[\vp ] = H_0[\vp ] + \ep \Dlt H \qquad ( \ep \ra 1 ) \; ,
\ee
with the perturbative term
\be
\label{2.63}
 \Dlt H = H[\vp ] - H_0[\vp ] = ( 1 - \om^2) \vp^2 + g\vp^4 \;  .
\ee

We expand the free energy (\ref{2.60}) in powers of $\varepsilon$. Starting with the
zero-order free energy
\be
\label{2.64}
 F_0(g,\om) = \ln \om \;  ,
\ee
and introducing the notations
\be
\label{2.65}
 \al \equiv 1 \; - \; \frac{1}{\om^2} \; ,\qquad
\bt \equiv  \frac{3g}{\om^4} \; ,
\ee
we find in the higher orders
\be
\label{2.66}
 F_k(g,\om) = F_{k-1}(g,\om) + \sum_{n=0}^k c_{kn} \al^{k-n} \bt^n \;  ,
\ee
where $k \geq 1$. The coefficients $c_{kn}$, depending on the order $k$, are here
as follows: to first order,
$$
 c_{10} = -\; \frac{1}{2} \; , \qquad  c_{11} = \frac{1}{4} \qquad ( k = 1 ) \; ,
$$
to second order,
$$
 c_{20} = -\; \frac{1}{4} \; , \qquad  c_{21} = \frac{1}{2} \; ,   \qquad
c_{22} = -\; \frac{1}{3}         \qquad ( k = 2 ) \;   ,
$$
to third order,
$$
  c_{30} = -\; \frac{1}{6} \; , \qquad  c_{31} = \frac{3}{4} \; , \qquad
c_{32} = -\; \frac{4}{3} \; , \qquad  c_{33} = \frac{11}{12} \qquad ( k = 3 ) \; ,
$$
to fourth order,
$$
c_{40} = -\; \frac{1}{8} \; , \qquad c_{41} = 1 \; , \qquad
c_{42} = -\; \frac{10}{3} \; , \qquad  c_{43} = \frac{11}{2} \; , \qquad
c_{44} = -\; \frac{34}{9}  \qquad ( k = 4 ) \;   ,
$$
and so on.

The control functions $\omega_k = \omega_k(g)$ for odd orders are defined by the
optimization condition
\be
\label{2.67}
  \frac{\prt F_k(g,\om_k)}{\prt \om_k} = 0 \qquad ( k = 1,3,\ldots ) \;  .
\ee
For even orders, the similar condition has no real solutions, and we set
$u_k=u_{k-1}(g)$ for $k = 2,4,\ldots$.

Equation (\ref{2.67}) yields
\be
\label{2.68}
 \om_k(g) = \left [ \frac{1}{2}\;\left ( 1 + \sqrt{1+ 12s_k g} \right ) \right ]^{1/2} \; ,
\ee
where
$$
s_1 = s_2 = 1 \; , \qquad s_3 = s_4 = 2.239674 \; .
$$
For large orders $k$, we have
$$
 \om_k(g) \sim g^{1/4} k^\gm \qquad
\left ( \frac{1}{4} \leq \gm \leq \frac{1}{2} \; , ~ k \ra \infty \right ) \;  .
$$

The optimized approximant is
\be
\label{2.69}
 \tilde f_k(g) = F_k(g,\om_k(g)) \;  .
\ee
Its accuracy can be characterized by the maximal error (\ref{2.59}) giving
$$
\ep_1 = 7\% \; , \qquad  \ep_2 = 4\% \; , \qquad \ep_3 = 0.2\% \; , \qquad
\ep_4 = 0.2\% \; .
$$

\subsection{Eigenvalue Problem}

One often needs to find the eigenvalues of Hermitian operators representing observable
quantities. Most often, one looks for the eigenvalues of Hamiltonians, describing energy
spectra and satisfying the Schr\"{o}dinger equation
\be
\label{2.70}
 \hat H | n \rgl =  E_n | n \rgl \;  .
\ee

The common touchstones for checking the accuracy of calculational procedures are
anharmonic models, such as the anharmonic oscillator for a single particle and
anharmonic crystals for many-body systems. The simplest case corresponds to the
one-dimensional anharmonic oscillator with the Hamiltonian
\be
\label{2.71}
\hat H = - \; \frac{1}{2} \; \frac{d^2}{dx^2} + \frac{1}{2} \; x^2 + g x^4
\ee
in dimensionless units, where $x\in (-\infty,\infty)$ and the anharmonicity parameter
$g \in [0, \infty)$. Optimized perturbation theory has been applied to both anharmonic
crystals \cite{Yukalov_3,Yukalov_4,Yukalov_5,Yukalov_6,Yukalov_7,Yukalov_8,Yukalov_9}
and to the simple anharmonic oscillator \cite{Kleinert_2,Caswell_14,Seznec_15,Halliday_16,
Killinbeck_17,Stevenson_18,Feranchuk_19,Okopinska_20,Dineykhan_21,Dineykhan_92,Feranchuk_22}.
Usually, one considers only the ground-state energy of Hamiltonian (\ref{2.71}). Here we
show the application of the theory to the whole spectrum of this Hamiltonian, as is done
in Ref. \cite{Yukalov_23}.

Choosing for the initial approximation the Hamiltonian
\be
\label{2.72}
 \hat H_0 = - \; \frac{1}{2} \; \frac{d^2}{dx^2} + \frac{\om^2}{2} \; x^2 \; ,
\ee
we have the perturbative term
\be
\label{2.73}
 \Dlt H \equiv \hat H - \hat H_0 = \frac{1-\om^2}{2} \; x^2 + g x^4 \; .
\ee
So, the total Hamiltonian can be written as
\be
\label{2.74}
 \hat H = \hat H_0 +\ep \Dlt H \qquad ( \ep \ra 1 ) \; .
\ee

Employing the Rayleigh-Scr\"{o}dinger perturbation theory with respect to the dummy
parameter $\varepsilon$, which afterwards is set to one, we find the spectrum of the
$k$-th order
\be
\label{2.75}
  E_{kn} = E_{kn}(g,\om) \qquad ( k = 0,1,2,\ldots ) \; ,
\ee
with the quantum index $n = 0,1,2,\ldots$ enumerating the energy levels. The spectrum
can be represented as the sum
\be
\label{2.76}
  E_{kn} = E_{0n}  + \Dlt_{1n} + \Dlt_{2n} + \ldots + \Dlt_{kn} \; ,
\ee
starting from the zero approximation
\be
\label{2.77}
 E_{0n} = \left ( n + \frac{1}{2} \right ) \om \;   ,
\ee
and where
\be
\label{2.78}
\Dlt_{kn} =  \Dlt_{kn}(g,\om)
\ee
are the perturbative corrections.

In the first order, we have
\be
\label{2.79}
 \Dlt_{1n} = V_{nn} \; , \qquad V_{mn} \equiv \lgl m | \Dlt H | n \rgl \;  .
\ee
The second order correction is
\be
\label{2.80}
 \Dlt_{2n}  = \sum_{m(\neq n)} \frac{|V_{nm}|^2}{E_{0n}-E_{0m} } \; .
\ee
The third order correction writes as
\be
\label{2.81}
 \Dlt_{3n}  = \sum_{m,i(\neq n)}
\frac{V_{nm}V_{mi}V_{in}}{(E_{0n}-E_{0m})(E_{0n}-E_{0i}) } - V_{nn} \sum_{m(\neq n)}
\frac{|V_{nm}|^2}{(E_{0n}-E_{0m})^2 } \;  .
\ee
And to fourth order, we get
$$
\Dlt_{4n}  = \sum_{m,i,j(\neq n)}
\frac{V_{nm}V_{mi}V_{ij}V_{jn}}{(E_{0n}-E_{0m})(E_{0n}-E_{0i})(E_{0n}-E_{0j}) } -
$$
\be
\label{2.82}
-
\sum_{m,i(\neq n)} \frac{|V_{nm}|^2 |V_{ni}|^2+ 2V_{nn}V_{nm}V_{mi}V_{in} }
{(E_{0n}-E_{0m})^2 (E_{0n}-E_{0i})^2}  +
| V_{nn} |^2 \sum_{m(\neq n)} \frac{|V_{nm}|^2}{(E_{0n}-E_{0m})^3} \;  .
\ee

It is convenient to introduce the notation
\be
\label{2.83}
 E_{kn}(g,\om) = \left ( n + \frac{1}{2} \right ) F_k(g,\om) \;  ,
\ee
dealing with the function $F_k(g, \omega)$, whose zero order is
\be
\label{2.84}
 F_0(g,\om) = \om \qquad ( k = 0) \; .
\ee
For the orders $k \geq 1$, we have
\be
\label{2.85}
 F_k(g,\om) = F_{k-1}(g,\om) - \;
\frac{\om}{2^{k+1}} \sum_{n=0}^k d_{kn} \al^{k-n} \bt^n \;  ,
\ee
where
\be
\label{2.86}
 \al \equiv 1 - \; \frac{1}{\om^2} \; , \qquad \bt \equiv 6\gm \; \frac{g}{\om^3} \; ,
\qquad \gm \equiv \frac{n^2+n+1/2}{n+1/2} = n+ \frac{1}{2} + \frac{1}{4(n+1/2)} \; ,
\ee
and the coefficients, depending on the order, are as follows. In the first order,
$$
d_{10} = 2 \; , \qquad d_{11} = -1 \qquad ( k = 1 ) \;   ,
$$
in the second order,
$$
d_{20} = 1 \; , \qquad d_{21} = -2 \; \qquad d_{22} = 2a \qquad ( k = 2 ) \; ,
$$
in the third order,
$$
d_{30} = 1 \; , \qquad d_{31} = -4 \; \qquad d_{32} = 10a \; , \qquad
d_{33} = -3b \qquad ( k = 3 ) \;    ,
$$
and in the fourth order,
$$
 d_{40} = \frac{5}{4} \; , \qquad d_{41} = -8 \; \qquad d_{42} = 35a \; , \qquad
d_{43} = -24b \; , \qquad d_{44} = 4c \qquad ( k = 4 ) \;   ,
$$
where the notation is used:
$$
a \equiv \frac{17n^2+17n+21}{(6\gm)^2} \; ,  \qquad
b \equiv \frac{125n^4 +250n^3 +472n^2+347n+111}{(n+1/2)(6\gm)^3} \; ,
$$
$$
c \equiv \frac{1}{8 (6\gm)^4} \left (
10689 n^4 + 21378 n^3 + 60616 n^2 + 49927 n + 30885 \right ) \; .
$$

The first-order control function is found from the optimization condition
\be
\label{2.87}
  \frac{\prt}{\prt\om} \; F_{1n}(g,\om) = 0 \; , \qquad \om = \om_1(g) \;  ,
\ee
which gives the equation
\be
\label{2.88}
 \om_1^3 - \om_1 - 6\gm g = 0 \;  .
\ee
The values
\be
\label{2.89}
 \al_k \equiv \al(\om_k) \; , \qquad   \bt_k \equiv \bt(\om_k)
\ee
are connected as
\be
\label{2.90}
 \al_1 = \bt_1 = 1 - \; \frac{1}{\om_1^2} \;  .
\ee
In the second order, the equation similar to (\ref{2.87}) does not have real
solutions, so we set $\om_2=\om_1$. Thus, for the first and second order, function
(\ref{2.85}) becomes
\be
\label{2.91}
F_{k+1}(g,\om_1) =
F_k(g,\om_1) - \; \frac{\om_1}{2^{k+2}} A_{1k} \al_1^{k+1} \;   ,
\ee
where $k=0,1$ and
$$
 A_{10} = 1 \; , \qquad A_{11} = 2a -1 \;  .
$$

In the third order, the optimization condition
\be
\label{2.92}
\frac{\prt}{\prt\om} \; F_3(g,\om) = 0 \; , \qquad \om = \om_3(g)
\ee
yields the equation
\be
\label{2.93}
 \om_3^3 - \om_3 - 6\gm\lbd  g = 0 \;  ,
\ee
in which the quantity
\be
\label{2.94}
\lbd \equiv \frac{\al_3}{\bt_3} = \frac{\om_3^2-1}{6\gm g} \; \om_3
\ee
satisfies the equation
\be
\label{2.95}
  5\lbd^3 - 24\lbd^2 +70 a\lbd - 24 b = 0 \; .
\ee
Now the parameters from Eq. (\ref{2.86}) are connected as
\be
\label{2.96}
  \al_3 = \al(\om_3) = 1 - \; \frac{1}{\om_3^2} \; , \qquad
\bt_3 \equiv \bt(\om_3) =  \frac{\al_3}{\lbd} \; .
\ee
In the fourth order, the equation similar to condition (\ref{2.92}) does not have real
solutions, because of which we can set $\omega_4 = \omega_3$.

Then, for the third and fourth orders, function (\ref{2.85}) reduces to the form
\be
\label{2.97}
F_{k+1}(g,\om_3) =
F_k(g,\om_3) - \; \frac{\om_3}{2^{k+2}} A_{3k} \al_3^{k+1} \;   ,
\ee
in which $k = 2,3$ and
$$
 A_{32} = 1 - \; \frac{4}{\lbd} + \frac{10a}{\lbd^2} \; - \; \frac{3b}{\lbd^3} \; ,
\qquad
A_{33} = \frac{5}{4} \; - \; \frac{8}{\lbd} + \frac{35a}{\lbd^2} \; - \;
\frac{24b}{\lbd^3} + \frac{4c}{\lbd^4}\; .
$$

The optimized approximant for the spectrum reads as
\be
\label{2.98}
\tilde E_{kn}(g) = \left ( n + \frac{1}{2} \right ) F_k(g,\om_k(g)) \; .
\ee
The accuracy of a $k$-th order approximant can be characterized by the percentage
error
\be
\label{2.99}
  \ep_{kn}(g) \equiv \left [ \frac{ \tilde E_{kn}(g)}{ E_n(g) } \; - \; 1
\right ] \times 100 \%
\ee
compared with the numerical solution of the eigenvalue problem \cite{Hioe_24}
giving $E_n(g)$. It is also instructive to define the maximal error of a given energy
level
\be
\label{2.100}
\ep_{kn}(g) \equiv \sup_g | \ep_{kn}(g) |
\ee
and the maximal percentage error for all levels
\be
\label{2.101}
 \ep_k \equiv \sup_n \ep_{kn} = \sup_{n,g} | \ep_{kn}(g) | \;  .
\ee
The latter errors are
$$
 \ep_1 = 2\% , \qquad \ep_2 = 0.8\% , \qquad \ep_3 = 0.8\% , \qquad
\ep_4 = 0.5\% \; .
$$

The ground-state level is often of special interest. For this level, we get
$$
 \ep_{10} = 2\% , \qquad \ep_{20} = 0.8\% , \qquad \ep_{30} = 0.04\% , \qquad
\ep_{40} = 0.03\%  \; .
$$

Thus, the optimized approximants provide good accuracy for all coupling
parameters and all energy levels. This accuracy is essentially better than that
of quasi-classical approximation, whose maximal error is close to $20\%$. Pad\'e
approximants are also much less accurate. For instance the best Pad\'e approximant
of $10$-th order has the error of $3\%$.

\subsection{Exact Solutions}

When the studied problem enjoys an exact solution, the optimized perturbation theory
can recover this solution. To illustrate this, let us consider the case treated in
Ref. \cite{Yukalov_25}.

Let us consider a two-frequency oscillator, with the Hamiltonian
\be
\label{2.102}
 \hat H = -\; \frac{1}{2m} \; \frac{d^2}{dx^2} + \frac{m\om_0^2}{2} \; x^2 +
\frac{m\om_1^2}{2} \; x^2 \;  ,
\ee
in which $x \in (-\infty, \infty)$, and $\omega_0$ and $\omega_1$ are different
frequencies. We can treat the term with $\omega_0$ as the main, while that with
$\om_1$ as a perturbation. The related perturbation parameter is defined as
\be
\label{2.103}
  g \equiv \frac{\om_1^2}{\om_0^2} \; .
\ee
The Hamiltonian (\ref{2.102}) takes the form
\be
\label{2.104}
  \hat H = -\; \frac{1}{2m} \; \frac{d^2}{dx^2} +
\frac{m\om_0^2}{2} \; (1+g) x^2 \;  .
\ee

Representing the spectrum as
\be
\label{2.105}
 E_n(g) = \left ( n + \frac{1}{2} \right ) \om_0 f(g) \;  ,
\ee
we need to find the function $f(g)$. The result of perturbation theory can be
compared with the exact solution
\be
\label{2.106}
 f(g) = \sqrt{1+g} \;  .
\ee

The first and second approximations of the simple perturbation theory in powers
of $g$ are
\be
\label{2.107}
 f_1(g) = 1 + \frac{g}{2} \; , \qquad
f_2(g) = 1 + \frac{g}{2} \; - \;  \frac{g^2}{8} \; ,
\ee
which is quite different from the exact solution (\ref{2.106}).

In the optimized perturbation theory, we start with the Hamiltonian
\be
\label{2.108}
 \hat H_0 = -\; \frac{1}{2m} \; \frac{d^2}{dx^2} + \frac{m\om^2}{2}\; x^2
\ee
containing a trial parameter $\omega$. Then the perturbative term is
\be
\label{2.109}
 \Dlt H \equiv \hat H - \hat H_0 = \frac{m\om_0^2}{2}\; (1 + g - u^2 ) x^2 \; ,
\ee
where
\be
\label{2.110}
  u \equiv \frac{\om}{\om_0}
\ee
is the dimensionless control parameter. Perturbation theory in powers of term
(\ref{2.109})
yields the spectrum
\be
\label{2.111}
 E_{kn}(g,u) = \left ( n + \frac{1}{2} \right ) \om_0 F_k(g,u) \; ,
\ee
with the perturbative expressions
$$
 F_0(g,u) = u \; , \qquad   F_1(g,u) = \frac{1+g+u^2}{2u} \; ,
$$
\be
\label{2.112}
 F_2(g,u) =  F_1(g,u) - \; \frac{(1+g-u^2)^2}{8u^3} \; ,
\ee
and so on.

The control function $u_k(g)$ can be found either from the minimal derivative
condition
\be
\label{2.113}
 \frac{\prt F_k(g,u_k)}{\prt u_k} = 0 \; , \qquad u_k = u_k(g) \;  ,
\ee
or from the minimal difference condition
\be
\label{2.114}
 F_{k+1}(g,u_k) -  F_k(g,u_k) = 0 \; , \qquad u_k = u_k(g) \;  .
\ee
In both the cases and in all orders, we get
\be
\label{2.115}
 u_k(g) = \sqrt{1+g} \;  .
\ee
Then the optimized approximants of any order
\be
\label{2.116}
\tilde E_{kn}(g,u) = \left ( n + \frac{1}{2} \right ) \om_0 F_k(g,u_k(g))
\ee
coincide with the exact solution
\be
\label{2.117}
 E_n(g) = \left ( n + \frac{1}{2} \right ) \om_0 \; \sqrt{1+g} \;  .
\ee

From this example we see that an optimized approximant of a finite order $k$
corresponds to the standard perturbation theory of infinite order.

\subsection{Sequence Convergence}

As is shown in Sec. 2.2, control functions in the optimized perturbation theory
are defined by optimization conditions following from the fastest convergence
criterion. One therefore could expect that the resulting sequence of optimized
approximants should converge. But, since the optimization conditions are anyway
approximate, the convergence is not guaranteed for sure and one should check
numerical convergence for each concrete case.

For some simple models, for which the structure of approximants can be found
for large orders, it is possible to prove convergence analytically. These simple
models are the partition function integral of Sec. 2.5 and the one-dimensional
anharmonic oscillator of Sec. 2.7. In papers
\cite{Buckley_26,Duncan_27,Bender_28,Guida_29} the following has been proved.

\vskip 2mm
(i) Consider the partition function integral
\be
\label{2.118}
 Z(g) = \int_{-\infty}^\infty \exp ( - H[\vp] ) \;
\frac{d\vp}{\sqrt{2\pi} } \; ,
\ee
with the Hamiltonian
\be
\label{2.119}
 H[\vp] = \frac{1}{2}\; \om_0^2 \vp^2 + \frac{1}{4} \; g \vp^4 \;  .
\ee
The optimized perturbation theory, starting with the initial Hamiltonian
\be
\label{2.120}
 H_0[\vp] = \frac{1}{2}\;\om^2 \vp^2 \;  ,
\ee
with the trial parameter $\omega$ generating control functions $\omega_k(g)$, yields
the sequence of optimized approximants $\Tilde{Z}_k(g)$ converging to the exact values
defined by the expression
\be
\label{2.121}
  Z(g) =  \frac{1}{(2g)^{1/4}}\; \exp \left (  \frac{\om_0^4}{8g}  \right )
D_{-1/2} \left (  \frac{\om_0^4}{\sqrt{2g}}  \right ) \; ,
\ee
provided that either
\be
\label{2.122}
\frac{\om_k(g)}{\om_0} \simeq C k^\gm \qquad
\left ( \frac{1}{4} < \gm < \frac{1}{2}\; , ~ C > 0 , ~ k \ra \infty \right )
\ee
or
\be
\label{2.123}
 \frac{\om_k(g)}{\om_0} \simeq C k^{1/4} \qquad
\left (  C \geq \al_c g^{1/4}  , ~ k \ra \infty \right ) \; .
\ee
The minimal derivative optimization condition gives $\alpha_c = 0.972780$, and
the minimal difference condition, $\alpha = 1.072986$.

\vskip 2mm
(ii) Let us consider the one-dimensional anharmonic oscillator defined by the
Hamiltonian
\be
\label{2.124}
  \hat H = -\; \frac{1}{2} \; \frac{d^2}{dx^2} +
\frac{\om_0^2}{2} \; x^2 + \frac{g}{4} \; x^4 \;  .
\ee
The optimized perturbation theory, starting with the initial Hamiltonian
\be
\label{2.125}
  \hat H_0 = -\; \frac{1}{2} \; \frac{d^2}{dx^2} +
\frac{\om^2}{2} \; x^2 \;  ,
\ee
with the trial parameter $\omega$ generating control functions $\om_k(g)$,
results in the sequence of energy levels $\Tilde{E}_k(g)$ converging to the exact
numerical values $E_k(g)$, provided that either
\be
\label{2.126}
 \frac{\om_k(g)}{\om_0} \simeq C k^\gm \qquad
\left ( \frac{1}{3} < \gm < \frac{1}{2}\; , ~ C > 0 , ~ k \ra \infty \right )\;  ,
\ee
or
\be
\label{2.127}
\frac{\om_k(g)}{\om_0} \simeq C k^{1/3} \qquad
\left (  C \geq \al_c g^{1/3}  , ~ k \ra \infty \right )   .
\ee
Here $\alpha_c = 0.570875$ found from the minimal derivative condition.

Thus, in the simple cases, when the structure of large-order perturbative terms
is known, it is straightforward to prove the convergence of optimized approximants
to the exact numerical values:
\be
\label{2.128}
  \lim_{k\ra\infty} \tilde f_k(g) =  f(g) \;  .
\ee
Unfortunately, for realistic physical problems the large-order expansion terms
practically always are unknown. Moreover the exact solution may not be available.

\subsection{Reexpansion Trick: Induced Form}

In Sec. 2.1, it was mentioned that control functions can be introduced using
a reexpansion trick. Sometimes such a trick is not arbitrary but is directly
motivated by the chosen initial conditions.

For example, let us consider Hamiltonian (\ref{2.119}) from the previous section
and take for the initial Hamiltonian form (\ref{2.120}). According to the standard
procedure, one defines the Hamiltonian
\be
\label{2.129}
 H_\ep[\vp ] = H_0[\vp] + \ep( H[\vp] - H_0[\vp] ) \;  ,
\ee
which can be written as
\be
\label{2.130}
  H_\ep[\vp ] = \frac{1}{2} \left [ \ep \om_0^2 + ( 1 - \ep) \om^2
\right ] \vp^2  + \ep \; \frac{g}{4} \; \vp^4 \;  .
\ee
Comparing Hamiltonians (\ref{2.119}) and (\ref{2.130}), we see that passing from
$H[\varphi]$ to $H_\varepsilon[\varphi]$ implies the replacement
\be
\label{2.131}
 \om_0^2 \ra \ep \om_0^2 + ( 1  - \ep) \om^2 \; , \qquad g \ra \ep g \;  .
\ee
Therefore, if the terms $f_k(\om_0^2, g)$ of an expansion, derived for $H[\vp]$,
are known, then the related terms for $H_\varepsilon[\varphi]$ are given by the
transformation
\be
\label{2.132}
 f_k(\om_0^2,g) \ra f_k(\ep\om_0^2+(1-\ep)\om^2,\ep g) \;  .
\ee
Then the optimized perturbation theory is equivalent to the reexpansion procedure
in powers of $\varepsilon$,
\be
\label{2.133}
 f_k(\ep\om_0^2+(1-\ep)\om^2,\ep g) \simeq
\sum_{n=0}^k c_n(\om_0,g,\om) \ep^n \; .
\ee
Setting here $ \varepsilon = 1$, we get
\be
\label{2.134}
 F_k(\om_0,g,\om) = \sum_{n=0}^k c_n(\om_0,g,\om) \; .
\ee
Defining control functions $\omega_k(\omega_0, g)$, we come to the optimized
approximants
\be
\label{2.135}
 \tilde f_k(\om_0,g) = F_k(\om_0,g,\om_k(\om_0,g) ) \;  .
\ee
The same reexpansion trick can be used for the anharmonic oscillator with
Hamiltonian (\ref{2.124}).

In these examples, the reexpansion trick is motivated by the chosen initial
Hamiltonians. And the introduction of control functions through the reexpansion
trick is equivalent to their introduction through initial conditions. Reexpansion
trick can also be motivated by selecting the initial conditions taking account of
the symmetry properties of the Hamiltonian \cite{Weniger_91}.

\subsection{Reexpansion Trick: Arbitrary Form}

Generally, as is mentioned in Sec. 2.1, one can introduce control functions by
reexpansion tricks that are not motivated by the choice of initial conditions,
but whose form is rather arbitrary. We shall illustrate here this trick by the
Kleinert reexpansion \cite{Kleinert_2}.

Suppose the sought function is approximated by an expansion
\be
\label{2.136}
f_k(x) = \sum_{n=0}^k c_n x^n
\ee
derived for an asymptotically small variable $x \ra 0$. It is possible to invent
an identity rewriting $x$ in a different form, for instance as
\be
\label{2.137}
 x = \frac{u}{(1-z)^q} \;  ,
\ee
where
\be
\label{2.138}
  z \equiv 1 - \left ( \frac{u}{x} \right )^{1/q} \; .
\ee
Identity (\ref{2.137}) slightly differs form the original one \cite{Kleinert_2},
but completely equivalent to it \cite{Kleinert_30}. Thus we consider
\be
\label{2.139}
  F_k(x,u,q) \equiv f_k \left ( \frac{u}{(1-z)^q} \right ) \; ,
\ee
where $u$ and $q$ are trial parameters for generating control functions.

Expression (\ref{2.139}) is expanded in powers of $z$ up to order $k-n$, with
the use of the binomial expansion
$$
 (1-z)^p \simeq \sum_{m=0}^{k-n} C_m^p (-z)^m \qquad ( z \ra 0 ) \;  ,
$$
where $u$ and $q$ are fixed and
$$
C_m^p \equiv \frac{\Gm(p+1)}{\Gm(m+1)\Gm(p-m+1)} \;   .
$$
This yields
\be
\label{2.140}
  F_k(x,u,q)  \simeq \sum_{n=0}^k \;
\sum_{m=0}^{k-n}  (C_m^{-n} )^q (-z)^m c_n u^n \; ,
\ee
where $z$ is given by definition (\ref{2.138}). Then one should define control
functions $u_k(x)$ replacing the trial parameter $u$. And the parameter $q$ is
usually found from asymptotic conditions, as is exemplified below.

Suppose we need to find only the value of the sought function at infinity, where
this function is finite. And assume that the values of control functions $u_k(\infty)$
at infinity are finite. Because of this,
\be
\label{2.141}
 \lim_{x\ra\infty} z = \lim_{x\ra\infty}  \left [ 1 -
\left ( \frac{u_k}{x} \right )^{1/q}  \right ] = 1 \;  ,
\ee
where $u_k = u_k(\infty)$. Then Eq. (\ref{2.140}) gives
$$
F_1(\infty,u,q) = c_0 + c_1 u \; , \qquad
F_2(\infty,u,q) = F_1(\infty,u,q) + c_1 q u + c_2 u^2 \; ,
$$
\be
\label{2.142}
F_2(\infty,u,q) = F_2(\infty,u,q) + \frac{c_1}{2} \;  q(1+q) u + 2c_2q u^2
+ c_3 u^3 \;   ,
\ee
and so on.

The optimization condition
\be
\label{2.143}
\frac{\prt}{\prt u_2}\; F_2(\infty,u_2,q) = 0
\ee
defines the second-order control function
\be
\label{2.144}
 u_2 = -\; \frac{c_1}{2c_2} \; (1 + q) \; .
\ee

The optimization condition, similar to (\ref{2.143}), for the third order term
$F_3(\infty, u, q)$ displays two solutions. In order to avoid the multiplicity of
solutions, we can resort to the condition
\be
\label{2.145}
\frac{\prt^2}{\prt u_3^2}\; F_3(\infty,u_3,q) = 0
\ee
giving the unique solution
\be
\label{2.146}
 u_3 = -\; \frac{c_2}{3c_1} \; (1 + 2q) \;   .
\ee
In this way, we get
$$
F_2(\infty,u_2,q) = c_0 -\; \frac{c_1^2}{4c_2} \; (1 + q)^2 \; ,
$$
\be
\label{2.147}
 F_3(\infty,u_3,q) = c_0 -\; \frac{c_1c_2}{6c_3} \; (1 + q)(1+2q)(2+q)
+  \frac{2c_2^3}{27c_3^2} \; (1 + 2q)^2 \;  .
\ee

As is said above, the parameter $q$ should be found form asymptotic conditions.
Assume that we know the asymptotic behaviour of the sought function at infinity,
where
\be
\label{2.148}
 f(x) \simeq B x^\bt \qquad ( x \ra \infty ) \;  .
\ee
Hence the power-law at infinity is described by the exponent
\be
\label{2.149}
 \bt = \lim_{  x \ra \infty} \; \frac{d\ln f(x)}{d\ln x} \; .
\ee
Respectively, for each approximation of order $k$, we can introduce the function
\be
\label{2.150}
  \bt_k(x) \equiv  \frac{d\ln f_k(x)}{d\ln x}
\ee
reminding us the Gell-Mann-Low function (see Sec. 4). Expanding this in powers of
$x$ yields
\be
\label{2.151}
  \bt_k(x) \simeq \sum_{n=0}^k b_n x^n \qquad ( x \ra 0 ) \; ,
\ee
where the coefficients are defined through the coefficients $c_n$ of expansion
(\ref{2.136}), depending on whether $c_0 \neq 0$,
$$
b_0 = 0\; , \qquad b_1 = \frac{c_1}{c_0} \; , \qquad
b_2 = \frac{2c_2}{c_0}\; - \; \frac{c_1^2}{c_0^2} \;  ,   \qquad
b_3 = \frac{c_1^3}{c_0^3}\; - \; \frac{3c_1 c_2}{c_0^2}  + \frac{3c_3}{c_0}
\qquad ( c_0\neq 0) \;  ,
$$
or if $c_0 = 0$,
$$
 b_0 = 1\; , \qquad b_1 = \frac{c_2}{c_1} \; , \qquad
b_2 = \frac{2c_3}{c_1}\; - \; \frac{c_2^2}{c_1^2} \;  ,   \qquad
b_3 = \frac{c_2^3}{c_1^3}\; - \; \frac{3c_2 c_3}{c_1^2}  + \frac{3c_4}{c_1}
\qquad ( c_0 = 0) \;  .
$$

Then expansion (\ref{2.151}) is treated with the reexpansion trick, in order to
derive $B_k(\infty, q_k)$, in the same way as is described above, when obtaining
Eqs. (\ref{2.147}).
To second order, the result is
\be
\label{2.152}
 B_2(\infty,q) = b_0 - \; \frac{b_1^2}{4b_2} \; ( 1 + q )^2 \;  .
\ee
By assumption, we know the behaviour of the sought function at infinity, which
is prescribed by Eq. (\ref{2.148}). Therefore we have to set
\be
\label{2.153}
 B_2(\infty,q) = \bt \; , \qquad q = q_2 \;  ,
\ee
which results in
\be
\label{2.154}
 q_2 = 2 \; \sqrt{(b_0 - \bt)\; \frac{b_2}{b_1^2} } \; - \; 1 \;  .
\ee

Thus, the optimized approximant at infinity reads as
\be
\label{2.155}
  \tilde f_k(\infty) = F_k(\infty,u_k,q_k) \; .
\ee

Since the accepted representation (\ref{2.137}) is not unique, but suggested more
or less arbitrarily, one could take another representation. When the exponent $q$
cannot be defined from asymptotic conditions, it is admissible to fix it. As examples
of other representations that could be taken, it is possible to mention the conformal
map \cite{Parwani_31}
$$
 x = \frac{4uz}{(1-z)^2} \; , \qquad
z = \frac{\sqrt{x+u}-\sqrt{u}}{\sqrt{x+u}+\sqrt{u}} \;  ,
$$
or the more general form
$$
 x = u \left [  \left ( \frac{1+z}{1-z} \right )^q -1 \right ] \; , \qquad
 z = \frac{(x+u)^{1/q}-u^{1/q}}{(x+u)^{1/q}+u^{1/q}} \; .
$$
Of course, the results can strongly depend on the chosen representation.

\subsection{Critical Temperature: Initial Conditions}

Defining phase-transition temperature is one of the most important problems in
statistical physics. Here we discuss the use of the optimized perturbation theory
for calculating the Bose-Einstein condensation temperature of a three-dimensional
weakly nonideal homogeneous Bose gas. This problem turned out to be highly
nontrivial, since Bose-Einstein condensation is accompanied by strong fluctuations
making impossible the use of perturbation theory with respect to atomic interactions
(see review articles \cite{Yukalov_32,Andersen_33,Yukalov_34}).

The Bose-Einstein condensation temperature of an ideal gas is known to be
\be
\label{2.156}
 T_0 = \frac{2\pi}{m} \left [ \frac{\rho}{\zeta(3/2)} \right ]^{2/3} \;  ,
\ee
where $m$ is mass, $\rho$ is average atomic density, and $\zeta(\cdot)$ is the
Riemann zeta function. The problem of interest is how switching on weak interactions
changes the transition temperature $T_0$ of an ideal gas to the condensation
temperature $T_c$ of an interacting gas. This change can be described by the relative
shift
\be
\label{2.157}
 \frac{\Dlt T_c}{T_0} \equiv \frac{T_c - T_0}{T_0} \;  .
\ee

Atomic interactions in dilute Bose gas can be characterized by the local potential
\be
\label{2.158}
\Phi(\br) = 4\pi\; \frac{a_s}{m} \; \dlt(\br) \;   ,
\ee
in which $a_s$ is scattering length. A convenient dimensionless interaction
characteristic is the gas parameter
\be
\label{2.159}
 \gm \equiv \rho^{1/3} a_s \;  .
\ee
At asymptotically small gas parameter, the critical temperature shift behaves as
\be
\label{2.160}
   \frac{\Dlt T_c}{T_0} \simeq c_1 \gm + (c_2 + c_2'\ln \gm ) \gm^2 \qquad
(\gm \ra 0 ) \;  .
\ee
The coefficient $c_2^\prime$ can be found by means of perturbation theory
\cite{Arnold_35} giving
\be
\label{2.161}
 c_2' = - \; \frac{64\pi\zeta(1/2)}{3[ \zeta(3/2)]^{5/3}} = 19.7518 \; ,
\ee
while $c_1$ and $c_2$ cannot be found by perturbation theory. Numerical Monte Carlo
simulations for three dimensional lattice $O(2)$ field theory give $c_1=1.32\pm 0.02$
\cite{Arnold_36,Arnold_37} and $c_1=1.29\pm 0.05$ \cite{Kashurnikov_38,Prokofev_39},
while $c_2 = 75.7 \pm 0.4$ \cite{Arnold_35}.

Optimized perturbation theory, by introducing control functions into an initial
approximation, has been employed in Refs.
\cite{Cruz_40,Cruz_41,Kneur_42,Braaten_43,Braaten_44,Kneur_45,Kneur_46,Kneur_47,Farias_48}.

The problem of Bose-Einstein condensation in a homogeneous Bose gas can be reduced
to considering the phase transition in the three-dimensional $O(N)$ scalar field
theory with the Euclidean action
\be
\label{2.162}
 S[\vp] = \int \left [ \frac{1}{2} \left ( \frac{\prt\vp}{\prt x}\right )^2 +
\frac{m^2}{2} \; \vp^2 + \frac{\lbd}{4!}\; \vp^4 \right ] \; dx \;  ,
\ee
in which
$$
 \vp = \vp(x) = \{ \vp_n(x) : \; n = 1,2,\ldots, N\} \;  , \qquad
x =  \{ x_\al : \; \al = 1,2,\ldots, d\} \;  ,
$$
with $d = 3$ for three dimensions.

For an initial approximation, one takes the action
\be
\label{2.163}
 S_0[\vp] = \int \left [ \frac{1}{2} \left ( \frac{\prt\vp}{\prt x}\right )^2 +
\frac{\om^2}{2} \; \vp^2  \right ] \; dx \;   ,
\ee
with a trial parameter $\omega$. Then the total action can be written as
\be
\label{2.164}
 S[\vp] = S_0[\vp] + \Dlt S \;  ,
\ee
where
\be
\label{2.165}
 \Dlt S \equiv S[\vp] - S_0[\vp] = \int \left ( \frac{m^2 -\om^2}{2} \; \vp^2 +
\frac{\lbd}{4!}\; \vp^4 \right ) \; dx \; .
\ee
Introducing a dummy parameter $\varepsilon$, one defines
\be
\label{2.166}
  S[\vp] = S_0[\vp] + \ep \Dlt S \qquad (\ep \ra 1 ) \;  .
\ee

The optimization is done with respect to the quantity
\be
\label{2.167}
 F_k(\lbd,\om) \equiv \lgl \vp^2 \rgl_k  \;  ,
\ee
in which $k$ is a perturbation order with respect to $\varepsilon$. The control
functions are defined either by the minimal derivative condition
\be
\label{2.168}
\frac{\prt F_k(\lbd,\om)}{\prt \om} = 0
\ee
or by the minimal difference condition $F_k-F_{k-1}=0$ both of which give close
results \cite{Kneur_42}. When there arise multiple solutions for $\omega$, one
chooses that which provides the better convergence. Thus one finds \cite{Cruz_41}
the coefficients $c_1^{(k)}$ in the order $k$ as $c_1^{(2)}=3.08$, $c_1^{(3)}=2.45$,
$c_1 ^{(4)}=1.48$, while for the coefficient $c_2^{(k)}$ one gets $c_2^{(2)}=101.4$,
$c_2^{(3)}=98.2$, $c_2^{(4)}=82.9$. These results show the convergence to the Monte
Carlo simulations for the three dimensional lattice $O(2)$ field theory giving
$c_1=1.32\pm 0.02$ \cite{Arnold_36,Arnold_37} and $c_1=1.29\pm 0.05$
\cite{Kashurnikov_38,Prokofev_39}, while $c_2=75.7\pm 0.4$ \cite{Arnold_35}.

It is clear that introducing several control functions allows for the improvement
of results. This can be done as follows. First, let us notice that when we pass from
action (\ref{2.162}) to action (\ref{2.166}), we actually make the replacement
\be
\label{2.169}
 m^2 \ra \ep m^2 + ( 1 - \ep) \om^2 \; , \qquad \lbd \ra \ep\lbd \;  .
\ee
It is admissible to introduce one more control function $a$ replacing in the
right-hand side of the above substitution $\omega^2$ by
\be
\label{2.170}
  \om^2 \ra [ 1 + ( 1 - a)\ep) ] \om^2 \;  ,
\ee
with defining the control functions by the conditions
\be
\label{2.171}
 \frac{\prt F_k(\lbd,\om)}{\prt \om} = 0 \; , \qquad
\frac{\prt^2 F_k(\lbd,\om)}{\prt \om^2} = 0 \;  .
\ee
Similarly, one can introduce, in addition to $\omega$, two control functions $a$
and $b$ through the replacement
\be
\label{2.172}
  \om^2 \ra \left [ 1 + ( 1 - a)\ep + b \ep^2 \right ] \om^2 \;  ,
\ee
imposing three optimization conditions
\be
\label{2.173}
 \frac{\prt F_k(\lbd,\om)}{\prt \om} = 0 \; , \qquad
\frac{\prt^2 F_k(\lbd,\om)}{\prt \om^2} = 0 \; , \qquad
\frac{\prt^3 F_k(\lbd,\om)}{\prt \om^3} = 0 \;  .
\ee
In this way, for the two-component theory, to the sixth order of optimized
perturbation theory, one gets $c_1=1.3$ and in the fifth order, $c_2=73.46$
\cite{Kneur_46}.

The coefficient $c_1$ has also been found by Monte Carlo simulations for $N=1$,
being $c_1=1.09\pm 0.09$ and for $N=4$, being $c_1=1.60\pm 0.10$ \cite{Sun_49}. The
six-th order optimized perturbation theory gives $c_1=1.11$ for $N=1$ and $c_1=1.56$
for $N = 4$ \cite{Kneur_46}.

\subsection{Critical Temperature: Reexpansion Trick}

The other way to find the coefficient $c_1$ in the critical temperature shift
(\ref{2.160}) is to consider the loop expansion
\cite{Kastening_50,Kastening_51,Kastening_52}
\be
\label{2.174}
c_1(x) \simeq \sum_{n=1}^k a_n x^n \qquad ( x \ra 0 )
\ee
in powers of the variable
\be
\label{2.175}
 x = \frac{\lbd_{eff}}{\mu_{eff}} \; (N+2) \;  ,
\ee
in which $\lambda_{eff}$ is the effective interaction parameter and $\mu_{eff}$ is
the effective chemical potential. The known coefficients $a_n$ can be found in the
articles \cite{Yukalov_34,Yukalov_53}.

The difficulty in using the loop expansion is that at the point of the phase
transition $\mu_{eff} \ra 0$, hence $x \ra \infty$. That is, we need to find the
limit
\be
\label{2.176}
 c_1 = \lim_{x\ra\infty} c_1(x) \;  .
\ee
As is evident, the asymptotic series (\ref{2.174}), derived for small $x \ra 0$, has
no sense for $x\ra\infty$. But it is necessary to define the effective limit of these
series.

In its turn, the coefficient $c_2$ in shift (\ref{2.160}) can be written as
$$
 c_2 = \frac{32\pi\zeta(1/2)}{3[\zeta(3/2)]^{5/3}} \; \left \{
\ln\; \frac{[\zeta(3/2)]^{2/3}}{128\pi^2} +
\frac{\ln 2}{\sqrt{\pi}}\;\zeta\left (\frac{1}{2}\right ) - \;
\frac{\sqrt{\pi}+0.270166}{\zeta(1/2)} \; - 1\right\} +
$$
\be
\label{2.177}
 + \frac{7}{4}\; c_1^2 +
\frac{8\zeta(1/2)}{[\zeta(3/2)]^{1/3}} \; (3d_2 - 2c_1) \; ,
\ee
where the loop expansion yields
\be
\label{2.178}
 d_N(x) = \frac{1}{x^2} \left ( \sum_{n=0}^7 b_n x^n + b_2' x^2 \ln x
\right ) \;  ,
\ee
with the several first coefficients given explicitly,
$$
b_0 = (N+2)A_0 \; , \qquad b_1 = \frac{N+2}{24\pi}\; A_0 \; , \qquad
b_2 = \frac{1-4\ln 6}{576\pi^2}\; A_0 = - 2.393297 \; ,
$$
$$
  b_2' = \frac{A_0}{288\pi^2} = 0.776158\; , \qquad
A_0 = \frac{256\pi^3}{[\zeta(3/2)]^{4/3}} = 2206.18611757 \; ,
$$
and the numerical values of the other coefficients can be found in review
\cite{Yukalov_34}. Again, at the critical point, the variable $x\ra\infty$, so
that one has to find
\be
\label{2.179}
 d_N = \lim_{x\ra\infty} d_N(x) \;  .
\ee

The effective summation of the divergent series has been done
\cite{Kastening_50,Kastening_51,Kastening_52} using the optimized perturbation
theory, with introducing control functions by means of the Kleinert reexpansion
trick (see Sec. 2.11). The results for $c_1$ and $d_N$ are close to those calculated
by Monte Carlo simulations that are summarized in Table 1. For the coefficient $c_2$
Monte Carlo calculations give \cite{Arnold_37} the value $c_2 = 75.7 \pm 0.4$.

\begin{table}
\renewcommand{\arraystretch}{1.25}
\centering
\caption{ Monte Carlo results for the coefficients $c_1$ and $d_N$ for
a different number of field components $N$.}
\label{tab-1}
\vskip 5mm
\begin{tabular}{|c|c|c|} \hline
$N$ &     $c_1$                             &   $d_N$                  \\ \hline
1   &  1.09$\pm$ 0.09 \cite{Sun_49}         & 0.898$\pm$ 0.004 \cite{Sun_49}  \\ \hline
2   &  1.32$\pm$ 0.02 \cite{Arnold_36}      & 1.059$\pm$ 0.001 \cite{Arnold_37}  \\
    &  1.29$\pm$ 0.05 \cite{Kashurnikov_38} &                       \\ \hline
4   &  1.60$\pm$ 0.10 \cite{Sun_49}         & 1.255$\pm$ 0.006 \cite{Sun_49}  \\ \hline
\end{tabular}
\end{table}

\subsection{Critical Temperature: Trapped Bosons}

Finite quantum systems are often considered as being trapped in an external confining
potential \cite{Blaizot_54,Birman_55}. Deep potentials are usually modelled by the
harmonic potential
\be
\label{2.180}
 U(\br) = \frac{m}{2} \; \left ( \om_x^2 x^2 +  \om_y^2 y^2 +
\om_z^2 z^2 \right ) \; .
\ee
Genuine phase transitions, as a rule, happen only in thermodynamic limit, except some
shape transitions \cite{Birman_55} and nonequilibrium transitions \cite{Yukalov_56}.
For a trapped system, effective thermodynamic limit implies \cite{Yukalov_57,Yukalov_58}
that any extensive observable quantity $A_N$ increases proportionally to the number of
particles $N$ in the system,
\be
\label{2.181}
  N \ra \infty \; , \qquad  A_N \ra \infty \; , \qquad
\frac{A_n}{N} \ra const \; .
\ee
For a system trapped in the harmonic potential, this reduces to
\be
\label{2.182}
 N \ra \infty \; , \qquad  \om_0 \ra 0 \; , \qquad  N\om_0^3 \ra const \;  ,
\ee
where
\be
\label{2.183}
 \om_0 \equiv ( \om_x \om_y \om_z )^{1/3} \;  .
\ee
In a finite system, there happens not the genuine condensation, but quasi-condensation,
although for large $N$ it is hardly distinguishable from the genuine condensation. So,
for short, we also call it Bose-Einstein condensation.

The ideal Bose gas condenses at the critical temperature
\be
\label{2.184}
 T_0 = \left [ \frac{N}{\zeta(3)} \right ]^{1/3}\; \om_0 \;  .
\ee
The dimensionless interaction parameter is conveniently defined as the ratio
\be
\label{2.185}
\al \equiv \frac{a_s}{\lbd_0} \qquad
\left ( \lbd_0 = \sqrt{\frac{2\pi}{m T_0} } \right )
\ee
of the scattering length to the thermal wavelength at the critical point.

Switching on weak atomic interactions shifts the critical temperature according to
the formula
\be
\label{2.186}
  \frac{\Dlt T_c}{T_0} \simeq \overline c_1 \al +
( \overline c_2 + \overline c_2' \ln \al ) \al^2 \qquad ( \al \ra 0) \; .
\ee
The coefficients $\bar{c}_1$ and $\bar{c}_2^\prime$ can be found from perturbative
calculations \cite{Arnold_59},
\be
\label{2.187}
 \overline c_1 = -3.426032 \; , \qquad
\overline c_2' = - \; \frac{32\pi\zeta(2)}{3\zeta(3)} = -45.856623 \; .
\ee
The coefficient $\bar{c}_2$ can be connected to lattice simulations in three-dimensional
$O(2)$ field theory \cite{Arnold_59}, resulting in the form
\be
\label{2.188}
\overline c_2 = 21.4 - \; \frac{16\pi\zeta(2)}{3\zeta(3)}
\left [ \ln (32\pi^3)  + 3.522272 \right ] +
12\zeta(2) \left [ \zeta \left ( \frac{3}{2}\right ) \right ]^{4/3} \; d_2 \; ,
\ee
with the same $d_2$ as in (\ref{2.178}). Invoking the value of $d_2 = 1.059$ from
Monte Carlo simulations \cite{Arnold_37} results in $\bar{c}_2 = -155.0 \pm 0.1$.
Calculations of Kastening \cite{Kastening_52}, employing the introduction of control
functions through the Kleinert reexpansion give a close result.

\subsection{Fluid String}

Membranes are frequent structures in different chemical and biological systems.
An important class of membranes are the so-called fluid membranes, whose constituent
molecules are able to move inside these membranes. The problem of interest is the
calculation of the pressure of a membrane between two rigid walls.

The calculational procedure starts with considering soft walls modelled by a smooth
potential. The dimensionless pressure $\alpha(x)$ is expressed as a function of the
expansion parameter $x$ and determined by perturbation theory in  powers of this
parameter. Passing to rigid walls implies the limit $x \ra \infty$.

In the one-dimensional case of a fluctuating string between the rigid walls, the
problem reduces to finding the ground state energy of a particle moving between hard
walls. This problem serves as a touchstone for checking the accuracy of calculational
methods, since its exact solution is known. In dimensionless units, the pressure of
a string between soft walls reads as
\be
\label{2.189}
 \al(x) = \frac{\pi^2}{128} \left ( \frac{1}{2} + \frac{16}{\pi^4x^2} +
\frac{1}{2} \; \sqrt{ 1 + \frac{64}{\pi^4 x^2} }  \right ) \;  .
\ee
From here, it is straightforward to find the limit of hard walls
\be
\label{2.190}
  \lim_{x\ra\infty} \al(x) = \frac{\pi^2}{128} = 0.0771063 \; .
\ee

By perturbation theory in powers of $x$, expression (\ref{2.189}) results in the
expansion
\be
\label{2.191}
 \al(x) \simeq \frac{1}{8\pi^2 x^2}\; f_k(x) \qquad ( x\ra 0) \;  ,
\ee
where
\be
\label{2.192}
 f_k(x) = 1 + \sum_{n=1}^k a_n x^n \;  .
\ee
The first several coefficients here are
$$
a_1 = \frac{\pi^2}{4} = 2.467401 \; , \qquad
a_2 = \frac{\pi^4}{32} = 3.044034 \; , \qquad
a_3 = \frac{\pi^6}{512} = 1.877713 \; , \qquad a_4 = 0 \; ,
$$
$$
a_5 = -\; \frac{\pi^{10}}{131072} = -0.7144779 \; , \qquad  a_6 = 0\; ,
\qquad  a_7 = \frac{\pi^{14}}{16777216} = 0.5437238 \; ,
$$
$$
a_8 = 0 \; , \qquad
a_9 = -\; \frac{5\pi^{18}}{8589934592} = -0.5172230 \; , \qquad  a_{10} = 0\; ,
$$
$$
a_{11} = \frac{7\pi^{22}}{10995116127776} = 0.5510556 \; , \qquad
a_{12} = 0 \; ,
$$
$$
 a_{13} = -\; \frac{21\pi^{26}}{281474976710656} = -0.6290370 \; , \qquad
a_{14} = 0 \; .
$$
At the end, one needs to find the limit
\be
\label{2.193}
 \al = \lim_{x\ra\infty} \al(x) \;  .
\ee

Kastening \cite{Kastening_60} has found the effective limit for $x \ra \infty$ for
the series (\ref{2.192}) of sixth order, by introducing control functions through
the Kleinert reexpansion. However, he confronted the problem that the consecutive
determination of the power $q$ in the change of the variable (see Sec. 2.11) turned
out to yield imprecise results. Because of this, the parameter $q$ was fixed by
hands. Of course, varying $q$ influences the sought answer. Thus for $q = 1$, he
got $\alpha = 0.0769910$, while for $q = 2$, $\alpha = 0.0770973$.

\subsection{Fluid Membrane}

A two-dimensional fluid membrane between two walls is a problem, similar to that
of the fluctuating string. The pressure of the membrane reads as
$$
 p = \frac{8T^2}{\varkappa L^3} \; \al \;  ,
$$
where $T$ is temperature, $\varkappa$ is bending rigidity, $L$ is the distance
between the walls, and $\alpha$ is a dimensionless function that can be found by
perturbation theory for the case of soft walls as the expansion
\be
\label{2.194}
  \al(x) \simeq \frac{1}{8\pi^2 x^2} \; f_k(x) \qquad ( x \ra 0) \;  ,
\ee
with the same form of $f_k(x)$ as in Eq. (\ref{2.192}), but with the coefficients
$$
a_1 = \frac{\pi^2}{4} = 2.467401 \; , \qquad a_2 = \frac{\pi^4}{32} = 3.044034 \; ,
\qquad  a_3 = 2.092317 \;  ,
$$
$$
a_4 = 0.524451 \; , \qquad a_5 = -0.675653 \; , \qquad a_6 = -1.643210 \; .
$$
While the case of hard walls corresponds to $x \ra \infty$. So that we need to find
the limit (\ref{2.193}).

Introducing control functions through the Kleinert reexpansion, with fixing the
power $q$, Kastening \cite{Kastening_60} has found $\alpha = 0.0820175$ for $q=1$
and $\alpha = 0.0815743$ for $q = 2$. This is a bit larger than the value
$\alpha = 0.0798 \pm 0.0003$ obtained by Monte Carlo simulations \cite{Gompper_61}.

\subsection{Linear Hamiltonians}

The optimized perturbation theory has been applied to many quantum-mechanical
problems with linear Hamiltonians. A great number of papers is devoted to the
one-dimensional anharmonic oscillator, as in Sec. 2.7, considering its eigenvalues
\cite{Kleinert_2,Caswell_14,Seznec_15,Halliday_16,Killinbeck_17,Stevenson_18,
Feranchuk_19,Okopinska_20,Dineykhan_21,Dineykhan_92,Feranchuk_22,Yukalov_23,
Yukalova_62,Feranchuk_63,Janke_64,Dobrovolska_65} and wave functions \cite{Hatsuda_66}.

The three-dimensional spherical anharmonic oscillator, with the Hamiltonian
\be
\label{2.195}
 \hat H = -\; \frac{1}{2m} \; \frac{d^2}{dr^2} + \frac{l(l+1)}{2mr^2} +
\frac{m\om_0^2}{2}\; r^2 + \lbd m^2 r^4 \;  ,
\ee
where $r \in [0, \infty)$ and $l = 0,1,2, \ldots$, has also been treated
\cite{Dineykhan_21,Dineykhan_92,Yukalova_67}, as well as the anharmonic oscillator
of arbitrary dimensionality, with the Hamiltonian
\be
\label{2.196}
 \hat H = -\; \frac{1}{2m} \; \frac{d^2}{dr^2} +
\frac{1}{2mr^2} \; \left (l + \frac{d-3}{2} \right )
\left (l + \frac{d-1}{2} \right ) +
\frac{m\om_0^2}{2}\; r^2 + \lbd m^2 r^4 \;   ,
\ee
where $d = 1,2,3, \ldots$ is space dimensionality \cite{Yukalova_68}.

Different power-law potentials have been considered, with one-dimensional
Hamiltonians \cite{Yukalov_13,Yukalova_69}
\be
\label{2.197}
 \hat H = -\; \frac{1}{2m} \; \frac{d^2}{dx^2} + A x^\nu \;  ,
\ee
spherically symmetric power-law potentials in three dimensions
\cite{Yukalov_13,Dineykhan_21,Dineykhan_92,Imbo_70,Coleman_71,Yukalov_72},
\be
\label{2.198}
 \hat H = -\; \frac{1}{2m} \; \frac{d^2}{dr^2} + \frac{l(l+1)}{2mr^2} +
 A r^\nu \; ,
\ee
and spherically-symmetric Hamiltonians of arbitrary dimensionality
\cite{Yukalov_13,Yukalova_69},
\be
\label{2.199}
  \hat H = -\; \frac{1}{2m} \; \frac{d^2}{dr^2} +
\frac{1}{2mr^2} \; \left (l + \frac{d-3}{2} \right )
\left (l + \frac{d-1}{2} \right )  + A r^\nu \; .
\ee
Two cases can occur in the power-law potentials, when both $A$ and $\nu$ are
positive, or when both of them are negative.

Quasistationary states, characterized by the Hamiltonian
\be
\label{2.200}
  \hat H = -\; \frac{1}{2m} \; \frac{d^2}{dx^2} + \frac{m\om_0^2}{2}\; x^2 -
 \lbd m^2 x^4 \;  ,
\ee
are studied \cite{Feranchuk_63,An_73,Karrlein_74,Yukalov_75}.

A double-well oscillator, described by the Hamiltonian
\be
\label{2.201}
 \hat H = -\; \frac{1}{2m} \; \frac{d^2}{dx^2} - \frac{m\om_0^2}{2}\; x^2 +
 \lbd m^2 x^4 \;
\ee
has been considered \cite{Kleinert_2,Feranchuk_63,Yukalov_75,Yukalov_76}.

The Hamiltonians with the Yukawa potential
\cite{Kleinert_2,Dineykhan_21,Dineykhan_92,Yukalov_77},
\be
\label{2.202}
 \hat H = -\; \frac{1}{2m} \; \frac{d^2}{dr^2} + \frac{l(l+1)}{2mr^2} -
\frac{A}{r}\; e^{-\al r} \;  ,
\ee
and with the logarithmic potential \cite{Yukalov_13,Dineykhan_21,Dineykhan_92},
\be
\label{2.203}
 \hat H = -\; \frac{1}{2m} \; \frac{d^2}{dr^2} + \frac{l(l+1)}{2mr^2} +
B\; \ln \; \frac{r}{b}\;   ,
\ee
have been treated. An interesting point in the case of the Yukawa potential is
to find the critical value of $\alpha$, when no bound states exist \cite{Luo_78}.

A hydrogen-like atom in an external magnetic field, described by the Hamiltonian
\cite{Kleinert_2,Dineykhan_79}
\be
\label{2.204}
 \hat H = -\; \frac{\nabla^2}{2m}  + \frac{m\om_0^2}{2}\; ( x^2 + y^2) -
\frac{e^2}{r} -  \om_0 L_z \;  ,
\ee
in which $e$ is the electron charge,
$$
L_z = [ \br \times \bp ]_z \; , \qquad \bp = - i\nabla \; , \qquad
\om_0 = \frac{eB_0}{2mc} \;   ,
$$
and in an external electric field, with the Hamiltonian \cite{Feranchuk_80}
\be
\label{2.205}
  \hat H = -\; \frac{\nabla^2}{2m}  - \; \frac{e^2}{r} + Az \;  ,
\ee
have been considered.

A semi-relativistic Schr\"{o}dinger equation, with the Hamiltonian
$$
 \hat H = \sqrt{ -\nabla^2 + m^2} \; + U(r) \; ,
$$
in which $U(r)$ is the Cornell potential
$$
U(r) = - \; \frac{A}{r} + Br \; ,
$$
and three-body Coulomb systems have also been considered \cite{Dineykhan_92}.
For the initial approximation, one usually takes a harmonic oscillator or the
Coulomb potential. Although one can also take other Hamiltonians, for which
exact solutions are available \cite{Znojil_93,Znojil_94}.

For a one-dimensional anharmonic oscillator, the related free energy and
propagators have been studied \cite{Okopinska_81,Weissbach_82}.

Quite a number of other papers have been devoted to the use of optimized
perturbation theory in quantum mechanics. Here some typical examples are
given, without the intension of listing all numerous publications.

Optimized perturbation theory yields accurate energy spectra, with a much higher
accuracy than other methods, such as quasiclassical approximation. Employing just
a couple of perturbative terms, it is straightforward to get accurate expressions
of energy levels for all quantum numbers and all coupling (anharmonicity)
parameters. Recall that the method of Pad\'{e} approximants is not applicable
at all, when only a couple of terms in an expansion are available.

\subsection{Nonlinear Hamiltonians}

A system of bosonic atoms in Bose-Einstein condensate is described by the
Nonlinear Schr\"{o}dinger (NLS) equation
\cite{Yukalov_32,Andersen_33,Yukalov_58,Pethick_83,Yukalov_84}. For a system of
$N$ condensed atoms, interacting through the local interaction potential
\be
\label{2.206}
 \Phi(\br) = \Phi_0 \dlt(\br) \qquad
\left ( \Phi_0 \equiv 4\pi\; \frac{a_s}{m} \right ) \;  ,
\ee
where $a_s$ is scattering length, the stationary NLS equation is
\be
\label{2.207}
  \hat H [\psi]\psi(\br)  = E \psi(\br) \;  ,
\ee
with the nonlinear Hamiltonian
\be
\label{2.208}
  \hat H [\psi] = - \; \frac{\nabla^2}{2m} + U(\br) + N\Phi_0 | \psi|^2 \; .
\ee
Here the external potential $U(\bf r)$ plays the role of a confining potential
trapping the atoms. Usually, it is modeled by a cylindric harmonic potential
\be
\label{2.209}
  U(\br) = \frac{m}{2} \; \om^2_\perp ( x^2 + y^2 + \al^2 z^2 ) \; ,
\ee
in which
\be
\label{2.210}
\al \equiv \frac{\om_z}{\om_\perp} = \left ( \frac{l_\perp}{l_z}\right )^2
\ee
is called aspect ratio, where $\omega_z$ and $\omega_\perp$ are the longitudinal
and transverse trapping frequencies, respectively, and
$$
 l_\perp \equiv \frac{1}{\sqrt{m \om_\perp} } \; , \qquad
 l_z \equiv \frac{1}{\sqrt{m \om_z} }
$$
are the oscillator lengths.

It is convenient to pass to dimensionless units, defining the dimensionless
coupling parameter
\be
\label{2.211}
 g \equiv 4\pi \; \frac{a_s}{l_\perp} \; N \; ,
\ee
dimensionless spatial variables
\be
\label{2.212}
 r \equiv \frac{\sqrt{x^2+y^2}}{l_\perp} \; , \qquad
\overline z \equiv \frac{z}{l_\perp} \;  ,
\ee
and the dimensionless wave function
\be
\label{2.213}
 \overline\psi(r,\vp,\overline z) \equiv l^{3/2}_\perp \psi(\br) \; ,
\ee
in which $\varphi$ is a polar angle. With the nonlinear Hamiltonian in units
of $\omega_\perp$, the NLS equation reads as
\be
\label{2.214}
 \hat H[  \overline\psi_{nmj} ]  \overline\psi_{nmj} =
E_{nmj}  \overline\psi_{nmj} \; ,
\ee
where $n = 0,1,2,\ldots$ is the radial quantum number, $m = 0,1,2,\ldots$ is the
azimuthal quantum number, and $j = 0,1,2,\ldots$ is the axial quantum number. The
eigenvalue $E_{nmj}$ is measured in units of $\omega_\perp$.

In what follows, for the simplicity of notations, we omit the bars above $z$ and
$\psi$. Then the dimensionless nonlinear Hamiltonian becomes
\be
\label{2.215}
 \hat H[\psi] = -\; \frac{\nabla^2}{2} +
\frac{1}{2} \; ( r^2 + \al^2 z^2 ) + g | \psi|^2 \; .
\ee
The ground state, with $n = m = j = 0$ describes the equilibrium Bose-Einstein
condensate, while the higher energy levels correspond to coherent topological
modes \cite{Yukalov_85,Yukalov_86,Yukalov_87,Yukalov_88}.

Solving the NLS equation by the optimized perturbation theory, we start with the
harmonic oscillator
\be
\label{2.216}
\hat H_0[\psi] = -\;\frac{\nabla^2}{2} + \frac{1}{2} \; ( u^2r^2 + v^2 z^2 ) \; ,
\ee
in which $u$ and $v$ are control parameters. The zero-order wave function is given
by the Hermite-Laguerre modes
$$
\psi_{nmj}(r,\vp,z) = \left [ \frac{2n! u^{|m|+1}}{(n+|m|)!} \right ]^{1/2} \;
r^{|m|} \exp \left ( - \; \frac{u}{2} \; r^2 \right ) L_n^{|m|}(ur^2) \;
\frac{e^{im\vp}}{\sqrt{2\pi}} \; \left ( \frac{v}{\pi}\right )^{1/4} \times
$$
\be
\label{2.217}
\times
\frac{1}{\sqrt{2^j j!} } \; \exp \left ( - \; \frac{v}{2} \; z^2 \right )
H_j(\sqrt{v}\; z) \; .
\ee
So that the zero-order spectrum is
\be
\label{2.218}
 E_{nmj}^{(0)} \equiv E_0(g,u,v) = ( 2n + |m| +1 )u +
\left ( \frac{1}{2} + j\right ) v \;  .
\ee

The first-order approximation for the energy levels can be written as
\be
\label{2.219}
  E_{nmj}^{(1)} \equiv E_1(g,u,v) =
\frac{p}{2}\; \left ( u + \frac{1}{u} \right )  +
\frac{q}{4}\; \left ( v + \frac{\al^2}{v} \right ) + u \sqrt{v}\; I_{nmj} g \; ,
\ee
where
$$
p \equiv 2n + |m| + 1 \; , \qquad q \equiv 2j + 1
$$
and
$$
 I_{nmj} \equiv
\frac{1}{u\sqrt{v}} \int |\psi_{nmj}(r,\vp,z)|^4 \; r\; dr d\vp dz \;  .
$$
Control functions can be found from the optimization conditions
\be
\label{2.220}
 \frac{\prt}{\prt u} \; E_1(g,u,v) = 0 \; , \qquad
\frac{\prt}{\prt v} \; E_1(g,u,v) = 0 \;  .
\ee
This gives the equations
\be
\label{2.221}
  p \left ( 1 - \; \frac{1}{u^2}  \right ) +
\frac{J}{\al p}\; \sqrt{\frac{v}{q} } = 0 \; , \qquad
q \left ( 1 - \; \frac{\al^2}{v^2}  \right ) +
\frac{uJ}{\al p\sqrt{vq} } = 0 \; ,
\ee
defining the control functions $u = u(g)$ and $v = v(g)$, where
\be
\label{2.222}
 J \equiv J_{nmj}(g) = 2\al p \sqrt{q}\; I_{nmj} g \;  .
\ee
Substituting these control functions into expression (\ref{2.219}) yields the
optimized approximant
\be
\label{2.223}
 \tilde E_1(g) \equiv E_1(g,u(g),v(g)) \;  .
\ee

The found approximant (\ref{2.223}) is valid for all quantum numbers and for any
strength of the coupling parameter $g$. Keeping in mind that $J \ra 0$ together
with $g \ra 0$, we have for the control functions
$$
u \simeq 1 - \; \frac{J}{2p^2\sqrt{\al q} } \; , \qquad
v \simeq \al - \; \frac{\al J}{2p^2(\al q)^{3/2} } \qquad ( J \ra 0) \;  .
$$
Then energy (\ref{2.223}) behaves as
\be
\label{2.224}
  \tilde E_1(g) \simeq p + \frac{\al q}{2} + \frac{J}{2p\sqrt{\al q } } \qquad
( J \ra 0 ) \;  .
\ee
In the opposite limit, when $J \ra \infty$ together with $g \ra \infty$, the
control functions diminish as
$$
  u \simeq \frac{p}{J^{2/5} } \; , \qquad v \simeq \frac{\al^2 q}{J^{2/5}}
\qquad ( J \ra \infty) \; .
$$
In this strong-coupling limit, energy (\ref{2.223}) tends to
\be
\label{2.225}
   \tilde E_1(g) \simeq  \frac{5}{4} \; J^{2/5} +
\frac{2p^2 +(\al q)^2}{4} \; J^{-2/5} \qquad ( J \ra \infty ) \; .
\ee
The first term here corresponds to the Thomas-Fermi approximation that is
asymptotically exact in the strong-coupling limit.

More information on the use of optimized perturbation theory to the problems with
nonlinear Hamiltonians can be found in Refs.
\cite{Courteille_178,Yukalov_179,Yukalov_180,Yukalov_181}.

\subsection{Hamiltonian Envelopes}

The general idea in choosing an initial Hamiltonian $\hat{H}_0$ in the optimized
perturbation theory is that this Hamiltonian should provide an explicit solution,
at the same time modeling the considered Hamiltonian $\hat{H}$. Keeping in mind
that the Hamiltonians have the forms
\be
\label{2.226}
 \hat H = -\; \frac{\nabla^2}{2m} + V(\br) \; , \qquad
\hat H_0 = -\; \frac{\nabla^2}{2m} + V_0(\br) \;  ,
\ee
in which $V({\bf r})$ and $V_0({\bf r})$ are the corresponding potential energies,
it is clear that $\hat{H}_0$ models $\hat{H}$, if the potential $V_0({\bf r})$ in
some sense models the potential $V({\bf r})$. However, it is not always possible to
have such a potential $V_0({\bf r})$ that would be similar to $V({\bf r})$ and at
the same time allowing for an exact solution. To overcome this difficulty, the
choice of initial Hamiltonians can be extended by resorting to the method of
Hamiltonian envelopes \cite{Yukalov_89}.

Recall that if a Hamiltonian $\hat{H}_0$ satisfies the eigenproblem
\be
\label{2.227}
 \hat H_0 \psi_n = E_n \psi_n \;  ,
\ee
then a function $h(\hat{H}_0)$ of this Hamiltonian satisfies the eigenproblem
\be
\label{2.228}
h(\hat H_0) \psi_n = h(E_n) \psi_n
\ee
with the same eigenfunctions. When the Hamiltonian $\hat{H}_0$ is not a good
initial choice, then it is possible to find a function $h(\hat{H}_0)$ that could
play the role of a better initial Hamiltonian. Therefore, one can consider the
Hamiltonian
\be
\label{2.229}
 \hat H_\ep = h(\hat H_0) + \ep [ \hat H - h(\hat H_0 ) ] \qquad (\ep \ra 1) \;  ,
\ee
with a more appropriate initial approximation and a smaller perturbative term.
The function $h(\hat{H}_0)$ can be called the {\it Hamiltonian envelope}.

The envelope $h(\hat{H}_0)$ models well the Hamiltonian $\hat{H}$, provided that
the effective potential $h(V_0(\br))$ is of the order of the potential $V(\br)$,
which can be denoted as the relation
\be
\label{2.230}
 h(V_0(\br)) \sim V(\br) \;  .
\ee
To illustrate how the Hamiltonian envelope can be chosen, let us consider the case
of power-law potentials of Sec. 2.17. For the potentials with positive power $\nu$,
one can take as $V_0({\bf r})$ the harmonic potential, so that we have
\be
\label{2.231}
 V(\br) \propto r^\nu \; , \qquad V_0(\br) \propto r^2  \qquad ( \nu > 0 ) \; .
\ee
To satisfy relation (\ref{2.230}), we need to have
$$
 h(V_0(\br)) \propto [ V_0(\br) ]^{\nu/2} \propto r^\nu \;  .
$$
Thence the Hamiltonian envelope should be chosen such that
\be
\label{2.232}
 h(\hat H_0) \propto (\hat H_0)^{\nu/2} \; .
\ee

For the potentials with negative powers, one takes as $V_0({\bf r})$ the Coulomb
potential,
\be
\label{2.233}
 V(\br) \propto r^{-\nu} \; , \qquad V_0(\br) \propto r^{-1}  \qquad
( \nu > 0 ) \;  .
\ee
Relation (\ref{2.230}) is valid when
$$
h(V_0(\br)) \propto [ V_0(\br) ]^\nu \propto r^{-\nu}  \;   .
$$
Hence the Hamiltonian envelope is to be such that
\be
\label{2.234}
  h(\hat H_0) \propto  (\hat H_0)^\nu \;  .
\ee

\subsection{Magnetic Systems}

Optimized perturbation theory can be applied to any system. The present section
demonstrates its application to the two-dimensional Ising model, whose Hamiltonian
is
\be
\label{2.235}
 \hat H = - J \sum_{\lgl ij\rgl} \sgm_i \sgm_j -  B_0 \sum_j \sgm_j \;  ,
\ee
in which $J > 0$, the notation $\lgl ij \rgl$ implies the nearest neighbors, and
$\sgm_i=\pm 1$, with $i = 1,2,\ldots,N_L$. Introducing the dimensionless coupling
parameter $g$ and magnetic
field $h_0$ as
\be
\label{2.236}
  g \equiv \bt J \; , \qquad h_0 \equiv \bt B_0 \; ,
\ee
where $\beta$ is inverse temperature, we have the partition function
\be
\label{2.237}
 Z = {\rm Tr} e^{-\bt \hat H} =  {\rm Tr} \exp \left (
- g   \sum_{\lgl ij\rgl} \sgm_i \sgm_j - h_0 \sum_j \sgm_j  \right ) \; .
\ee
The free energy per site is
\be
\label{2.238}
 f(g,h_0) \equiv -\; \frac{1}{N_L} \; \ln Z \;  .
\ee
And the dimensionless magnetization is defined as
\be
\label{2.239}
 m(g,h_0) \equiv -\; \frac{\prt f(g,h_0)}{\prt h_0} \;  .
\ee

The initial Hamiltonian can be chosen, following Ref. \cite{Aoyama_90}, as
\be
\label{2.240}
  \hat H_0 = - B \sum_j \sgm_j \; ,
\ee
with a control parameter $B$. Then for the Hamiltonian
\be
\label{2.241}
  \hat H_\ep = \hat H_0 + \ep ( \hat H - \hat H_0 ) \qquad ( \ep \ra 1 ) \; ,
\ee
we have
\be
\label{2.242}
 \hat H_\ep = - \ep J \sum_{\lgl ij\rgl} \sgm_i \sgm_j -
[ B + \ep ( B_0 - B ) ] \sum_j \sgm_j \;  .
\ee
This shows that passing form $\hat{H}$ to $\hat{H}_\varepsilon$ is equivalent to
the replacement
\be
\label{2.243}
 J \ra \ep J \; , B_0 \ra B + \ep ( B_0 - B ) \; ,
\ee
or, in dimensionless units, to the replacement
\be
\label{2.244}
g \ra \ep g \; , \qquad h_0 \ra h + \ep ( h_0 - h ) \;   ,
\ee
where
\be
\label{2.245}
 h \equiv \bt B \;  .
\ee
Therefore, it is possible to use the reexpansion trick, as described in Sec. 2.10.
To this end, we can derive an expansion of free energy (\ref{2.238}) in powers of $g$,
after which to invoke replacement (\ref{2.244}). The weak-coupling expansion of free
energy (\ref{2.238}) reads as
\be
\label{2.246}
 f_k(g,h_0) = \sum_{n=0}^k c_n(h_0) g^n \;  ,
\ee
with the coefficients
$$
c_0 = -\ln (2\cosh h_0 ) \; , \qquad c_1 =  - 2t_0^2 \; , \qquad
c_2 = - 1 - 6t_0^2 + 7t_0^4 \; ,
$$
$$
 c_3 =  - \; \frac{52}{3} \; t_0^2 + 56 t_0^4 -\; \frac{116}{3} \; t_0^6\; ,
\qquad
c_4 =  - \; \frac{5}{6} - 46 t_0^2 + \frac{937}{3} \; t_0^4 - 526 t_0^6 +
\frac{521}{2} \; t_0^8 \;  ,
$$
where
\be
\label{2.247}
 t_0 = \tanh(h_0 ) \;  .
\ee
The free energy
\be
\label{2.248}
f_\ep(g,h_0,h) \equiv - \; \frac{1}{N_L} \; \ln {\rm Tr} \exp( -\bt \hat H_\ep)
\ee
is obtained from expression (\ref{2.238} by making replacement (\ref{2.244}),
\be
\label{2.249}
  f_\ep(g,h_0,h) = f(\ep g,h+\ep(h_0-h))\; .
\ee
Accomplishing the reexpansion in powers of $\varepsilon$ in sum (\ref{2.246}), we
have
\be
\label{2.250}
f_k(\ep g,h+\ep(h_0-h)) = \sum_{n=0}^k c_n(h+\ep(h_0-h)) (\ep g)^n \simeq
\sum_{n=0}^k b_n(g,h_0,h) \ep^n \;  .
\ee
Setting here $\varepsilon = 1$, we get
\be
\label{2.251}
 F_k(g,h_0,h) \equiv \sum_{n=0}^k  b_n(g,h_0,h) \;  .
\ee
Thus, in the first two approximations, we find
\be
\label{2.252}
  F_0(g,h_0,h) = - \ln [ 2 \cosh(h) ] \; , \qquad
F_1(g,h_0,h) = F_0(g,h_0,h) + t(h-h_0) - 2t^2 g \;  ,
\ee
where
\be
\label{2.253}
 t \equiv \tanh(h ) \;   .
\ee
The related magnetization
\be
\label{2.254}
M_k(g,h_0,h) \equiv -\; \frac{\prt F_k(g,h_0,h)}{\prt h_0} \; ,
\ee
becomes
\be
\label{2.255}
  M_0(g,h_0,h) = t \; , \qquad
M_1(g,h_0,h) = t + ( t^2 - 1 )( h - h_0 ) + 4t( 1 - t^2 ) g \;   .
\ee

Control functions $h_k(g,h_0)$ are found from the optimization condition
\be
\label{2.256}
 \frac{\prt^2 F_k(g,h_0,h)}{\prt h_k^2} = 0 \; , \qquad h_k = h_k(g,h_0) \; .
\ee
This gives the optimized approximant for the free energy
\be
\label{2.257}
 \tilde f_k(g,h_0) = F_k(g,h_0,h_k(g,h_0) ) \; .
\ee

Condition (\ref{2.256}) is valid for all $g$, including the critical value $g_c$,
where external magnetic fields are zero. Hence the critical $g_c$ can be found from
the condition
\be
\label{2.258}
 \lim_{h\ra 0} \; \frac{\prt^2 F_k(g,0,h)}{\prt h^2} = 0 \;  .
\ee
In the fourth order, this yields $g_c = 0.37681$ \cite{Aoyama_90}, which is close
to the known exact value
$$
g_c = \frac{1}{2} \; \ln ( 1 + \sqrt{2} ) =0.44067 \; .
$$

It is important to stress that the reexpansion trick leads to reasonable results,
provided the reexpansion is motivated by the choice of an initial Hamiltonian,
as is explained in Sec. 2.10. If the reexpansion is arbitrary, but not induced
by initial conditions, it may result in inappropriate conclusions. To illustrate
this, let us consider the free energy (\ref{2.238}) as a function of the variables
\be
\label{2.259}
 \lbd \equiv e^{-4g} \; , \qquad x \equiv e^{-2h_0} \;  .
\ee
And let us examine the strong-field expansion
\be
\label{2.260}
f_k(x,\lbd) = \sum_{n=0}^k d_n(\lbd) x^n \;   .
\ee
To fourth order, this reads as
$$
f_4(x,\lbd) = \frac{1}{2} \; \ln(\lbd x) - \lbd^2 x - 2 \left (
1 - \; \frac{5}{4}\;\lbd \right ) \lbd^3 x^2 -
$$
$$
 - 6 \left (
1 - \; \frac{8}{3}\;\lbd +  \frac{31}{18}\;\lbd^2 \right ) \lbd^4 x^3 -
\left ( 1 + 18 \lbd - 85 \lbd^2 + 96 \lbd^3 - \; \frac{121}{4}\; \lbd^4
\right ) \lbd^4 x^4 \; .
$$
If we make here the replacement
$$
x \ra \ep x \; , \qquad \lbd \ra u + \ep (\lbd - u ) \; ,
$$
that is not motivated by initial conditions, and follow the same way as in the
case of the initial-condition induced replacement (\ref{2.244}), then we come to
the conclusion that there is no critical $g_c$, that is, there is no phase
transition \cite{Aoyama_90}.

\subsection{Field-Theory Models}

Optimized perturbation theory has been used for considering several field-theory
models. Among them: the $O(N)$- symmetric $\varphi^4$ theory
\cite{Farias_48,Stancu_91,Stancu_92,Haugerud_93,Okopinska_94,Kleinert_95,Chiku_96,Honkonen_97,
Kleinert_98,Honkonen_99,Strosser_100,Rosa_102}, charged scalar field in
an external field \cite{Duarte_107}, Walecka nuclear-matter model \cite{Krein_108},
massive Schwinger model on a lattice \cite{Byrnes_109}, $U(2)$ theory on a lattice
\cite{Evans_110}, and Gross-Neveu model
\cite{Kneur_111,Kneur_112,Kneur_271,Kneur_113,Kneur_114}.
Note that the so-called screened perturbation theory and hard-thermal-loop
resummation, used in field theory are just variants of optimized perturbation
theory \cite{Kraemmer_115}.

Optimized perturbation theory has been applied as well in quantum electrodynamics
and quantum chromodynamics
\cite{Field_116,Mattingly_117,Arvanitis_118,Kneur_119,Kneur_120,Sissakian_121,
Inoui_122,Stevenson_123}.

The choice of scheme and scale parameters in renormalization group
\cite{Litim_124,Litim_125,Litim_126,Litim_127,Litim_128}
can also be treated as an application of optimized perturbation theory, where
the scheme and scale parameters play the role of control functions \cite{Wu_129}.

Optimized perturbation theory can be combined with renormalization group providing
fast convergent scale invariant sequence of optimized approximants
\cite{Kneur_130,Kneur_131,Kneur_132}. For example, considering a field theory with
mass $m$ and coupling parameter $g$, it is possible to use the reexpansion trick
described in Secs. 2.10 and 2.11. This can be accomplished as follows. Starting with
an initial Lagrangian, one introduces mass, coupling, and vacuum energy counterterms
canceling the original divergences. This renormalization defines the running
coupling parameter $g(\mu)$ as a function of an arbitrary renormalization scale
$\mu$, usually appearing in the process of dimensional regularization in the
$\overline{\rm{MS}}$ scheme. Then, for a quantity of interest, one can employ
the usual perturbation theory in powers of $g$ getting an expansion
\be
\label{2.261}
 f_k(g,\mu,m) = \sum_{n=0}^k c_n(\mu,m) g^n \;  .
\ee
This is reorganized by means of the substitution
\be
\label{2.262}
 m^2 \ra ( 1  - \ep)^{2a} m^2 \; , \qquad g \ra \ep g \;  ,
\ee
keeping in mind that $m$ represents a control parameter. The value of $a$ is
fixed by the condition of preserving the renormalization group invariance. Then
one reexpands the studied quantity in powers of the dummy parameter $\ep$,
\be
\label{2.263}
 f_k(\ep g,\mu,(1-\ep)^a m) \simeq \sum_{n=0}^k b_n(g,\mu,m) \ep^n \;  .
\ee
Setting here $\varepsilon = 1$, one obtains
\be
\label{2.264}
  F_k(g,\mu,m) = \sum_{n=0}^k b_n(g,\mu,m) \;  .
\ee
The standard renormalization group equation reads as
\be
\label{2.265}
 \mu \; \frac{d}{d\mu} \;  F_k(g,\mu,m) = 0 \; ,
\ee
where the renormalization group operator is
\be
\label{2.266}
 \mu \; \frac{d}{d\mu} \equiv \mu \; \frac{\prt}{\prt \mu} +
\bt(g) \; \frac{\prt}{\prt g} - \gm(g) m \; \frac{\prt}{\prt m} \; .
\ee
The control parameter $m$ is to be turned into a control function by an optimization
condition. For instance, we can accept the minimal derivative condition
\be
\label{2.267}
 \frac{\prt}{\prt m} \;  F_k(g,\mu,m) = 0 \;  ,
\ee
as a result of which the renormalization group equation simplifies to
\be
\label{2.268}
\left ( \mu \; \frac{\prt}{\prt \mu} + \bt(g) \; \frac{\prt}{\prt g} \right )
 F_k(g,\mu,m) = 0 \;   .
\ee
Equations (\ref{2.267}) and (\ref{2.268}) define the control functions
\be
\label{2.269}
 g_k = g_k(\mu) \; , \qquad m_k = m_k(\mu) \;  .
\ee
So that the optimized approximant becomes
\be
\label{2.270}
 \tilde f_k(\mu) \equiv F_k(g_k(\mu),\mu,m_k(\mu) ) \;  .
\ee

The renormalization scale is often chosen as $\mu = 2 \pi T$ to avoid the
logarithmic terms $\ln (\mu/2 \pi T)$ coming from remnant scale dependence. This
method was used for considering the $O(N)$- symmetric $\vp^4$ theory, a three-color
Nambu-Jona-Lasinio model, and quantum chromodynamics
\cite{Kneur_133,Kneur_134,Kneur_135,Duarte_136,Ferrari_137}.

\section{Self-Similar Approximation Theory}

Despite very wide and successful application of Optimized Perturbation Theory, some
questions have remained:

\begin{enumerate}[label=(\Roman{*})]
\item
Is it possible to improve the accuracy of the optimized approximants within the
given number of perturbative terms?

\item
How to select the best way of introducing control functions, when there are
several such ways, but the exact solution is not known?

\item
How to prove the stability of a calculational procedure, if the exact solution of
the considered problem is not available, hence the accuracy of approximants cannot
be defined by the direct comparison with the exact solution?

\item
When the asymptotic behaviour of a sought function is known at the boundaries
of an interval, what is the best explicit analytical interpolation of the function
between these two limits?

\item
When the asymptotic behaviour of a function is known only at one boundary, what
is the best explicit analytical extrapolation of the function to its whole domain?

The answers to these questions have been given by Self-Similar Approximation Theory
advanced in Refs.
\cite{Yukalov_138,Yukalov_139,Yukalov_140,Yukalov_141,Yukalov_142,Yukalov_143,
Yukalov_144,Yukalov_145}.
\end{enumerate}

\subsection{Approximation Cascade}

The main idea of self-similar approximation theory is to reformulate perturbation
theory in terms of dynamical theory, treating the approximation order as discrete
time, so that the approximation sequence be isomorphic to the trajectory of a
discrete-time dynamical system, called cascade. Then the effective limit of the
perturbative sequence corresponds to a fixed point of the cascade. Analyzing the
stability of the constructed dynamical system gives information on the stability
of the calculational procedure. Reformulating perturbation theory to the language
of dynamical theory makes it straightforward to invoke the rich mathematical
techniques of the latter \cite{Guckenheimer_146,Hale_147,Sinai_148}.

At the first step, we follow the idea of optimized perturbation theory by
reorganizing a perturbative sequence of terms $f_k(x)$ to a sequence of terms
$F_k(x,u)$ through incorporating control parameters $u$. Introducing, instead of
the parameters $u$, control functions $u_k=u_k(x)$, as is explained in the previous
Chapter, we obtain optimized approximants
$$
\widetilde{f}_k(x) = F_k(x,u_k(x)) \; .
$$
Then we define an {\it expansion function} $x_k(f)$ by the reonomic constraint
\be
\label{3.1}
 F_0(x,u_k(x)) = f \; , \qquad x = x_k(f) \;  ,
\ee
whose inverse is
\be
\label{3.2}
x_k(  F_0(x,u_k(x)) ) = x \;  .
\ee

The endomorphism
\be
\label{3.3}
y_k(f) \equiv \widetilde f_k(x_k(f) )
\ee
acts in the range of all possible values of $f$ composing a measurable space. From
this definition it follows that
\be
\label{3.4}
 y_k(  F_0(x,u_k(x)) ) = \widetilde f_k(x) \;  ,
\ee
and the initial condition, with respect to the varying $k$, is
\be
\label{3.5}
 y_0(f) = f \;  .
\ee
By this construction, the sequence of the approximants $\{\tilde{f}_k(x)\}$ is
bijective to the sequence of the endomorphisms $\{y_k(f)\}$.

Because the sequences $\{\tilde{f}_k(x)\}$ and $\{y_k(f)\}$ are bijective, they
converge together. If a limit $\tilde{f}(x)$ of the sequence $\{\tilde{f}_k(x)\}$
exists, then a limit $y^*(f)$ of the sequence of the endomorphisms, corresponding
to $\tilde{f}(x)$ also exists. The sequence $\{y_k(f)\}$ satisfies the Cauchy
criterion following from Eq. (\ref{2.11}): For any $\ep>0$, there exists a number
$n_\varepsilon$, such that
$$
 | y_{k+p}(f) - y_k(f) | < \ep \;  ,
$$
for all $k \geq n_\varepsilon$ and $p \geq 0$. The limit of a sequence of endomorphisms
is an attractive fixed point, for which
\be
\label{3.6}
 y_k(y^*(f) ) = y^*(f) \;  .
\ee
In the vicinity of a fixed point, the endomorphism enjoys the property of
{\it functional self-similarity}
\cite{Yukalov_138,Yukalov_139,Yukalov_142,Yukalov_144,Yukalov_145,Yukalov_149,Yukalov_150}
\be
\label{3.7}
 y_{k+p}(f) = y_k(y_p(f) ) \;  ,
\ee
which is easy to check noticing that relation (\ref{3.7}) reduces to the identity
$y^*(f)=y^*(f)$, when $f \ra y^*(f)$. Functional self-similarity \cite{Bogolubov_151}
is a generalization of scale self-similarity \cite{Sedov_152}.

The property of self-similarity is equivalent to the semigroup property
$$
y_k \cdot y_p = y_{k+p} \; , \qquad y_0 = 1 \;   .
$$
The semigroup of endomorphisms, with discrete time, forms a dynamical system in
discrete time, termed cascade \cite{Smale_153,Saperstone_154}. In the present case,
the semigroup of endomorphisms
\be
\label{3.8}
\{ y_k : ~ \mathbb{Z}_+ \times \mathbb{R} \ra \mathbb{R} \}
\ee
is called {\it approximation cascade}, since its trajectory is bijective with the
approximation sequence,
\be
\label{3.9}
 \{ y_k(f) : ~ k \in \mathbb{Z}_+ \} \longleftrightarrow
\{ \widetilde f_k(x) : ~ k \in \mathbb{Z}_+ \} \;  .
\ee

Summarizing, a sequence of the approximants $F_k(x,u)$, with introduced control
parameters $u$, can be transformed into a sequence of endomorphisms. Assuming the
existence of a fixed point, the sequence of these endomorphisms exhibits the property
of self-similarity, which results in a dynamical system in discrete time, whose role
is played by the approximation order $k$. This dynamical system is named approximation
cascade. The trajectory of the cascade, by construction, is bijective with the
approximation sequence. And the fixed point of the cascade is isomorphic to the sought
function $f(x)$. Hence, to find the latter, one needs to study the evolution of the
cascade for defining its fixed point. Being limited by a finite number of perturbative
terms, usually we can find only an approximate fixed point $y^*_k(f)$. Respectively,
the latter yields a self-similar approximant $f^*_k(x)$ for the sought function,
so that
\be
\label{3.10}
  f_k^*(x) \equiv y_k^*( F_0(x,u_k(x) )) \; .
\ee

Note that, in addition to, or instead of control functions defined by optimization
conditions, it is possible to introduce control functions prescribed by asymptotic
conditions, as will be shown in the following sections.

\subsection{Approximation Flow}

It is more convenient to deal with a dynamical system in continuous time then with
a system in discrete time. To pass to a system in continuous time, we embed the
approximation cascade into a flow,
\be
\label{3.11}
 \{ y_k(f) : ~ k \in \mathbb{Z}_+ \} \subset
\{ y(t,f) : ~ t \in \mathbb{R}_+ \} \;  .
\ee
The flow is a semigroup of endomorphisms, transforming a measurable space into
itself,
\be
\label{3.12}
 \{ y(t,f) : ~  \mathbb{R}_+ \times \mathbb{R} \ra \mathbb{R} \} \; ,
\ee
and satisfying the semigroup property, that is, the same self-similarity relation
\be
\label{3.13}
y(t+t',f) = y(t,y(t',f) ) \;   .
\ee
And embedding requires that the trajectory of the flow passes through all points of
the cascade trajectory,
\be
\label{3.14}
y(t,f) = y_k(f) \qquad ( t = k ) \;  .
\ee
Thus the flow, embedding the approximation cascade, can be called the approximation
flow.

For a dynamical system in continuous time, with the self-similarity relation
(\ref{3.13}), it is possible to write the Lie equation
\be
\label{3.15}
 \frac{\prt}{\prt t}\; y(t,f) = v(y(t,f) ) \;  ,
\ee
with the velocity field
\be
\label{3.16}
  v(y(t,f) ) = \lim_{\tau\ra 0} \; \frac{\prt}{\prt \tau}\; y(\tau,y(t,f) )\;  .
\ee

The Lie evolution equation can be integrated between a point $y_k = y_k(f)$ of the
approximation cascade and an approximate fixed point $y^*_k = y^*_k(f)$ that can be
termed a quasi-fixed point. The integration gives the evolution integral
\be
\label{3.17}
 \int_{y_k}^{y_k^*} \frac{dy}{v(y)} = t_k \;  ,
\ee
where $t_k$ is the effective time needed for moving from the point $y_k$ to the
point $y^*_k$. The motion time $t_k$ is defined by additional conditions, e.g.,
from asymptotic conditions. The velocity in the interval between $y_k$ and $y^*_k$
is given by the cascade velocity $v_k$.

By the construction in the previous section, the cascade trajectory point $y_k(f)$
corresponds to the approximant $\tilde{f}_k(x)$, while the quasi-fixed point $y^*_k(f)$,
to a self-similar approximant  $f^*_k(x)$. Therefore the evolution integral can be
rewritten as
\be
\label{3.18}
 \int_{\tilde f_k}^{f_k^*} \frac{df}{v_k(f)} = t_k \;  ,
\ee
with $\tilde{f}_k = \tilde{f}_k(x)$ and $f^*_k = f^*_k(x)$.

If the cascade velocity were zero, then the self-similar approximant  $f^*_k(x)$
would coincide with the approximant $\tilde{f}_k(x)$. But generally, the cascade
velocity can be defined by the finite difference
\be
\label{3.19}
 v_k(f) = y_{k+1}(f) - y_k(f) \;  ,
\ee
which is equivalent to the difference
\be
\label{3.20}
 v_k(f) = F_{k+1}(x_{k+1},u_{k+1}) - F_k(x_k,u_k) \;  ,
\ee
in which $x_k=x_k(f)$ and $u_k = u_k(x_k)$. To reach faster the quasi-fixed point,
the cascade velocity is to be as small as possible, which can be achieved by choosing
the appropriate control functions. Thus we come to the minimization condition
\be
\label{3.21}
\min_u \left | F_{k+1}(x_{k+1},u_{k+1}) - F_k(x_k,u_k) \right | \;   .
\ee
This is nothing but the fastest convergence criterion (\ref{2.12}) of Sec 2.2. Using
here the expansion of $F_{k+1}(x_{k+1},u_{k+1})$ of Sec. 2.2 and limiting ourselves by
the first order, we get the cascade velocity
\be
\label{3.22}
 v_k(f) =  F_{k+1}(x_k,u_k) -  F_k(x_k,u_k) +
(u_{k+1} - u_k) \; \frac{\prt}{\prt u_k} \; F_k(x_k,u_k) \;  .
\ee
Defying the control functions from the optimization condition
\be
\label{3.23}
 (u_{k+1} - u_k) \; \frac{\prt}{\prt u_k} \; F_k(x_k,u_k) = 0 \; ,
\ee
we obtain the velocity
\be
\label{3.24}
 v_k(f) =  F_{k+1}(x_k,u_k) -  F_k(x_k,u_k) \; .
\ee

Note that the cascade velocity could be defined by the Euler discretization of the
evolution equation (\ref{3.15}) yielding either Eq. (\ref{3.19}) or
$$
 v_k(f) = y_k(f) - y_{k-1}(f) \;  ,
$$
with control functions given by some of the optimization conditions of Secs. 2.2
and 2.3.

\subsection{Stability Analysis}

For a dynamical system, we can accomplish its stability analysis, which in our case
is equivalent to studying the stability of the calculational procedure in the
corresponding approximation theory \cite{Yukalov_11,Yukalov_12,Yukalov_13,Yukalov_25,
Yukalov_145,Yukalov_150,Yukalov_155,Yukalov_156,Yukalov_157}.

The stability of a cascade is characterized by the local map multipliers
\be
\label{3.25}
\mu_k(f) \equiv \frac{\dlt y_k(f)}{\dlt y_0(f)} = \frac{\prt y_k(f)}{\prt f}
\ee
and the maximal map multipliers
\be
\label{3.26}
 \mu_k \equiv \sup_f | \mu_k(f) | \;  ,
\ee
where the supremum is taken with respect to the whole range of $\tilde{f}_k(x)$.
The stability of the quasi-fixed points is described by the local map multipliers
\be
\label{3.27}
\mu_k^*(f) \equiv \frac{\dlt y_k^*(f)}{\dlt y^*_0(f)}
\ee
and the maximal map multiplier
\be
\label{3.28}
  \mu_k^* \equiv \sup_f | \mu_k^*(f) | \; ,
\ee
with the supremum over the whole range of $f_k^*(x)$. The corresponding conditions
of local stability are
\be
\label{3.29}
| \mu_k(f) | < 1 \; , \qquad \mu_k < 1 \; ,
\ee
and
\be
\label{3.30}
| \mu_k^*(f) | < 1 \; , \qquad \mu^*_k < 1 \;    .
\ee
When a multiplier equals one, this corresponds to neutral stability.

Equivalently, we can define the local map multipliers
\be
\label{3.31}
m_k(x) \equiv \frac{\dlt \widetilde f_k(x)}{\dlt \widetilde f_0(x)} =
\mu_k(F_0(x,u_k(x) )
\ee
and the maximal map multipliers
\be
\label{3.32}
  m_k \equiv \sup_x | m_k(x) | \;  ,
\ee
with the supremum over the whole domain of $x$. And the stability of the quasi-fixed
points is defined by the local multipliers
\be
\label{3.33}
m^*_k(x) \equiv \frac{\dlt f^*_k(x)}{\dlt f^*_0(x)}
\ee
and the maximal multipliers
\be
\label{3.34}
 m_k^* \equiv \sup_x | m^*_k(x) | \;    ,
\ee
with the supremum over the domain of $x$. Under $f_0^*(x)$ we mean here the first
nontrivial approximant, since the zero, in the direct sense, self-similar approximant
maybe not defined. Respectively, the conditions of local stability are
\be
\label{3.35}
 | m_k(x) | < 1 \; , \qquad m_k < 1 \;
\ee
and
\be
\label{3.36}
 | m_k^*(x) | < 1 \; , \qquad m^*_k < 1 \;   .
\ee

The above multipliers characterize stability, with respect to the variation of
initial conditions, at a $k$-th step of the calculational procedure. It is also
possible to introduce the ultralocal multipliers
\be
\label{3.37}
 \overline \mu_k(f) \equiv \frac{\dlt y_k(f)}{\dlt y_{k-1}(f)}  =
\frac{\mu_k(f)}{\mu_{k-1}(f)}
\ee
and maximal ultralocal multipliers
\be
\label{3.38}
  \overline \mu_k \equiv \sup_f |  \overline \mu_k(f) | \;  ,
\ee
characterizing the stability at a $k$-th step with respect to the variation of
the previous step. Respectively, the images of these ultralocal multipliers are
\be
\label{3.39}
\overline m_k(x) \equiv \frac{\dlt \widetilde f_k(x)}{\dlt \widetilde f_{k-1}(x)}
 = \frac{m_k(x)}{m_{k-1}(x)} =  \overline \mu_k( F_0(x,u_k(x) ) )
\ee
and the maximal ultralocal multiplier
\be
\label{3.40}
 \overline m_k \equiv \sup_x |  \overline m_k(x) | \;  .
\ee
The procedure is ultralocally stable if either
\be
\label{3.41}
| \overline \mu_k(f) | < 1 \; , \qquad \overline \mu_k < 1 \; ,
\ee
or
\be
\label{3.42}
| \overline m_k(x) | < 1 \; , \qquad \overline m_k < 1 \;   .
\ee

Finally, we can define the ultralocal multipliers in the vicinity of quasi-fixed
points,
\be
\label{3.43}
\overline \mu_k^*(f) \equiv \frac{\dlt y^*_k(f)}{\dlt y^*_{k-1}(f)}
 = \frac{\mu^*_k(f)}{\mu^*_{k-1}(f)}
\ee
and the related maximal multipliers
\be
\label{3.44}
\overline \mu^*_k  \equiv \sup_f |  \overline \mu^*_k(f) | \;  .
\ee
The images of these multipliers are
\be
\label{3.45}
\overline m^*_k(x) = \frac{\dlt f^*_k(x)}{\dlt f^*_{k-1}(x)}
 = \frac{m^*_k(x)}{m^*_{k-1}(x)} = \overline\mu_k^*(F_0(x,u_k(x) ) )
\ee
and
\be
\label{3.46}
\overline m^*_k  \equiv \sup_x |  \overline m^*_k(x) | \;   .
\ee
The corresponding stability conditions are either
\be
\label{3.47}
|  \overline \mu^*_k(f) | < 1 \; , \qquad \overline\mu^*_k  < 1 \; ,
\ee
or
\be
\label{3.48}
 |  \overline m^*_k(x) | < 1 \; , \qquad \overline m^*_k  < 1 \;  .
\ee

It is also possible to introduce the local Lyapunov exponents
\be
\label{3.49}
 \lbd_k(f) \equiv \frac{1}{k} \; \ln | \mu_k(f) | \;  .
\ee
Then the local stability requires that $\lambda_k(f)<0$. Considering the
variation
$$
\dlt y_k(f) = y_k(f + \dlt f) - y_k(f) \simeq \mu_k(f) \dlt f
$$
for a small $\delta f$, one sees that the local stability implies the exponential
stability, since
$$
 | \dlt y_k(f) | \simeq | \dlt f | \exp\{ \lbd_k(f) k \} \;  .
$$
In rough words, one can say that when $\lambda_k(f)<0$, there is exponential
convergence.

A dynamical system not necessarily enjoys the existence of an attractive fixed point,
but may possess other kinds of attractors or repellers, including chaotic attractors,
periodic or quasi-periodic orbits, and like that \cite{Eckmann_158,Eckmann_159,Ford_160}.
If a dynamical system exhibits unstable motion, then its points of the trajectory may
fluctuate without approaching a limit. However convergence on average may exist,
when the effective limit is defined as an ergodic average \cite{Halmos_161}.
If such a case happens for an approximation cascade, then the resulting approximants
should be understood in the sense of the ergodic average
\be
\label{3.50}
 \overline f_k(x) \equiv \frac{1}{k} \sum_{n=1}^k f_n^*(x) \;  .
\ee

\subsection{Free Energy}

Self-similar approximation theory has been applied to solving numerous problems.
In the present review, we illustrate its application just for several typical cases.
One such a typical case is the zero-dimensional $\vp^4$ theory, with the partition
function (\ref{2.34}). We calculate the free energy
\be
\label{3.51}
 f(g) = -\ln Z(g) \;  .
\ee
Details of the procedure can be found in Refs. \cite{Yukalov_13,Yukalov_162,Yukalov_163}.

At the first step, we have to introduce control functions, which we can accomplish as
has been done in Secs. 2.5 and 2.6, starting from the initial approximating Hamiltonian
(\ref{2.61}). Then we follow the approach described in the previous sections.

The equation defining an expansion function (\ref{3.1}) now reads as
\be
\label{3.52}
F_0(g,\om_k(g) ) = f \; , \qquad g = g_k(f) \;   ,
\ee
which is equivalent to the equality
$$
 \ln \om_k(g) = f \;  .
$$
The control functions $\omega_k(g)$ are defined by condition (\ref{3.23}), as in
Sec. 2.6, which results in the expansion function
\be
\label{3.53}
 g_k(f) = \frac{e^{2f}}{3s_k} \; \left ( e^{2f} - 1 \right ) \;  ,
\ee
where $s_k$ are given in Sec. 2.6.

The endomorphism (\ref{3.3}) takes the form
\be
\label{3.54}
 y_k(f) = f + \sum_{n=1}^k A_{kn} \al^n(f) \;  ,
\ee
where
$$
A_{kn} = \sum_{m=0}^n \frac{C_{nm}}{s_k^m} \qquad ( n \leq k )
$$
and
$$
 \al(f) \equiv 1 - e^{-2f} \;  .
$$
The coefficients $A_{kn}$ are as follows: In first order,
$$
 A_{11} = -\; \frac{1}{4} \qquad ( k = 1 ) \;  .
$$
In second order,
$$
  A_{21} = -\; \frac{1}{4} \; , \qquad A_{22} = -\; \frac{1}{12}
\qquad ( k = 2 ) \; .
$$
In third order,
$$
  A_{31} = - 0.388377 \; , \qquad A_{32} = -0.093205 \; , \qquad
A_{33} = -0.016011 \qquad ( k = 3) \;  .
$$
And in fourth order,
$$
A_{41} =  A_{31} \; , \qquad A_{42} =  A_{32} \; , \qquad
A_{43} =  A_{33} \; , \qquad A_{44} = -0.003606 \qquad ( k = 4) \; .
$$
The cascade velocity (\ref{3.24}) reads as
\be
\label{3.55}
  v_k(f) = B_k \al^{k+1}(f) \qquad ( B_k \equiv A_{k+1,k+1} ) \;  .
\ee
By introducing the notations
\be
\label{3.56}
 f_k^*(g) \equiv \ln\; \sqrt{1+z_k^*(g)} \; , \qquad
f_k(g) \equiv \ln\; \sqrt{1+z_k(g)} \; ,
\ee
the evolution integral (\ref{3.18}) can be written as
\be
\label{3.57}
 \int_{z_k}^{z_k^*} \frac{(1+z)^k}{z^{k+1}} \; dz = 2B_k t_k \;  ,
\ee
where $z_k =z_k(g)$ and $z^*_k =z^*_k(g)$. After integrating, we get the expression
\be
\label{3.58}
 z_k^* = z_k \exp\left\{ P_k \left ( \frac{1}{z_k^*} \right ) -
P_k\left ( \frac{1}{z_k} \right ) + 2B_k t_k \right \} \;  ,
\ee
in which
$$
P_k(x) \equiv \sum_{n=0}^{k-1} C_k^n \; \frac{x^{k-n}}{k-n} \; , \qquad
C_k^n \equiv \frac{k!}{(k-n)! n!} \;   .
$$

The map multiplier (\ref{3.25}) is
\be
\label{3.59}
\mu_k(f) = 1 + 2 [ 1 - \al(f) ] \sum_{n=1}^k n A_{kn} \al^{n-1}(f) \; .
\ee
For $g \in [0,\infty)$, the function $f_k(g)$ is in the range $[0,\infty)$ and
$\al(f)\in [0, 1]$. Then the maximal multiplier (\ref{3.26}) is smaller than unity,
$\mu_k<1$. This means that the procedure is stable and we can set $t_k=1$.

The accuracy of the procedure is characterized by the percentage errors
\be
\label{3.60}
 \ep_k^*(g) \equiv \left [ \frac{f_k^*(g)}{f(g)} \; - \; 1 \right ] \times 100\%
\ee
and the maximal percentage error
\be
\label{3.61}
  \ep_k^* \equiv \sup_g |\ep_k^*(g) | \;  .
\ee
We find
$$
\ep_1^* = 3\% \; , \qquad  \ep_2^* = 2\% \; , \qquad  \ep_3^* = 0.1\% \; .
$$
Comparing with the optimized approximants of Sec. 2.6, we see that
$\varepsilon^*_k < \varepsilon_k$.

\subsection{Eigenvalue Problem}

As a typical example of an eigenvalue problem (\ref{2.70}), let us consider the
anharmonic oscillator of Sec. 2.7. We are looking for the energy levels
\be
\label{3.62}
E_{kn}(g,\om) = \left ( n + \frac{1}{2} \right ) F_k(g,\om) \;  ,
\ee
where the index $n$ enumerates the levels. Here we show the main steps of
calculations, while the details can be found in Refs.
\cite{Yukalov_13,Yukalov_23,Yukalov_157,Yukalova_164,Yukalov_165}.

The expansion function follows from the equation
\be
\label{3.63}
 F_0(g,\om_k(g)) = f \; , \qquad g = g_k(f) \;  ,
\ee
giving
$$
 \om_k(g) = f \;  .
$$
Control functions are defined as in Sec. 2.7, being the solutions to the
equation
\be
\label{3.64}
 \om_k^3 - \om_k - 6\gm \lbd_k g = 0 \;  ,
\ee
in which $\lbd_1=1$, $\lbd_2=\lbd_3=\lbd$, with $\lbd$ given by equations (\ref{2.94})
and (\ref{2.95}).

Then the expansion function becomes
\be
\label{3.65}
 g_k(f) = \frac{f(f^2-1)}{6\gm\lbd_k} \;  .
\ee
And the cascade velocity (\ref{3.24}) is
\be
\label{3.66}
 v_k(f) = C_k f \left ( 1 - \; \frac{1}{f^2} \right )^{k+1} \;  ,
\ee
where
$$
C_0 = -\; \frac{A_{10}}{2^2} \; , \qquad C_1 = -\; \frac{A_{11}}{2^3} \; ,
\qquad
C_2 = -\; \frac{A_{32}}{2^4} \; , \qquad C_3 = -\; \frac{A_{33}}{2^5} \;  .
$$

The evolution integral, setting $t_k = 1$, reads as
\be
\label{3.67}
 \int_{f_k}^{f_k^*} \frac{f^{2k+1}}{(f^2-1)^{k+1}} \; df =  C_k \;  .
\ee
Using the notations
$$
f_k^*(g) \equiv \sqrt{1+z_k^*(g)} \; , \qquad f_k(g) \equiv \sqrt{1+z_k(g)} \; ,
$$
we obtain
\be
\label{3.68}
 z_k^* = z_k \exp \left\{ P_k\left ( \frac{1}{z_k^*}\right ) -
 P_k\left ( \frac{1}{z_k}\right ) + 2C_k \right \} \; ,
\ee
where the polynomial $P_k(x)$ is defined in the previous Sec. 3.4.

The endomorphism
$$
y_k(f) = F_k(g_k(f),\om_k(g_k(f)) = \widetilde f_k(g_k(f))
$$
takes the form
\be
\label{3.69}
 y_k(f) = f + f \sum_{j=1}^k C_{j-1}
\left ( 1 - \; \frac{1}{f^2}\right )^j \;  .
\ee

The local map multipliers are
\be
\label{3.70}
 \mu_k(f) = \frac{\prt y_k(f)}{\prt f} = 1 + \sum_{j=1}^k C_{j-1}
\left ( 1 - \; \frac{1}{f^2}\right )^{j-1}
\left ( 1 - \; \frac{1-2j}{f^2}\right ) \;  .
\ee
For all $f \in [1, \infty)$ and any $n$, we have $\mu_k < 1$, hence the procedure
is stable. But ultralocal multipliers (\ref{3.37}) for some energy levels become
larger then one, because of which the numerical convergence is not monotonic. The
related maximal errors are
$$
\ep_1^* = 0.4\% \; , \qquad \ep_2^* = 0.4\% \; , \qquad \ep_3^* = 0.65\% \;   .
$$
Nevertheless, for all considered $k$, the accuracy of the self-similar approximants
is better than that of the optimized approximants of Sec. 2.7, since $\ep_k^*<\ep_k$.

Several other eigenvalue problems have been treated employing the self-similar
approximation theory, as in the present section. Among them: the three-dimensional
spherical anharmonic oscillator, with Hamiltonian (\ref{2.195}) \cite{Yukalova_67},
the anharmonic oscillator of arbitrary dimensionality, with Hamiltonian (\ref{2.196})
\cite{Yukalova_68}, power-law potentials, with one-dimensional Hamiltonians (\ref{2.197})
\cite{Yukalov_13,Yukalova_69}, spherically symmetric power-law potentials in three
dimensions, with Hamiltonians (\ref{2.198}) \cite{Yukalov_13,Coleman_71,Yukalov_72},
spherically-symmetric power-law Hamiltonians (\ref{2.199}) of arbitrary dimensionality
\cite{Yukalov_13,Yukalova_69}, quasistationary states characterized by the Hamiltonian
(\ref{2.200}) \cite{Yukalov_75}, double-well oscillator, described by Hamiltonian
(\ref{2.201}) \cite{Yukalov_75,Yukalov_76}, Hamiltonians (\ref{2.202}) with the
Yukawa potential \cite{Yukalov_77}, and logarithmic potential \cite{Yukalov_13},
with Hamiltonian (\ref{2.203}). These examples show that the accuracy of self-similar
approximants is higher than that of optimized approximants.

\subsection{Choice of Initial Approximation}

Self-similar approximation theory makes it possible to improve the accuracy of
optimized approximants and analyze the stability of the procedure. Also, it allows
us to choose the best initial approximation, when several such approximations are
admissible. Here we illustrate this by considering the spherically symmetric
Hamiltonian with a logarithmic potential,
\be
\label{3.71}
\hat H = -\; \frac{1}{2} \; \frac{d^2}{dr^2} +
\frac{l(l+1)}{2r^2} + g\ln r \;  ,
\ee
where $r > 0$ and $l = 0, 1, 2, \ldots$. Perturbation theory for this Hamiltonian
can be started either with the harmonic-oscillator Hamiltonian
\be
\label{3.72}
 \hat H_0 = -\; \frac{1}{2} \; \frac{d^2}{dr^2} + \frac{l(l+1)}{2r^2} +
\frac{\om^2}{2} \; r^2 \;  ,
\ee
in which $\omega$ is a control parameter, or with the Coulomb Hamiltonian
\be
\label{3.73}
 \hat H_0 = -\; \frac{1}{2} \; \frac{d^2}{dr^2} + \frac{l(l+1)}{2r^2} +
\frac{u}{r} \; ,
\ee
where $u$ is a control parameter.

Accomplishing calculations for both these cases \cite{Yukalov_13}, we find that the
procedure starting from the harmonic oscillator is more stable than that starting
from the Coulomb potential, since the map multipliers $\mu_k$ are about twice smaller
for the former procedure than that for the latter. In the present case, the higher
stability is in agreement with the better accuracy, since the errors $\varepsilon_k$
of the energy levels for the procedure, starting with the harmonic oscillator, are
about twice smaller than those for the procedure starting with the Coulomb potential.

Thus, when it is possible to take several initial approximations, and the exact
answers are not known, one should select that initial approximation that provides
the most stable calculational procedure. In the considered case, it was possible to
also check the accuracy of the results for the energy levels, comparing them with
numerical calculations based on the direct solution of the Schr\"{o}dinger equation,
which confirmed that the higher accuracy is in line with the higher stability.

\subsection{Fractal Transform}

In Sec. 2.1, it was mentioned that control functions can be introduced through
series transformations. For this purpose, we opt for a transformation that, by
the use of the self-similar approximation theory, could extract a hidden similarity
between the sequence of perturbative terms and would allow for the derivation of
recurrence relations making it possible to obtain an explicit sequence of self-similar
approximants. Here we show that this can be done by using {\it fractal transform}
\cite{Yukalov_166,Gluzman_167,Yukalov_168}.

First, let us define the {\it fractal transform}. For a function $f(x)$, the fractal
transform is given by the equation
\be
\label{3.74}
 F(x,s) \equiv x^s f(x) \;  ,
\ee
with the inverse transformation
\be
\label{3.75}
 f(x) = x^{-s} F(x,s) \;  .
\ee
Here $s$, playing the role of a control function, is called {\it scaling exponent},
since it enters the following scaling relation:
\be
\label{3.76}
\frac{F(\lbd x,s)}{f(\lbd x)} = \lbd^s \; \frac{F(x,s)}{f(x) } \; .
\ee
Scaling relations of this type appear in several physical problems considering
fractal systems \cite{Meakin_169}. This is why, we call the expression $F(x,s)$,
defined by equation (\ref{3.74}), fractal transform.

Suppose we are looking for a function $f(x)$ for which we can get only an
asymptotic expansion at small $x \ra 0$,
\be
\label{3.77}
 f(x) \simeq f_k(x) \qquad ( x \ra 0 ) \;  ,
\ee
where
\be
\label{3.78}
 f_k(x) = f_0(x) \left ( 1 + \sum_{n=1}^k a_n x^{\al_n} \right ) \;  ,
\ee
with the powers arranged in increasing order,
$$
 \al_n < \al_{n+1} \qquad ( n = 1,2,\ldots,k-1) \;  ,
$$
and $f_0(x)$ being a known function.

For what follows, it is sufficient to consider the expansions in the form
\be
\label{3.79}
 f_k(x) = 1 + \sum_{n=1}^k a_n x^{\al_n} \;  ,
\ee
since, the initial form (\ref{3.78}) is easily obtained by the substitution
$$
 f_k(x) \ra f_0(x) f_k(x) \;  .
$$
The fractal transform of expansion (\ref{3.79}) is
\be
\label{3.80}
 F_k(x,s_k) = x^{s_k} f_k(x) \;  .
\ee

In order to show the result of the application of self-similar approximation theory
to asymptotic expansions, let us consider a first-order expansion
\be
\label{3.81}
  f_1(x) = 1 + a_1 x^{\al_1} \;  ,
\ee
whose fractal transform is
\be
\label{3.82}
 F_1(x,s_1) = x^{s_1} + a_1 x^{\al_1+s_1} \;  .
\ee

The equation
\be
\label{3.83}
F_0(x,s_k) = x^{s_k} = f
\ee
gives the expansion function
\be
\label{3.84}
 x_k(f) = f^{1/s_k} \;  .
\ee
Then we get the first-order endomorphism
\be
\label{3.85}
 y_1(f) \equiv F_1(x_1(f),s_1) = f + a_1 f^{1+\al_1/s_1} \;  ,
\ee
with the initial condition
\be
\label{3.86}
  y_0 = f \; .
\ee

The evolution integral (\ref{3.17}) reads as
\be
\label{3.87}
 \int_{y_0}^{y_1^*} \frac{df}{v_1(f)} = t_1 \;  ,
\ee
with the cascade velocity
\be
\label{3.88}
 v_1(f) = y_1(f) - y_0(f) = a_1 f^{1+\al_1/s_1} \;  .
\ee
From here, the first-order fixed point is
\be
\label{3.89}
y_1^*(f) = \left ( f^{1/n_1} + A_1 \right )^{n_1} \; ,
\ee
where
$$
n_1 \equiv - \; \frac{s_1}{\al_1} \; , \qquad
A_1 \equiv \frac{a_1 t_1}{n_1} \;   .
$$
And the corresponding first-order self-similar approximant
\be
\label{3.90}
f_1^*(x) = x^{-s_1} y_1^*(x^{s_1} )
\ee
becomes
\be
\label{3.91}
 f_1^*(x) = \left ( 1 + A_1 x^{\al_1}\right )^{n_1} \;  .
\ee
The control functions $A_1$ and $n_1$ can be found from asymptotic conditions, as
will be explained below. The procedure of using the fractal transform and constructing
self-similar approximants, as is done above, will be called, for short, self-similar
renormalization.

\subsection{Self-Similar Root Approximants}

Applying the techniques of the previous section to the asymptotic expansion (\ref{3.79}),
it is possible to derive different variants of self-similar approximants. In the present
section, we derive root approximants
\cite{Yukalov_11,Yukalov_75,Gluzman_170,Yukalov_171,Yukalov_172}.

Accomplishing the fractal transform (\ref{3.80}) for the asymptotic series (\ref{3.79}),
we have
\be
\label{3.92}
 F_k(x,s_k) = x^{s_k} + \sum_{n=1}^k a_n x^{\al_n+ s_k} \;  .
\ee
With the reonomic constraint (\ref{3.83}), we get the expansion function (\ref{3.84}).
Then the cascade endomorphism reads as
\be
\label{3.93}
  y_k(f) = f + \sum_{n=1}^k a_n f^{1+\al_n/s_k} \;   .
\ee

Integrating the evolution equation (\ref{3.15}) between $y_{k-1}^*(f)$ and $y_k^*(f)$,
we get the integral
\be
\label{3.94}
 \int_{y_{k-1}^*}^{y_k^*} \frac{df}{v_k(f)} = t_k \;  .
\ee
The cascade velocity can be taken as
\be
\label{3.95}
 v_k(f) = y_k(f) - y_{k-1}(f) = a_n f^{1+\al_k/s_k} \;  .
\ee
Hence integral (\ref{3.94}) yields the recurrent relation
\be
\label{3.96}
 y_k^*(f) = \left\{ \left[ y_{k-1}^*(f) \right ]^{1/m_k} + A_k
\right \}^{m_k} \;  ,
\ee
in which
$$
m_k \equiv -\; \frac{s_k}{\al_k} \; , \qquad
A_k \equiv \frac{a_k t_k}{m_k} \; .
$$
This, for the self-similar approximant
\be
\label{3.97}
f_k^*(x) = x^{-s_k} y_k\left ( x^{s_k} \right )
\ee
transforms into the recurrent relation
\be
\label{3.98}
 f_k^*(x) = \left\{ \left[ f_{k-1}^*(x) \right]^{1/m_k} + A_k x^{\al_k}
\right \}^{m_k} \;  .
\ee
Iterating this relation $k-1$ times and using the notation
$$
 n_j \equiv \frac{m_j}{m_{j+1}} \qquad ( j = 1,2,\ldots,k-1) \;  ,
$$
we come to the {\it self-similar root approximant}
\be
\label{3.99}
  f_k^*(x) = \left ( \left ( \left ( 1 + A_1 x^{\al_1} \right )^{n_1} +
A_2 x^{\al_2}\right )^{n_2} + \ldots + A_k x^{\al_k} \right )^{m_k} \;  .
\ee
The control parameters $A_j$, $n_j$, and $m_k$ can be found from asymptotic
conditions. There can happen several cases.

\vskip 2mm

(i) {\it Large-variable expansion with $k$ terms is available}.

\vskip 2mm

This expansion can be written as
\be
\label{3.100}
 f(x) \simeq \sum_{n=1}^k b_n x^{\bt_n} \qquad ( x \ra \infty) \;  ,
\ee
where $b_1 \neq 0$, $\beta_1 \neq 0$, and the powers are arranged in the descending
order,
$$
\bt_{n+1} < \bt_n \qquad ( n = 1,2,\ldots,k-1 ) \;   .
$$
All control parameters can be found by equating the large-variable expansion of the
self-similar root approximant (\ref{3.99}) with the given expansion (\ref{3.100}),
that is, by equating
\be
\label{3.101}
  \sum_{n=1}^k b_n x^{\bt_n} \simeq f_k^*(x) \qquad ( x \ra \infty) \; .
\ee
The powers $n_j$ and $m_k$ satisfy the equations \cite{Yukalov_11}
\be
\label{3.102}
\al_k m_k = \bt_1 \; ,  \qquad
  \al_j n_j = \al_{j+1} - \bt_{k-j} + \bt_{k-j+1} \qquad
( j = 1,2,\ldots,k-1 ) \; .
\ee

In the particular case of expansion (\ref{3.79}) with integer powers $\al_n=n$,
\be
\label{3.103}
 f_k(x) = 1 + \sum_{n=1}^k a_n x^n \;  ,
\ee
the self-similar root approximant (\ref{3.99}) becomes
\be
\label{3.104}
 f_k^*(x) = \left ( \left ( \left ( 1 + A_1 x \right )^{n_1} +
A_2 x^2 \right )^{n_2} + \ldots + A_k x^k \right )^{m_k} \;  .
\ee
Then its powers are defined by the equations
\be
\label{3.105}
 k m_k = \bt_1 \; , \qquad  j n_j = j+1 - \bt_{k-j} + \bt_{k-j+1} \qquad
( j = 1,2,\ldots,k-1 ) \; .
\ee
The parameters $A_j$ are defined through the values of $b_j$. For instance
\be
\label{3.106}
 \left ( \left ( \left ( A_1^{n_1} + A_2 \right )^{n_2} +
A_3 \right )^{n_3} + \ldots + A_k \right )^{m_k} = b_1 \;  .
\ee

\vskip 2mm

(ii) {\it Only one term of the large-variable expansion is known, while $k-1$
terms of the small-variable expansion are available}.

\vskip 2mm

Let the known term of the large-variable expansion be
\be
\label{3.107}
 f(x) \simeq B x^\bt \qquad ( x \ra \infty ) \;  .
\ee
Notice that the previous case reduces to the present one by setting $\bt_n=\bt$.
Then Eqs. (\ref{3.102}) reduce to the equalities
\be
\label{3.108}
 m_k = \frac{\bt}{\al_k} \; , \qquad n_j =   \frac{\al_{j+1}}{\al_j} \qquad
( j = 1,2, \ldots, k-1 ) \; .
\ee
The $k-1$ parameters $A_j$ can be found by expanding the self-similar root approximant
(\ref{3.99}) in powers of small $x$ and equating the like terms with those of the
small-variable expansion (\ref{3.79}),
$$
 f_k(x) \simeq f_k^*(x) \qquad ( x \ra 0 ) \;  .
$$
And the additional equation is Eq. (\ref{3.106}).

In the case of integer $\alpha_j = j$, we have
\be
\label{3.109}
 m_k = \frac{\bt}{k} \; , \qquad n_j =   \frac{j+1}{j} \qquad
( j = 1,2, \ldots, k-1 ) \;  .
\ee
So that the root approximant takes the form
\be
\label{3.110}
  f_k^*(x) = \left ( \left ( \left ( 1 + A_1 x \right )^2 +
A_2 x^2 \right )^{3/2} + \ldots + A_k x^k \right )^{\bt/k} \; .
\ee
The control parameters $A_j$ are defined by the asymptotic condition at small variable
plus the equation
\be
\label{3.111}
 \left ( \left ( \left ( A_1^2 + A_2 \right )^{3/2} +
A_3 \right )^{4/3} + \ldots + A_k \right )^{\bt/k} = b_1 \;  .
\ee

\vskip 2mm

(iii) {\it The known large-variable expansion contains $p$ terms and the small-variable
expansion, $k-p$ terms}.

\vskip 2mm

When the large-variable expansion is
\be
\label{3.112}
  f(x) \simeq \sum_{n=1}^p b_n x^{\bt_n} \qquad ( x \ra \infty) \;  ,
\ee
with $p < k$, then the lower powers are given by the expressions
\be
\label{3.113}
  n_j =   \frac{\al_{j+1}}{\al_j} \qquad ( j = 1,2, \ldots, k-p ) \;   ,
\ee
while the higher powers, by the equations
$$
\al_k m_k = \bt_1 \; ,
$$
\be
\label{3.114}
 \al_j n_j = \al_{j+1} - \bt_{k-j} + \bt_{k-j+1} \quad
 ( j = k-p+1, k-p+2, \ldots, k-1 ) \; .
\ee

\vskip 2mm

(iv) {\it Only a small-variable expansion is given}

\vskip 2mm

In this case, we set the internal powers according to the rule
$$
 n_j =   \frac{\al_{j+1}}{\al_j} \qquad ( j = 1,2, \ldots, k-1 ) \;  ,
$$
while the external power $m_k$ and all parameters $A_j$ are defined by the
accuracy-through-order procedure \cite{Yukalov_276}.

\vskip 2mm

In a particular situation of case (i), it may happen that
$n_j = 1$ for $j = 1, 2, \ldots, k-1$, which occurs if
$$
\al_{j+1} - \al_j = \bt_{k-j} - \bt_{k-j+1} \qquad ( j = 1,2, \ldots, k-1 ) \; ,
$$
$$
\al_k m_k = \bt_1 \qquad ( n_j = 1 )  \; .
$$
In such a case, the root approximant reads as
\be
\label{3.115}
 f_k^*(x) = \left ( 1 + A_1 x^{\al_1} + A_2 x^{\al_2} + \ldots +
A_k x^{\al_k} \right )^{m_k} \;  .
\ee
In the case of integer $\alpha_j = j$, $n_j$ can be equal to one, if
$$
\bt_{k-j} - \bt_{k-j+1} = 1 \qquad ( n_j = 1 , ~ \al_j = j ) \; ,   .
$$
Then approximant (\ref{3.115}) becomes
\be
\label{3.116}
 f_k^*(x) = \left ( 1 + A_1 x + A_a x^2 + \ldots + A_k x^k \right )^{m_k} \; .
\ee

From their structure, it is clear that the root approximants are convenient for the
problem of interpolation between two asymptotic limits of the sought function.

Also, notice that setting in approximant (\ref{3.104}) all powers $n_j = \pm 1$, we
get different variants of Pad\'{e} approximants $P_{M/N}(x)$, with $M + N = k$. And
setting the internal powers $n_j = \pm 1$, while keeping the external power $m_k$
such that the asymptotic condition (\ref{3.107}) be valid, we get different variants
$[P_{M/N}(x)]^{m_k}$ of the modified Pad\'{e} approximants of Baker and Gammel
\cite{Baker_173}, where $(M - N)m_k = \beta$. So, Pad\'{e} approximants, including
the modified Pad\'{e} approximants, can be considered as just a particular case of
the root approximants. However, the root approximants are essentially more general,
allowing for the use of more information on the studied function and, as has been
checked \cite{Yukalov_11,Yukalov_75,Gluzman_170,Yukalov_171,Yukalov_172}, guaranteeing
much better accuracy.

\subsection{Self-Similar Nested Approximants}

Applying the self-similar renormalization, discussed in Sec. 3.7, in a bit different
way, it is straightforward to derive other types of self-similar approximants. For
instance, expansion (\ref{3.79}) can be rewritten in the form
\be
\label{3.117}
 f_k(x) = 1 + p_1(x) \;  ,
\ee
in which
\be
\label{3.118}
  p_1(x) = a_1 x^{\al_1} ( 1 + p_2(x) )\; , \qquad
p_2(x) = \frac{a_2}{a_1} \; x^{\al_2-\al_1} \; ( 1 + p_3(x) ) \;  ,
\ee
and so on till the last term
\be
\label{3.119}
 p_k(x) = \frac{a_k}{a_{k-1}} \; x^{\al_k-\al_{k-1} } \; .
\ee
The general relation between the terms $p_j(x)$ reads as
\be
\label{3.120}
 p_j(x) = \frac{a_j}{a_{j-1}} \; x^{\al_j-\al_{j-1} } ( 1 + p_{j+1}(x) )\;  ,
\ee
where
$$
 j = 1,2, \ldots, k \; ; \qquad a_0 \equiv 1 \; , ~ \al_0 \equiv 0 \; , ~
a_{k+1} \equiv 0 \;   .
$$

Applying the self-similar renormalization to $f_k(x)$, with $p_1(x)$ as a variable,
and to $p_j(x)$, with $p_{j+1}(x)$ as a variable, we obtain
\be
\label{3.121}
 f_k^*(x) =  \left ( 1 + A_1 p_1^*(x) \right )^{n_1}
\ee
and the recurrence relation
\be
\label{3.122}
 p_j^*(x) = \frac{a_j}{a_{j-1}} \; x^{\al_j-\al_{j-1} }
 \left ( 1 + A_{j+1}p^*_{j+1}(x) \right )^{n_{j+1}}\;   ,
\ee
where
$$
 A_j \equiv \frac{t_j}{n_j} \; , \qquad n_j \equiv - s_j \quad
( j = 1,2,\ldots, k ) \;  .
$$

Iterating relation (\ref{3.120}) $k-1$ times, and employing the notation
$$
 B_j \equiv \frac{a_j}{a_{j-1}} \; A_j = - \; \frac{a_j t_j}{a_{j-1} s_j} \;  ,
$$
we come to the self-similar nested approximant
\be
\label{3.123}
 f_k^*(x) = \left ( 1 + B_1 x^{\al_1} \left ( 1 + B_2 x^{\al_2-\al_1} \ldots
\left ( 1 + B_k x^{\al_k - \al_{k-1}} \right )^{n_k} \right )^{n_{k-1}} \ldots
\right )^{n_1}  \;  .
\ee
The parameters $B_j$ and $n_j$ are control parameters that need to be defined from
additional conditions. These parameters, of course, depend on the order of the
considered approximation $k$, so that, $B_j = B_j(k)$ and $n_j = n_j(k)$. We do not
use the complicated notations showing this dependence in order to keep formulas more
compact.

The form of approximants (\ref{3.123}) resembles that of nested radicals
\cite{Landau_174,Landau_175}, because of which we call them nested approximants.

At large $x$, approximant (\ref{3.123}) behaves as
\be
\label{3.124}
 f_k^*(x) \simeq B_k^* x^{n_1 S_k} \qquad ( x \ra \infty ) \;  ,
\ee
with the amplitude
\be
\label{3.125}
B_k^* = \left ( B_1 B_2^{n_2} B_3^{n_2 n_3} \ldots B_k^{n_2 n_3 \ldots n_k}
\right )^{n_1}
\ee
and the notation
\be
\label{3.126}
 S_k \equiv \al_1 + (\al_2 - \al_1) n_2 +  (\al_3 - \al_2) n_2 n_3 +
\ldots + (\al_k - \al_{k-1}) n_2 n_3 \ldots n_k \; .
\ee

If the behavior of the sought function at large variable is known, being
\be
\label{3.127}
 f(x) \simeq B x^\bt \qquad  ( x \ra \infty ) \; ,
\ee
then we have two conditions defining
\be
\label{3.128}
 B_k^* = B \; , \qquad n_1(k) = \frac{\bt}{S_k} \;  .
\ee

In the case, when the powers $\alpha_j = j$ in expansion (\ref{3.117}) are
integers, approximant (\ref{3.123}) simplifies to the form
\be
\label{3.129}
f_k^*(x) = \left ( 1 + B_1 x \left ( 1  + B_2 x \ldots \left ( 1 + B_k x
\right )^{n_k} \right )^{n_{k-1}}  \ldots \right )^{n_1}
\ee
and sum (\ref{3.126}) becomes
\be
\label{3.130}
S_k = 1 + n_2 + n_2 n_3 + \ldots + n_2 n_3 \ldots n_k \;   .
\ee

Generally, we can try to define the control parameters $B_j$ from the
accuracy-through-order procedure at small variable by equating the expansion of
approximant (\ref{3.123}) with expansion (\ref{3.117}). For example, for $B_j$ of
low orders this gives
$$
B_1 = \frac{a_1}{n_1} = \frac{a_1}{\bt} \; S_k \; , \qquad
B_2 = \frac{a_1^2 + n_1(2a_2-a_1^2)}{2a_1n_1n_2} =
\frac{a_1^2S_k + \bt(2a_2-a_1^2)}{2a_1\bt n_2} \; .
$$
However the powers $n_j$ cannot be uniquely defined from such a procedure, either
at small or large variables. To define the powers $n_j$, some additional conditions
are needed. Also, to avoid complex numbers the coefficients $B_j$ should be positive
for keeping the function $f_k^*(x)$ real valued. But $B_j$ can be negative, if the
described problem has critical points.

Notice that setting in Eq. (\ref{3.129}) all powers $n_j=\pm 1$, we get Pad\'{e}
approximants $P_{M/N}(x)$ or continued fractions, while setting the internal powers
$n_j=\pm 1$, with the external power $n_1$ given by condition (\ref{3.128}), we
come to the modified Pad\'e approximants $[P_{M/N}(x)]^{\beta/(M-N)}$. However, such
a choice is not uniquely defined, since it is possible to construct several Pad\'e
approximants satisfying the condition $M+N=k$. In addition, dealing with  Pad\'{e}
approximants, we confront the known difficulty of unphysical poles.

The other way could be by setting all powers, including $n_1$, equal, $n_j=q$,
and defining $q$ by the expression $q = \beta/ S_k$ following from condition
(\ref{3.128}). Then
$$
S_k = \sum_{j=0}^{k-1} q^j = \frac{1-q^k}{1-q}  \qquad ( k \geq 1 ) \;
$$
But in this way, to define $q$ uniquely, it is necessary to neglect the term $q^k$,
which disturbs condition (\ref{3.128}) (see Ref. \cite{Gluzman_176}).

One more possibility could be as follows. We can take the same number $q$ only for
all internal powers $n_j$, with $j = 2,3, \ldots, k$. But the external power $n_1(k)$
should be defined, as earlier, by condition (\ref{3.128}), so that $n_1(k) S_k=\bt$.
This leads to the self-similar nested approximant
$$
f_k^*(x) = \left ( 1 + B_1 x \left ( 1  + B_2 x \ldots \left ( 1 + B_k x
\right )^{q} \right )^{q}  \ldots \right )^{m_k} \;  .   ,
$$
in which
$$
m_k \equiv n_1(k) = \frac{\bt}{S_k} = \frac{1-q}{1-q^k} \; \bt \; .
$$
This way does not impose constraints on the value of $\beta$. The power $q$ can be
defined from additional conditions that would guarantee good numerical convergence
of the sequence $\{f_k^*(x)\}$.

To make the internal expressions nonnegative, we may set $q = 2$, which gives
$$
m_k = \frac{\bt}{2^k-1} \qquad ( q = 2 ) \; .
$$

\subsection{Self-Similar Exponential Approximants}

An important particular case of the nested approximants can be derived by setting the
powers $s_j$ of the fractal transform (\ref{3.80}) as $s_j\ra\infty$. Consequently,
$n_j\ra -\infty$, $A_j\ra 0$, and $B_j\ra 0$. Then expression (\ref{3.123}) reduces to
a self-similar exponential approximant \cite{Yukalov_168,Yukalov_177}
\be
\label{3.131}
 f_k^*(x) = \exp \left ( C_1 x^{\al_1} \exp \left ( C_2 x^{\al_2-\al_1}\ldots
\exp\left ( C_k x^{\al_k - \al_{k-1}} \right ) \right ) \right ) \;  ,
\ee
in which
\be
\label{3.132}
 C_j \equiv \frac{a_j}{a_{j-1}} \; t_j \qquad ( j = 1,2,\ldots, k ) \;  ,
\ee
where $a_0=1$. For integer $\al_j=j$, this leads to the expression
\be
\label{3.133}
f_k^*(x) = \exp \left ( C_1 x \exp \left ( C_2 x \ldots
\exp\left ( C_k x \right ) \right ) \right ) \; .
\ee

It is useful to stress that the obtained self-similar exponential approximants are
rather different from the iterated Euler \cite{Euler_182} exponentials (see also
\cite{Knoebel_183,Bender_184}). The latter have the same radius of convergence as
the initial series (\ref{3.103}). Hence, if the series (\ref{3.103}) are divergent,
the Euler exponentials do not allow for an effective extrapolation. This is because
the coefficients in the Euler exponentials are defined by the accuracy-through-order
procedure. While the parameters of the self-similar exponentials are given by
definition (\ref{3.132}).

The control functions $t_j$ should be found from additional conditions. For example
by minimizing a cost functional \cite{Yukalov_11,Yukalov_185}. The other way could
be as follows \cite{Yukalov_168,Yukalov_177}. Let us set in function (\ref{3.133})
the control parameters $t_j = 1$ for $j = 0, 1, 2, \ldots, k-1$, denoting the result
as $F_k(x,t_k)$. The control parameters can be defined by the minimal difference
optimization condition
\be
\label{3.134}
 F_k(x,t_k) - F_{k-1}(x,t_k) = 0 \; ,
\ee
which yields the equation for the control function $t_k=t_k(x)$,
\be
\label{3.135}
  t_k = \exp \left ( \frac{a_k}{a_{k-1}} \; t_k x \right ) \; .
\ee
This defines the exponential approximant
$$
f_k^*(x) = F_k(x,t_k(x)) \; .
$$

As is mentioned above, it is possible to define the parameters $C_n$ by minimizing
some cost functionals \cite{Yukalov_11,Yukalov_185}, which gives other forms of the
control parameters, for instance
$$
 C_n = \frac{a_n(1+a_1^2)}{na_{n-1}(1+a_n^2) } \qquad ( n = 1,2,\ldots, k) \; .
$$

\subsection{Self-Similar Additive Approximants}

The small-variable expansion (\ref{3.103}) can be reorganized using the asymptotic
relation
\be
\label{3.136}
 1 + \sum_{n=1}^k a_n x^n \simeq \sum_{j=1}^k d_j ( 1 + \lbd x)^j
\qquad ( x \ra 0 ) \;  ,
\ee
where $d_j$ are defined through $a_n$, with the condition
$$
 \sum_{j=1}^k d_j = 1 \;  .
$$
Accomplishing the self-similar renormalization of the linear terms, as in Sec. 3.7,
we get the approximant
\be
\label{3.137}
f_k^*(x) = \sum_{j=1}^k d_j ( 1 + \lbd_j x)^{n_j} \; .
\ee
Here $d_j$ are known from relation (\ref{3.136}), while $2k$ parameters $\lbd_j$
and $n_j$ are control parameters. Rewriting the above approximant in a bit different
way, as
\be
\label{3.138}
f_k^*(x) = \sum_{j=1}^k D_j ( 1 + \lbd x)^{n_j} \;   ,
\ee
we have $2k + 1$ control parameters $D_j$, $\lambda$, and $n_j$. But only $2k$ of
them are independent, since there is the additional condition
$$
 \sum_{j=1}^k D_j = 1 \;  .
$$

Let the large-variable expansion be
\be
\label{3.139}
 f(x) \simeq \sum_{j=1}^k b_j x^{\bt_j} \qquad ( x \ra \infty ) \; ,
\ee
where the powers are arranged in the descending order,
$$
 \bt_j > \bt_{j+1} \qquad ( j = 1,2, \ldots,k-1 ) \;  .
$$
Requiring that the approximant (\ref{3.138}) would reproduce the terms of the
large-variable expansion (\ref{3.139}), we set
\be
\label{3.140}
n_j = \bt_j  \qquad ( j = 1,2, \ldots,k ) \;   .
\ee
This gives
\be
\label{3.141}
 f_k^*(x) = \sum_{j=1}^k D_j ( 1 + \lbd x)^{\bt_j} \;  .
\ee
However, strictly speaking, expanding expression (\ref{3.141}) at a large variable
$x$, in addition to the terms with the powers $\beta_j$, we also will get the terms
with the powers $\bt_j-1$. If such terms pertain to the set $\{\bt_j\}$, then there
is no problem. However, if they are not in that set, then we need to add to expression
(\ref{3.141}) compensating terms canceling the terms of wrong powers arising at
$x\ra\infty$. Thus the general form of the additive approximant is
\be
\label{3.142}
f_k^*(x) = \sum_{j=1}^k D_j ( 1 + \lbd x)^{\bt_j} +
\sum_j C_j ( 1 + \lbd x)^{\bt_j-1}\; .
\ee

The parameters $D_j$ and $\lambda$ are defined by the accuracy-through-order
procedure either at small or at large variable, or at both, while the role of
$C_j$ is to cancel the terms yielding wrong powers at the large variable expansion.
Employing the accuracy-through-order procedure, we confront the problem of multiple
solutions. Among them there are real as well as complex-valued solutions. The latter
come in complex conjugate pairs, so that their sum is real. It is admissible to
either select only real solutions or to take the sum of all solutions. In both these
cases, the results are of comparable accuracy \cite{Gluzman_186}.

\subsection{Self-Similar Factor Approximants}

Self-similar renormalization of the expansion
\be
\label{3.143}
f_k(x) = 1 + \sum_{n=1}^k a_n x^n
\ee
can also be done in the following way
\cite{Yukalov_187,Gluzman_188,Yukalov_189,Yukalov_190}. The above sum can be
rewritten as
\be
\label{3.144}
 f_k(x) = \prod_{j=1}^k ( 1 + b_j x ) \;  ,
\ee
with the coefficients $b_j$ expressed through $a_n$. Applying the procedure of the
self-similar renormalization to each of the factors results in the transformation
$$
 1 + b_j x \longrightarrow ( 1 + A_j x )^{n_j} \;  .
$$
Then we get the self-similar factor approximant
\be
\label{3.145}
 f_k^*(x) = \prod_{j=1}^{N_k} ( 1 + A_j x )^{n_j} \;  ,
\ee
where the number of factors $N_k$ is prescribed by the number of terms in the initial
expansion (\ref{3.143}). This number $N_k$ will be specified below.

When the order $k$ of the expansion (\ref{3.143}) is even, then we can take $N_k=k/2$
and define all control parameters $A_j$ and $n_j$ from the accuracy-through-order
procedure at small variables,
\be
\label{3.146}
  f_k^*(x) \simeq  f_k(x)  \qquad ( x \ra 0 ) \; ,
\ee
resulting in the equations
\be
\label{3.147}
 \sum_{j=1}^{k/2} n_j A_j^n = D_n \qquad ( n = 1,2,\ldots, k ) \;  ,
\ee
where
\be
\label{3.148}
D_n \equiv \frac{(-1)^{n-1}}{(n-1)!} \; \lim_{x\ra 0} \; \frac{d^n}{dx^n} \;
\ln \left ( 1 + \sum_{m=1}^n a_m x^m \right ) \; .
\ee
Thus we have $k/2$ parameters $A_j$ and $k/2$ parameters $n_j$ that are uniquely
defined by $k$ equations (\ref{3.147}).

When $k$ is odd, we take $N_k=(k + 1)/2$, obtaining the equations
\be
\label{3.149}
 \sum_{j=1}^{(k+1)/2} n_j A_j^n = D_n \qquad ( n = 1,2,\ldots, k ) \; .
\ee
Now we have $(k + 1)/2$ parameters $A_j$ and $(k + 1)/2$ parameters $n_j$, hence
altogether $k + 1$ parameters, but only $k$ equations (\ref{3.149}). Thus all
parameters $A_j$ and $n_j$ can be expressed through one of the parameters, say
through $A_1$. The latter is to be defined by an additional condition.

The accuracy-through-order equations with respect to the control parameters
$A_j$ are polynomial, the first equation being of first order, the second, of
second order, and so on, with the last equation being of order $k$. Each polynomial
equation of order $n$ possesses $n$ solutions. So that the total number of solutions
is $1\times2\times3\times\ldots\times k=k!$. At the same time, from the form of the 
factor approximants it is evident that each of them is invariant with respect to the 
$k!$ permutations
$$
A_i \longrightarrow A_j \; , \qquad  n_i \longrightarrow n_j \;  .
$$
Therefore, the multiplicity of solutions for $A_j$ is trivial, being related
to the enumeration of the parameters. Up to this enumeration, the solutions are
unique. If the parameters $A_j$ are given, then we get the linear algebraic
equations with respect to $n_j$ that are uniquely defined \cite{Yukalov_34}.

In the large-variable limit, approximant (\ref{3.145}) behaves as
\be
\label{3.150}
 f_k^*(x) \simeq B_k x^{\bt_k} \qquad ( x \ra \infty ) \;  ,
\ee
with the amplitude and power given by the expressions
\be
\label{3.151}
 B_k = \prod_{j=1}^{N_k} A_j^{n_j} \; , \qquad
\bt_k = \sum_{j=1}^{N_k} n_j \;  ,
\ee
and with the number of factors
\begin{eqnarray}
\label{3.152}
N_k = \left\{ \begin{array}{ll}
k/2 \; , ~ & ~ k = 2,4,\ldots \\
(k+1)/2 \; , ~ & ~ k = 3,5,\ldots
\end{array} \; .
\right.
\end{eqnarray}

Note that the values of the parameters $A_j = A_j(k)$ and $n_j = n_j (k)$ depend
on the considered approximation order $k$. If the large-variable behavior of the
sought function is known, being
\be
\label{3.153}
 f(x) \simeq B x^\bt \qquad ( x \ra \infty ) \;  ,
\ee
then we have two additional equations
\be
\label{3.154}
 B_k = B \; , \qquad \bt_k = \bt \;  .
\ee

As is mentioned above, in the case of an odd approximation order $k$, one
of the parameters $A_j$ remains undefined. It would be possible to fix
$A_1=A_1(k)$ in the odd order $k$ equal to $A_1(k-1)$ from the previous even
order $k-1$ or to define $A_1$ in the odd order approximant by a variational
procedure \cite{Yukalov_191}. Another way could be based on scaling arguments
\cite{Yukalov_190}. One more possibility could be based on the assumption of
convergence.

Assume that the sequence $\{f_k^*(x)\}$ converges. This implies that the
sequence of the large-variable amplitudes $B_k$ also converges, so that there
exists a limit
\be
\label{3.155}
  \lim_{k\ra\infty} B_k = B^* \; .
\ee
For an even approximation order $k=2p$, we have
\be
\label{3.156}
B_{2p} = \prod_{j=1}^p A_j^{n_j} \;  ,
\ee
where $A_j=A_j(2p)$ and $n_j=n_j(2p)$. And for the next, odd, approximation order
$k=2p+1$, the amplitude is
\be
\label{3.157}
 B_{2p+1} = \prod_{j=1}^{p+1} A_j^{n_j} \;   ,
\ee
where $A_j=A_j(2p+1)$ and $n_j=n_j(2p+1)$. The latter expression can be represented
as
\be
\label{3.158}
 B_{2p+1} = \left ( \prod_{j=1}^p A_j^{n_j} \right ) A_{p+1}^{n_{p+1}}\;  .
\ee
Taking here the limit $p \ra \infty$ gives
\be
\label{3.159}
 B^* = B^* \lim_{p\ra\infty}  A_{p+1}^{n_{p+1}}\;  ,
\ee
from where
\be
\label{3.160}
 \lim_{p\ra\infty}  A_{p+1}^{n_{p+1}} = 1 \;  .
\ee
Generally, there are two possibilities: either $A_{p+1}$ tends to a finite value
and $n_{p+1}$ tends to zero, or $A_{p+1}$ tends to one, while $n_{p+1}$ tends to a
finite value. Aiming at using the limiting behavior as an approximation at finite
orders, we cannot take $n_{p+1}$ as zero, since then the term $(1+A_{p+1}x)^{n_{p+1}}$
disappears at al. In order that the term $(1+A_{p+1} x)^{n_{p+1}}$ in the general form
of approximant (\ref{3.145}) be nontrivial and unambiguously defined, the parameters
$A_{p+1}$ and $n_{p+1}$ have to be finite and nonzero. Therefore we have to accept
that limit (\ref{3.160}) yields
\be
\label{3.161}
 \lim_{p\ra\infty}  A_{p+1} = 1 \;  ,
\ee
which does not exclude that $n_{p+1}$ simultaneously tends to zero. Moreover,
if $n_{p+1}$ tends to zero, then a finite value of $A_{p+1}$ is not of importance,
hence can be set to one. Thus condition (\ref{3.161}) is sufficient for the validity
of limit (\ref{3.160}). This suggests that for large approximation orders one of the
parameters $A_j$ in the odd approximant can be set to one. Keeping in mind this
asymptotic equality, we can apply it, as an approximation, for finite orders $k$.
By re-enumerating the parameters, it is admissible to set to one the term $A_1$.
Since setting $A_1 = 1$ for finite odd approximation orders is an approximation, we
have to accept that the most reliable are the even factor approximants that do not
involve such an additional condition. But considering odd factor approximants gives
us information how close we are to the saturation limit of large $k$, where even and
odd approximants should coincide.

Obtaining the sequence of the factor approximants $f_k^*(x)$, it is possible to
estimate their accuracy by the error bar
\be
\label{3.162}
 \pm \frac{1}{2} \; \left | f_k^*(x) - f_{k-1}^*(x) \right | \; .
\ee

It may happen that the accuracy-through-order equations result in complex-valued
parameters $A_j$ and $n_j$. But this is not a problem, since such solutions come
in complex conjugated pairs, so that the total approximant is always real.

Factor approximants can provide good approximations for various functions, rational,
irrational, as well as transcendental
\cite{Yukalov_187,Gluzman_188,Yukalov_189,Yukalov_190}.
Moreover, there exists a large class of functions that can be reproduced exactly by
self-similar approximants of finite order. This class is composed of the functions
of the form
$$
R_k(x) = \prod_{j=1}^M P_{m_j}^{\al_j}(x) \qquad
\left ( k = \sum_{j=1}^M m_j \right ) \; ,
$$
where $P_m(x)$ is a polynomial and $\alpha_j$ are complex-valued numbers.
Transcendental functions arise as limits of the above form. For instance, applying
factor approximants to the series
$$
 e^x \simeq \sum_{n=0}^k \frac{x^n}{n!} \;  ,
$$
we exactly reconstruct the exponential for any order $k \geq 2$ because of the limit
$$
\lim_{a\ra 0} ( 1 + ax)^{1/a} = e^x \; .
$$

Note that the number of factors in a factor approximant, defined in Eq. (\ref{3.152}),
corresponds to the case when only a small-variable expansion (\ref{3.143}) is available.
In the case, when the large-variable asymptotic behavior (\ref{3.153}) is also given,
the number of the factors depends on what information is additionally used. Thus, if
both the amplitude $B$ as well as the power $\beta$ are used as known values, then
\begin{eqnarray}
\nonumber
N_k = \left\{ \begin{array}{ll}
(k+2)/2 \; , ~ & ~ k = 2,4,\ldots \\
(k+3)/2 \; , ~ & ~ k = 3,5,\ldots
\end{array} \qquad ( B_k = B \; , ~ \bt_k = \bt ) \; ,
\right.
\end{eqnarray}
with the additional condition $A_1 = 1$ for odd approximants.

And if only the power $\beta$ is employed, while the amplitude $B$ is not known, then
the number of factors is
\begin{eqnarray}
\nonumber
N_k = \left\{ \begin{array}{ll}
(k+2)/2 \; , ~ & ~ k = 2,4,\ldots \\
(k+1)/2 \; , ~ & ~ k = 3,5,\ldots
\end{array} \qquad ( \bt_k = \bt ) \; ,
\right.
\end{eqnarray}
with the additional condition $A_1=1$ for even approximants.

It is also useful to mention that, defining the parameters $A_j$ and powers $n_j$,
one can meet complex values. However, the related parameters and powers appear in
complex conjugate pairs, which makes the overall approximant real, since
$$
 \left | ( 1 + Ax)^m \right | = | 1 + Ax |^{{\rm Re}\; m} \exp\left\{ - ({\rm Im}\; m)
{\rm arg}(1 + Ax) \right \} \; .
$$

\subsection{Self-Similar Combined Approximants}

It is possible to combine different self-similar approximants in the following way
\cite{Gluzman_192}. Suppose we have expansion (\ref{3.143}) of order $k$. We may
consider a subexpansion of order $n < k$,
\be
\label{3.163}
 f_n(x) = 1 + \sum_{j=1}^n a_j x^j \qquad ( n < k ) \;  ,
\ee
for which we can construct a self-similar approximant $f_n^*(x)$. Then we introduce
the function
\be
\label{3.164}
 C_{k/n}(x) \equiv \frac{f_k(x)}{f_n^*(x)} \;  ,
\ee
which is expanded in powers of $x$, giving
\be
\label{3.165}
 C_{k/n}(x) \simeq 1 +  \sum_{j=n+1}^k b_j x^j \qquad ( x \ra 0 ) \; ,
\ee
with the coefficients $b_j$ expressed through $a_j$. On the basis of the latter
expansion, we construct another self-similar approximant $C_{k/n}^*(x)$. The result
is the combined approximant
\be
\label{3.166}
 f_k^*(x) = f_n^*(x) C^*_{k/n}(x) \; .
\ee
Such combined approximants, composed of two (or more) factors, can be used if one
of the factors catches better the behavior at small variables, while the other is
more accurate at large variables.

On the basis of expansion (\ref{3.165}), it is also admissible to construct a Pad\'e
approximant $P_{M/N}(x)$, with $M + N = k - n$, defining the Pad\'{e} approximant
parameters by the standard condition
$$
P_{M/N}(x) \simeq C_{k/n}(x) \qquad ( x \ra 0 ) \;  .
$$
In that way, we obtain the combined expression
\be
\label{3.167}
  f_k^*(x) = f_n^*(x) P_{M/N}(x)
\ee
that is a self-similarly corrected Pad\'{e} approximant. Such combined approximants
are useful when one tries to approximate an irrational function that cannot be well
approximated by Pad\'{e} approximants \cite{Gluzman_193} and when the latter are not
applicable because of the problems typical of Pad\'{e} approximants, such as the
appearance of spurious poles, incompatibility of small-variable and large-variable
behaviors, and like that \cite{Gluzman_194}.

\subsection{Acceleration of Convergence}

Obtaining a sequence $\{f_k^*(x)\}$ of self-similar approximants, it is possible
to accelerate its convergence by employing additional tricks. One method for this
purpose is by using splines based on several last approximants \cite{Yukalov_34}.

For example, we can treat the last approximants $f_{k-2}^*(x)$, $f_{k-1}^*(x)$,
and $f_k^*(x)$ as the terms of a new sequence. Then we define a polynomial spline,
say a quadratic one,
\be
\label{3.168}
 g(x,t) = a(x) + b(x) t + c(x) t^2 \;  ,
\ee
whose coefficients are prescribed by the conditions
\be
\label{3.169}
 g(x,0) = f_{k-2}^*(x) \; , \qquad  g(x,1) = f_{k-1}^*(x) \; , \qquad
 g(x,2) = f_k^*(x) \; ,
\ee
which give
$$
a(x) = f_{k-2}^*(x) \; , \qquad
b(x) = -\;\frac{1}{2}\; \left [ f_k^*(x) - 4f_{k-1}^*(x) + 3f_{k-2}^*(x) \right ] \; ,
$$
$$
c(x) = \frac{1}{2} \; \left [ f_k^*(x) - 2f_{k-1}^*(x) + f_{k-2}^*(x) \right ] \;.
$$
Considering spline (\ref{3.168}) as an expansion in powers of $t$, we construct
a self-similar approximant
\be
\label{3.170}
 g^*(x,t) = a(x) [ 1 + A(x) t ]^{n(x)} \;  ,
\ee
in which the control functions $A(x)$ and $n(x)$ are defined by the
accuracy-through-order procedure at small $t$, resulting in
$$
 A(x) = \frac{b^2(x) - a(x) c(x)}{a(x)b(x)} \; , \qquad
 n(x) = \frac{b^2(x)}{b^2(x)-a(x)c(x)} \; .
$$
The final answer is given by the expression
\be
\label{3.171}
 g^*(x) =  \frac{1}{2} \; \left [ g^*(x,2) + g^*(x,3) \right ] \; ,
\ee
with the error bar
\be
\label{3.172}
  \pm \; \frac{1}{2} \left | g^*(x,2) - g^*(x,3) \right ] \; .
\ee
Note that if $f_k^*(x)\propto x^\beta$, then also $g^*(x)\propto x^\beta$.

It is possible to employ some other tricks improving convergence. For example, one
can treat the sequence of self-similar approximants $\{f_k^*(x)\}$ similarly to a
set of expansions $f_k(x)$, defining a new approximation cascade, whose trajectory
is bijective to the sequence $\{f_k^*(x)\}$. Then, following the standard scheme,
one can construct a twice renormalized self-similar approximant \cite{Gluzman_194}.

It is also possible to introduce more control functions, for instance by using
power transforms \cite{Gluzman_195}
$$
P_k(x,m) \equiv \left [ f_k(x) \right ]^m \;   ,
$$
with the power $m$ playing the role of a control parameter. Then the transformed
expression is re-expanded in powers of $x$ resulting in
$$
P_k(x,m) \simeq \sum_{n=0}^k b_n(m) x^n \qquad ( x \ra 0 ) \; ,
$$
with $b_n(m)$ expressed through $a_n$. On the basis of the latter expansion, we
construct a self-similar approximant $P_k^*(x,m)$. Accomplishing the inverse power
transformation, we get
$$
 F^*_k(x,m) = \left [ P^*_k(x,m) \right ]^{1/m} \; .
$$
Defining control functions $m_k = m_k(x)$ by some optimization conditions, we
come to the self-similar approximant
$$
 f^*_k(x,m) = F^*_k(x,m_k(x) ) \;  .
$$
This way, however, is quite cumbersome. The method of using splines looks much
easier, being rather accurate.

\subsection{Electron Gas}

Self-similar approximation theory has been used for solving hundreds of problems,
as can be inferred from the cited references. Here we illustrate its application 
just for a few examples.

Let us consider the correlation energy of one-dimensional uniform electron gas as
a function of the Wigner-Seitz radius $r_s$ defined by the relation
$$
 r_s \equiv \frac{1}{2\rho} \qquad \left ( \rho \equiv \frac{N}{L} \right ) \; ,
$$
in which $\rho$ is mean density. The asymptotic behavior of the correlation energy
at small $r_s$, or high density, is given \cite{Loos_196} by the expansion
\be
\label{3.173}
 E_{cor}(r_s) \simeq -\; \frac{\pi^2}{360} +  0.00845r_s \qquad
( r_s \ra 0 ) \; .
\ee
At large $r_s$, hence low density, the behavior is also known \cite{Loos_196} being
\be
\label{3.174}
 E_{cor}(r_s) \simeq \frac{b_1}{r_s} +  \frac{b_2}{r_s^{3/2}} \qquad
( r_s \ra \infty ) \;  ,
\ee
with the coefficients
$$
 b_1 = - \left ( \ln\sqrt{2\pi} \; - \; \frac{3}{4} \right ) = - 0.168939 \; ,
\qquad b_2 = 0.359933 \; .
$$

When the small, as well as large, variable behaviors are known, it is reasonable
to resort to self-similar root approximants. The root approximant, satisfying both
the small and large $r_s$ expansions reads as \cite{Gluzman_197}
\be
\label{3.175}
 E_3^*(r_s) = -\; \frac{\pi^2}{360} \left ( \left (  ( 1 + A_1 r_s )^{3/2}
+ A_2 r_s^2 \right )^{5/4} + A_3 r^3 \right )^{-1/3} \;  ,
\ee
where
$$
 A_1 = 0.493150\; , \qquad A_2 = 0.056122 \; , \qquad A_3 = 0.004274 \;  .
$$

This correlation energy has been calculated by diffusion Monte Carlo simulations
\cite{Loos_196} in the interval of the Wigner-Seitz radius $r_s\in [0, 20]$. Comparing
the numerical results \cite{Loos_196} with the root approximant (\ref{3.175}), we find
that the maximal error of the latter is $8\%$ for $r_s$ in that interval.

\subsection{Harmonium Atom}

An $N$-electron harmonium atom is described by the Hamiltonian
\be
\label{3.176}
 \hat H = \frac{1}{2} \sum_{i=1}^N \left ( -\nabla^2_i + \om^2 r_i^2 \right )
+ \frac{1}{2} \sum_{i\neq j}^N \frac{1}{r_{ij} } \;  ,
\ee
in which dimensionless units are used and
$$
r_i \equiv |\br_i | \; , \qquad r_{ij} \equiv | \br_i -\br_j | \; .
$$
This Hamiltonian provides a realistic modeling of many finite quantum systems, such
as trapped ions, quantum dots, atomic nuclei, and metallic grains \cite{Birman_55}.
Here we consider the ground-state energy of a two-electron harmonium $(N=2)$.

When the trapping is weak, so that $\omega$ is small, the ground-state energy reads
as
\be
\label{3.177}
 E(\om) \simeq c_0 \om^{3/2} + c_1 \om + c_2\om^{4/3} \qquad ( \om \ra 0 ) \; ,
\ee
with the coefficients
$$
c_0 = \frac{3}{2^{4/3}} = 1.19055 \; , \qquad
c_1 = \frac{1}{2} \; ( 3 + \sqrt{3} ) = 2.36603 \; , \qquad
c_2 = \frac{7}{36\cdot 2^{2/3}} = 0.122492 \; .
$$
And for a hard trapping by a rigid potential with a large $\omega$, the energy
behaves as
\be
\label{3.178}
 E(\om) \simeq b_0 \om + b_1 \om^{1/2} + b_2 + b_3 \om^{-1/2} \qquad
( \om \ra \infty ) \;  ,
\ee
where
$$
 b_0 = 3 \; , \qquad b_1 =\sqrt{\frac{2}{\pi} } = 0.797885 \; , \qquad
b_2 = -\; \frac{2}{\pi} \; \left ( 1 -\; \frac{\pi}{2} + \ln 2 \right ) = -0.077891 \; ,
$$
$$
b_3 = \left ( \frac{2}{\pi} \right )^{3/2} \;
\left [ 2 - 2G -\; \frac{3}{2}\;\pi + ( 3+\pi) \ln 2 +
\frac{3}{2} \; (\ln 2)^2 -\; \frac{\pi^2}{24} \right ] = 0.0112528 \; ,
$$
and $G$ is the Catalan constant
$$
 G \equiv \sum_{n=0}^\infty \frac{(-1)^n}{(2n+1)^2} = 0.91596559 \; .
$$

Having both the small-variable and large-variable expansions, it is again reasonable
to employ self-similar root approximants. Such a root approximant, satisfying both
asymptotic expansions, at small as well as at large $\om$, is \cite{Gluzman_197}
$$
 E_6^*(\om) = c_0\om^{2/3} \left( \left( \left( \left( \left(
\left( 1 + A_1 \om^{1/3} \right)^{1/2} + A_2 \om^{2/3} \right)^{3/4} +
A_3 \om \right)^{5/6} + \right.   \right. \right.
$$
\be
\label{3.179}
\left. \left. \left.
 + A_4 \om^{4/3} \right)^{7/8} + A_5 \om^{5/3} \right)^{9/10}
+ A_6 \om^2 \right)^{1/6} \; ,
\ee
where
$$
A_1 =48.4532 \; , \qquad A_2 = 564.108\; , \qquad A_3=1088.39 \; ,
$$
$$
A_4 = 1221.08 \; , \qquad A_5 = 796.791 \; , \qquad A_6 = 256 \;  .
$$
The accuracy of this approximant, as compared with numerical calculations
\cite{Matito_198}, is very good, having an error of only $0.9 \%$. Note that Pad\'e
approximants cannot be used in the case of harmonium, since the small-variable and
large-variable asymptotic expansions are incompatible.

\subsection{Schwinger Model}

The Schwinger model \cite{Schwinger_199,Banks_200} is a lattice gauge theory in
$1 + 1$ dimensions representing Euclidean quantum electrodynamics with a Dirac
fermion field. It possesses many properties in common with QCD, such as confinement,
chiral symmetry breaking, and charge shielding. For this reason, it has become a
standard test bed for the study of numerical techniques in field theory. Here we
consider the ground state of the model, corresponding to a vector boson of mass
$M(x)$ as a function of the variable $x=m/g$, where $m$ is the electron mass and $g$
is the coupling parameter having the dimension of mass, so that $x$ is dimensionless.
The energy is given by the relation $E=M-2m$.

The small-$x$ expansion for the ground-state energy
\cite{Carrol_201,Vary_202,Adam_203,Striganesh_204} is
\be
\label{3.180}
 E(x) \simeq 0.5642 - 0.219 x + 0.1907 x^2 \qquad ( x \ra 0 ) \; .
\ee
While the large-$x$ expansion \cite{Striganesh_204,Coleman_205,Hamer_206,Hamer_207}
reads as
\be
\label{3.181}
 E(x) \simeq 0.6418 x^{-1/3} -\; \frac{1}{\pi}\; x^{-1} - 0.25208 x^{-5/3} 
  \qquad ( x \ra \infty ) \; .
\ee
Using again self-similar root approximants, we find the form agreeing with both
asymptotic expansions,
$$
 E_5^*(x) = A \left( \left( \left( \left(  ( 1 + A_1 x )^{4/3} + A_2x^2 \right)^{7/6}
+A_3 x^3 \right)^{10/9} \right. \right. +
$$
\be
\label{3.182}
 +
\left. \left.
A_4 x^4 \right)^{13/12} + A_5 x^5 \right)^{-1/15} \; ,
\ee
where
$$
A = 0.5642 \; , \qquad A_1 = 3.109547 \; , \qquad A_2 = 3.640565 \; ,
$$
$$
 A_3 = 4.028571 \; , \qquad A_4 = 1.070477 \; , \qquad A_5 = 0.144711 \;   .
$$

The root approximant (\ref{3.182}) provides a very good accuracy, as compared
with the calculations using a density-matrix renormalization group approach
\cite{Byrnes_208} and fast moving frame estimates \cite{Kroger_209}. Table 2 shows
that the root approximant $E_5^*(x)$ gives the results coinciding with those of
these numerical techniques, in the frame of their accuracy.

\begin{table}
\renewcommand{\arraystretch}{1.25}
\centering
\caption{ Ground-state energy of the Schwinger model, for varying dimensionless
parameter $x=m/g$, obtained by using a density matrix renormalization group,
$E_{DMRG}$; fast moving frame estimates, $E_{FMFE}$; and the self-similar root
approximants, $E_5^*$.}
\vskip 5mm
\label{tab-2}
\begin{tabular}{|c|c|c|c|} \hline
$x$   &  $E_{DMRG}$ &  $E_{FMFE}$ & $E^*_5$  \\ \hline
0.125 &  0.540      &  0.528      & 0.540    \\ \hline
0.25  &  0.519      &  0.511      & 0.519   \\ \hline
0.5   &  0.487      &  0.489      & 0.487    \\ \hline
1     &  0.444      &  0.455      & 0.444   \\ \hline
2     &  0.398      &  0.394      & 0.392   \\ \hline
4     &  0.340      &  0.339      & 0.337   \\ \hline
8     &  0.287      &  0.285      & 0.284    \\ \hline
16    &  0.238      &  0.235      & 0.235   \\ \hline
\end{tabular}
\end{table}

\subsection{Large-Variable Prediction: Partition Function}

It is often of great interest to learn the behavior of a function at large variables,
while only a small-variable expansion is available. This problem can be solved by
resorting to self-similar factor approximants.

To illustrate how it is possible to predict the large-variable behavior, let us
consider the partition function of the so-called zero-dimensional field theory
\be
\label{3.183}
 Z(g) = \frac{1}{\sqrt{\pi} }
\int_{-\infty}^{\infty} e^{-\vp^2-g\vp^4} \; d\vp \;  ,
\ee
whose structure is mathematically typical of many problems in statistical physics
and field theory.

The weak-coupling expansion yields
\be
\label{3.184}
 Z_k(g) = 1 + \sum_{n=1}^k a_n g^n \qquad ( g \ra 0 ) \; ,
\ee
with the coefficients
$$
 a_n = \frac{(-1)^n}{\sqrt{\pi}\; n!} \; \Gm \left( 2n + \frac{1}{2}\right)
$$
factorially diverging for large orders $n$. We wish to find out what would be the
behavior of the partition function in the strong-coupling limit $g \ra \infty$. No
information from this limit is supposed to be used, just the weak-coupling expansion.
The accuracy of the predicted result can be defined from the exact asymptotic limit
\be
\label{3.185}
 Z(g) \simeq 1.022765 g^{-1/4}  \qquad ( g \ra \infty ) \;  .
\ee

Constructing the factor approximants $Z^*_k(g)$, we find the strong-coupling limit
\be
\label{3.186}
  Z_k^*(g) \simeq B_k g^{-\bt_k}  \qquad ( g \ra \infty ) \;   .
\ee
The amplitude $B$ and power $\beta$ are predicted within the errors of the order
of $20\%$, starting from the seventh approximation \cite{Yukalov_190}.

If we are interested solely in the value of the power $\beta$, then it is possible
to employ another procedure. We define
\be
\label{3.187}
\bt_k(g) \equiv \frac{d\ln Z_k(g)}{d\ln g}
\ee
and expand it in powers of $g$,
\be
\label{3.188}
  \bt_k(g) \simeq a_1 g\left( 1 + \sum_{n=1}^k b_n g^n \right) \; .
\ee
On the basis of this expansion, we construct the corresponding factor approximants
\be
\label{3.189}
  \bt_k^*(g) \simeq a_1 g \prod_{j=1}^{N_k} ( 1 +  A_j g )^{n_j} \;  ,
\ee
imposing an additional constraint requiring the finiteness of the limit $g\ra\infty$,
\be
\label{3.190}
1 +  \sum_{j=1}^{N_k}  n_j = 0 \; .
\ee
Then the sought answer is given by the limit
\be
\label{3.191}
  \bt_k^* \equiv \lim_{g\ra\infty} \bt_k^*(g) =
a_1  \prod_{j=1}^{N_k}  A_j^{n_j} \;  .
\ee
Applying this method to the considered problem gives the exponent with an
error close to, although slightly larger, than the direct method of constructing
$Z^*_k(g)$.

\subsection{Large-Variable Prediction: Eigenvalue Problem}

As another example, illustrating the use of factor approximants for predicting the
large-variable behavior, we take the anharmonic oscillator with the Hamiltonian
\be
\label{3.192}
 \hat H = -\; \frac{1}{2} \; \frac{d^2}{dx^2} + \frac{1}{2}\; x^2 + g x^4 \; .
\ee
The weak-coupling expansion for the ground-state energy reads as
\cite{Hioe_24,Bender_210}
\be
\label{3.193}
  E_k(g) \simeq \frac{1}{2}\; \left( 1 + \sum_{n=1}^k a_n g^n \right) \;  ,
\ee
with the first ten coefficients
$$
a_1 = 1.5 \; , \qquad a_2 = -5.25 \; , \qquad a_3 = 41.625 \; , \qquad
a_4 = -482.578125 \; ,
$$
$$
a_5 = 7161.9609375 \; , \qquad a_6 = -127965.6269532 \; , \qquad
a_7 = 2659467.4541 \; ,
$$
$$
 a_8 = -62896429.3856 \; , \qquad a_9 = 1667083206.526 \; , \qquad
a_{10} = -48957881405.6 \;  .
$$
And the strong-coupling limit is known to be
\be
\label{3.194}
 E(g) \simeq 0.667986 g^{1/3} \qquad ( g \ra \infty ) \; .
\ee

Using factor approximants, we get
\be
\label{3.195}
 E^*_k(g) \simeq B_k g^{\bt_k} \qquad ( g \ra \infty ) \;  .
\ee
Starting from the seventh order, the amplitude and power are predicted
\cite{Yukalov_190} within the errors of about $10 \%$.

\subsection{Exact Solutions: Initial-Value Problem}

As has been mentioned above, in some cases self-similar factor approximants
provide exact answers. This is illustrated below by solving nonlinear differential
equations \cite{Yukalova_211}. An equation is termed singular, if it does not
allow for finding a solution by using perturbation theory in powers of $\ep$
\cite{Kevorkian_212,Hinch_213,Chen_214}.

\vskip 2mm

(i) {\it Linear singular problem}

\vskip 2mm

Consider the singular differential equation
\be
\label{3.196}
 \ep \; \frac{d^2 f}{dt^2} + 2 \; \frac{d f}{d t} + \frac{f}{\ep} = 0
\ee
for a function $f(t)$ with $t \geq 0$ and with the initial conditions
$$
f(0) = 1 \; , \qquad \left. \frac{df}{dt} \right \vert_{t=0} = -\; \frac{1}{\ep} \;  .
$$

Looking for the solution at small time $t$, we get the expansion
\be
\label{3.197}
 f(t) \simeq 1 - \; \frac{t}{\ep} + \frac{t^2}{2\ep^2} + \ldots \qquad
( t \ra 0) \; .
\ee
Constructing factor approximants, we find in any order, starting from $k=2$,
the form
\be
\label{3.198}
 f_k^*(t) = e^{-t/\ep} \qquad ( k \geq 2 ) \; ,
\ee
which is the exact solution of Eq. (\ref{3.196}).

\vskip 2mm

(ii) {\it Nonlinear singular problem}

\vskip 2mm

Let us consider the singular equation
\be
\label{3.199}
 (\ep f + t ) \; \frac{df}{dt} + f - 1 = 0 \;  ,
\ee
with the initial condition $f(0) = 2$.

Using again factor approximants, we get \cite{Yukalova_211}, in any order
starting from the fourth order, the exact solution
\be
\label{3.200}
 f^*_k(t) = \sqrt{ 4 + \frac{2t}{\ep} + \frac{t^2}{\ep^2}} \; - \;
\frac{t}{\ep} \qquad ( k \geq 4) \;  .
\ee

\vskip 2mm

(iii) {\it Singular logistic equation}

\vskip 2mm

Logistic equations are widespread in various applications \cite{Svirezhev_215}.
Here we consider a singular logistic equation
\be
\label{3.201}
 \ep \; \frac{df}{dt} = f ( 1 - f ) \; ,
\ee
with an initial condition $f(0) = f_0$.

Factor approximants for any order after the second, give \cite{Yukalova_211} the
exact solution
\be
\label{3.202}
 f_k^*(t) = \frac{f_0}{f_0 - (f_0-1)e^{-t/\ep} } \qquad ( k \geq 3) \; .
\ee

\subsection{Exact Solutions: Boundary-Value Problem}

Factor approximants also allow for finding exact solutions for some boundary-value
problems \cite{Yukalova_211}, as is shown below.

\vskip 2mm

(i) {\it Kink soliton equation}

\vskip 2mm

The $\varphi^4$ model, with particle mass $1/\varepsilon$, is described by the
nonlinear Schr\"{o}dinger equation
\be
\label{3.203}
  \frac{\ep}{2} \; \frac{d^2f}{dx^2} + f - f^3 = 0  \; .
\ee
As boundary conditions, we take
\be
\label{3.204}
f(-\infty) = -1 \; , \qquad f(\infty) = 1 \; .
\ee

Considering expansions near the boundaries and employing factor approximants,
we find \cite{Yukalova_211}, in any order starting from the fourth, the exact kink
soliton solution
\be
\label{3.205}
 f_k^*(x) = \tanh\left(  \frac{x}{\sqrt{\ep}} \right)
\qquad ( k \geq 4) \; .
\ee
Such solitons are called topological \cite{Lee_216}.

\vskip 2mm

(ii) {\it Bell soliton equation}

\vskip 2mm

A nonlinear Schr\"{o}dinger equation, with mass $-1/\varepsilon$ is
\be
\label{3.206}
   \frac{\ep}{2} \; \frac{d^2f}{dx^2} - f + f^3 = 0  \;  .
\ee
We assume the boundary conditions
\be
\label{3.207}
 f(-\infty) = 0 \; , \qquad f(\infty) = 0 \;  .
\ee

Again, we derive expansions near the boundaries and use factor approximants,
which results \cite{Yukalova_211}, in any order after the second, in the exact
bell soliton solution
\be
\label{3.208}
 f_k^*(x) = \sqrt{2}\; {\rm sech}\left(  \sqrt{ \frac{2}{\ep} }\; x \right)
\qquad ( k \geq 3)  \; .
\ee
This is an example of a nontopological soliton \cite{Lee_216}.

\subsection{Vortex-Line Equation}

The radial nonlinear Schr\"{o}dinger equation
\be
\label{3.209}
 \frac{d^2f}{dr^2} + \frac{1}{r}\; \frac{df}{dr} \; - \;
\frac{f}{r^2} + f - f^3 = 0\;  ,
\ee
where $r > 0$ is a dimensionless radial variable, can be met in different branches
of physics. Complimented by the boundary conditions
\be
\label{3.210}
  f(0) = 0 \; , \qquad f(\infty) = 1 \; ,
\ee
it describes vortex lines.

At short radius $r$, we can find the expansion
\be
\label{3.211}
 f_k(r) \simeq cr \left( 1 + \sum_{n=1}^k a_n r^{2n} \right ) \qquad
( r \ra 0 ) \;  ,
\ee
with the coefficients
$$
 a_1 = -\; \frac{1}{8} \; , \qquad a_2 = \frac{1+8c^2}{192} \; , \qquad
a_3 = - \; \frac{1+80c^2}{9216} \; , \qquad
a_4 = \frac{1+656c^2+1152c^4}{737280} \;  .
$$
The coefficient $c$ is defined from the boundary conditions.

On the basis of expansion (\ref{3.211}), we construct factor approximants
\be
\label{3.212}
  f_k^*(r) \simeq c_k r \prod_{j=1}^{N_k} ( 1 + A_j r^2  )^{n_j} \;  .
\ee
The boundary condition $f_k^*(0) = 0$ is automatically valid. And the boundary
condition $f_k^*(\infty)=1$ imposes the constraints
\be
\label{3.213}
 c_k  \prod_{j=1}^{N_k}  A_j^{n_j} = 1 \; , \qquad
1 + 2 \sum_{j=1}^{N_k} n_j = 0 \; ,
\ee
defining the coefficient $c_k$.

To characterize the accuracy of differential equations, it is possible to proceed
as follows \cite{Stetter_217}. Denoting the equation in the symbolic form as
\be
\label{3.214}
 \hat E[f(r) ] = 0 \;  ,
\ee
one defines the {\it solution defect} of $f_k^*(r)$ as
\be
\label{3.215}
 D [ f_k^*(r) ] \equiv | \hat E[f_k^*(r)] | \; .
\ee
The {\it maximal solution defect} is given by the maximal value
\be
\label{3.216}
D [ f_k^* ] \equiv \sup_r D [ f_k^*(r) ] \; .
\ee

The other characteristic of accuracy is the {\it solution error}
\be
\label{3.217}
 \Dlt [ f_k^*(r) ] \equiv | f_k^*(r) - f(r) |
\ee
defining the deviation of the approximant from the exact solution at each point
$r$. Respectively, the {\it maximal solution error} is
\be
\label{3.218}
\Dlt [ f_k^*] \equiv  \sup_r \Dlt [ f_k^*(r) ] \; .
\ee

The advantage of using as the characteristics of accuracy the solution defects
is that they can be defined without knowing the exact solution. For the factor
approximants (\ref{3.212}), the maximal solution defects (\ref{3.216}), are found
to be
$$
D[f_2^*] = 0.12 \; , \qquad  D[f_3^*] = 0.017 \; ,
$$
$$
 D[f_4^*] = 0.015 \; , \qquad  D[f_5^*] = 0.002 \; , \qquad D[f_6^*] = 0.0018 \;.
$$
These results demonstrate a high accuracy of the factor approximants and their
good numerical convergence.

If we wish to employ self-similar root approximants from Sec. 3.8 for approximating
the solutions to Eq. (\ref{3.209}), we need to find out the asymptotic expansion at
large $r$. The latter is
\be
\label{3.219}
 f(r) \simeq 1 - \; \frac{1}{2}\; r^{-2} -\; \frac{9}{8} \; r^{-4} -\;
\frac{161}{16}\; r^{-6} \qquad ( r \ra \infty) \;  .
\ee
The first several self-similar root approximants, that we denote as $R_k^*(r)$ to
distinguish them from the factor approximants, are
$$
R_2^*(r) = \frac{r}{2} \left( 1 + \frac{1}{4}\; r^2 \right)^{-1/2} \; ,
\qquad
R_3^*(r) = \frac{r}{\sqrt{2}} \left( 1 + \frac{1}{2}\; r^2  +
\frac{1}{4}\; r^4 \right)^{-1/4} \; ,
$$
$$
R_4^*(r) = \frac{r}{4^{1/3}} \left( 1 + \frac{3}{4}\; r^2  +
\frac{3}{16}\; r^4   + \frac{1}{16}\; r^6 \right)^{-1/6} \; ,
$$
$$
R_5^*(r) = \frac{r}{136^{1/8}} \left( 1 +  r^2  +
\frac{9}{68}\; r^4   + \frac{1}{34}\; r^6 + \frac{1}{136}\; r^8 \right)^{-1/8} \;   .
$$
Their maximal solution defects are close to $0.1$, which is worse than that of
the factor approximants.

\subsection{Critical Phenomena: Spin Systems}

Self-similar approximants allow for a straightforward description of critical
phenomena. As and example, let us consider the three-dimensional spin - $1/2$
cubic Ising model with the Hamiltonian
\be
\label{3.220}
 \hat H = - J \sum_{\lgl ij\rgl} \sgm_i^z \sgm_j^z \;  ,
\ee
where $\sgm_j=\pm 1$, the summation is over the nearest neighbors and $J>0$.
Thermodynamic characteristics of the model can be calculated by perturbation
theory with respect to the parameter
\be
\label{3.221}
  v = \tanh \left( \frac{J}{T} \right) \; ,
\ee
in which $T$ is temperature. Series in powers of $v$ correspond to the
weak-coupling or high-temperature asymptotic expansions.

One often considers the second derivative of the susceptibility
\be
\label{3.222}
f(v) \equiv \left. \frac{d^2\chi}{dB^2}\right \vert_{B=0} \;   .
\ee
The related weak-coupling expansion \cite{Guttmann_218} reads as
\be
\label{3.223}
 f_k(v) = -2 \left ( 1 + \sum_{n=1}^k a_n v^n \right) \;  ,
\ee
where the several first coefficients are
$$
 a_1 = 24 \; , \qquad a_2 = 318 \; , \qquad a_3 = 3240 \; , \qquad
a_4 = 28158 \; .
$$

The fourth-order factor approximant is
\be
\label{3.224}
  f_4^*(v) = -2 ( 1 + A_1 v)^{n_1} ( 1 + A_2 v)^{n_2} \; ,
\ee
with the parameters
$$
 A_1 = -4.602 \; , \quad n_1 = -4.30 \; , \qquad
A_2 = 7.373 \; , \quad n_1 = 0.571 \; , \qquad  .
$$
The critical point, corresponds to the closest to zero pole of the considered
function, with the related power defining the critical exponent, which gives
\be
\label{3.225}
 v_c = \frac{1}{|A_1|} = 0.217 \; , \qquad | n_1| = 4.3 \;  .
\ee
These results are in good agreement with the values found by Monte Carlo simulations
\cite{Guttmann_218,Barber_219} yielding $v_c=0.218092$ and $|n_1|=4.37$. The accuracy
of the best Pad\'e approximant of the same order is much worse, leading to $v_c=0.153$
and to the critical exponent equal to one.

Critical points and related critical exponents can also be found by employing
self-similar root approximants \cite{Yukalov_166,Gluzman_220}.

\subsection{Critical Phenomena: Bose Gas}

Calculation of the critical temperature shift of the Bose-Einstein condensation for
a weakly interacting Bose gas is discussed in Secs. 2.12 and 2.13, where optimized
perturbation theory is employed. Recall that the aim is to find the coefficient
$c_1$ in the temperature shift
\be
\label{3.226}
\frac{\Dlt T_c}{T_0} \simeq c_1 \gm \qquad ( \gm \ra 0 )
\ee
for an asymptotically weak gas parameter $\gm\equiv\rho^{1/3}a_s$. Resorting to loop
expansion in $O(N)$ field theory $\vp^4$ yields
\cite{Kastening_50,Kastening_51,Kastening_52} the expansion
\be
\label{3.227}
  c_1(x) \simeq \sum_{n=1}^5 a_n x^n \qquad ( x \ra 0 )
\ee
in powers of the variable (\ref{2.175}) and with the coefficients listed in Table 3.
The required answer is given by the limit
$$
 c_1 = \lim_{x\ra\infty} c_1(x) \; .
$$

Employing for this purpose self-similar factor approximants, we construct the
second-order approximant
\be
\label{3.228}
 f_2^*(x) = a_1 x ( 1 + x)^{n_1} (1 + A_2 x)^{n_2} \;  ,
\ee
with the condition
$$
 n_1 + n_2 + 1 = 0 \;  ,
$$
the third-order approximant
\be
\label{3.229}
 f_3^*(x) = a_1 x ( 1 + A_1 x)^{n_1} (1 + A_2 x)^{n_2} \;   ,
\ee
with the same condition $n_1+n_2+1=0$, and the fourth-order approximant
\be
\label{3.230}
f_4^*(x) = a_1 x ( 1 + x)^{n_1} (1 + A_2 x)^{n_2} (1 + A_3 x)^{n_3} \; ,
\ee
with the constraint
$$
 n_1 + n_2 + n_3 + 1 = 0  .
$$
Then, using the convergence acceleration scheme of Sec. 3.14, we find
\cite{Yukalov_34,Yukalov_53} the values of $c_1$ for a different number of field
components $N$. Our results are presented in Table 4 together with the results of
the known Monte Carlo simulations. As is seen, our results coincide with the Monte
Carlo ones, within the accuracy of the latter.

\begin{table}
\centering
\caption{Coefficients $a_n$ of the asymptotic expansion for $c_1(x)$
for a different number of field components $N$.}
\vskip 3mm
\label{tab-3}
\renewcommand{\arraystretch}{1.25}
\begin{tabular}{|c|c|c|c|c|c|} \hline
$N$   &         0    &       1      &    2          &    3    &    4     \\ \hline
$a_1$ &    0.111643  &    0.111643  &    0.111643   &    0.111643   &    0.111643 \\ \hline
$a_2$ & $-$0.0264412 & $-$0.0198309 & $-$0.0165258  & $-$0.0145427  & $-$0.0132206 \\ \hline
$a_3$ &    0.0086215 &    0.00480687&    0.00330574 &    0.00253504 &    0.0020754 \\ \hline
$a_4$ & $-$0.0034786 & $-$0.00143209& $-$0.000807353& $-$0.000536123& $-$0.000392939 \\ \hline
$a_5$ &    0.00164029&    0.00049561&    0.000227835&    0.000130398&    0.0000852025 \\ \hline
\end{tabular}
\end{table}

\vskip 5mm

\begin{table}
\centering
\caption{Coefficient $c_1$ of the critical temperature shift for a different number
of components $N$, found by using self-similar factor approximants, and compared with
the available Monte Carlo simulations.}
\vskip 3mm
\label{tab-4}
\renewcommand{\arraystretch}{1.25}
\begin{tabular}{|c|c|c|} \hline
$N$ &     $c_1$      &       Monte Carlo                       \\ \hline
0   & 0.77$\pm$ 0.03 &                                         \\ \hline
1   & 1.06$\pm$ 0.05 & 1.09$\pm$ 0.09  \cite{Sun_49}           \\ \hline
2   & 1.29$\pm$ 0.07 & 1.29$\pm$ 0.05 \cite{Kashurnikov_38}    \\
    &                & 1.32$\pm$ 0.02 \cite{Arnold_36}  \\ \hline
3   & 1.46$\pm$ 0.08 &                       \\ \hline
4   & 1.60$\pm$ 0.09 & 1.60$\pm$ 0.10 \cite{Sun_49}              \\ \hline
\end{tabular}
\end{table}

\subsection{Critical Exponents}

Critical exponents describe the behavior of thermodynamic characteristics in the
vicinity of second order phase transitions (see, e.g., Ref. \cite{Yukalov_221}
and also Sec. 4). Here we show how such exponents can be calculated by using
self-similar factor approximants. Asymptotic series for critical exponents can
be derived by employing the so-called epsilon expansion (see Sec. 4). One usually
derives the series in powers of $\ep=d-4$ for the critical exponents $\eta$,
$\nu^{-1}$, and $\om$. The other exponents can be found from the scaling relations
\be
\label{3.231}
 \al = 2 - \nu d \; , \qquad \bt = \frac{\nu}{2}\; ( d - 2 + \eta) \; ,
\qquad \gm = \nu (2 -\eta) \; , \qquad \dlt = \frac{d+2-\eta}{d-2+\eta} \;  ,
\ee
where $d$ is space dimensionality. For the three-dimensional space $(d = 3)$, we
have
$$
\al = 2 - 3\nu \; , \qquad \bt = \frac{\nu}{2}\; ( 1 + \eta) \; ,
$$
\be
\label{3.232}
\gm = \nu (2 -\eta) \; , \qquad \dlt = \frac{5-\eta}{1+\eta} \qquad
( d = 3 )\;   .
\ee

Different $N$ correspond to different physical systems. Thus $N = 0$ corresponds
to dilute polymer solutions, $N=1$, to the Ising model, $N = 2$, to magnetic $XY$
models and to superfluids, and $N=3$ characterizes the Heisenberg model. But, in
general, it is possible to consider other $N$.

For $N = -2$, the critical exponents for any $d$ are defined exactly:
$$
\al = \frac{1}{2} \; , \qquad \bt = \frac{1}{4}\; , 
\qquad
\gm = 1 \; , \qquad \dlt = 5 \; ,
$$
\be
\label{3.233}
\eta = 0 \; , \qquad \nu = \frac{1}{2} \qquad ( N = - 2) \;  .
\ee
In the limit of $N\ra\infty$, the exponents also are known exactly:
$$
\al = \frac{d-4}{d-2} \; , \qquad \bt = \frac{1}{2}\; , \qquad
\gm = \frac{2}{d-2} \; , \qquad \dlt = \frac{d+4}{d-2} \; ,
$$
\be
\label{3.234}
\eta = 0 \; , \qquad \nu = \frac{1}{d-2} \; , \qquad \om = 4 - d \qquad
( N \ra \infty) \; ,
\ee
which for the three dimensional space reduces to
$$
\al = -1 \; , \qquad \bt = \frac{1}{2}\; , \qquad
\gm = 2 \; , \qquad \dlt = 5 \; ,
$$
\be
\label{3.235}
  \eta = 0 \; , \qquad \nu = 1 \; ,
\qquad \om = 1 \qquad ( d = 3 , ~ N \ra \infty) \;  .
\ee

The $\ep$- expansions for the critical exponents $\eta$, $\nu^{-1}$, and $\om$, 
in three dimensions for arbitrary $N$ can be found in the book \cite{Kleinert_222}.
Applying to these expansions self-similar factor approximants for small $\ep$ and 
then setting $\ep=1$, we obtain \cite{Yukalov_223,Yukalov_224} the results summarized 
in Table 5. For the components $N=-2$ and $N\ra\infty$, factor approximants give the 
exact values. The results for all $N$ are in good agreement with the known experimental 
exponents, as well as with Monte Carlo simulations, Borel-Pad\'{e} summation, and other 
methods, as can be inferred from reviews
\cite{Kleinert_30,Kleinert_222,Kazakov_225,Pelissetto_226} (see also
\cite{Lipa_227,Garcia_228,Pogorelov_229,Pogorelov_230}).

\begin{table}
\centering
\caption{Critical exponents for the $N$-component $\vp^4$ field theory, obtained
by the summation of $\ep$-expansions using self-similar factor approximants.}
\vskip 3mm
\label{tab-5}
\renewcommand{\arraystretch}{1.25}
\begin{tabular}{|c|c|c|c|c|c|c|c|} \hline
$N$&   $\al$     & $\bt$   &   $\gm$ & $\dlt$ & $\eta$  & $\nu$   & $\om$ \\ \hline
-2 &    0.5      &  0.25   &  1      & 5      & 0       & 0.5     & 0.79838 \\
-1 &    0.36612  & 0.27742 & 1.0791  & 4.8897 & 0.01874 & 0.54463 & 0.79380  \\
 0 &    0.23466  & 0.30268 & 1.1600  & 4.8323 & 0.02875 & 0.58845 & 0.79048 \\
 1 &    0.10645  & 0.32619 & 1.2412  & 4.8050 & 0.03359 & 0.63118 & 0.78755 \\
 2 &   -0.01650  & 0.34799 & 1.3205  & 4.7947 & 0.03542 & 0.67217 & 0.78763 \\
 3 &   -0.13202  & 0.36797 & 1.3961  & 4.7940 & 0.03556 & 0.71068 & 0.78904 \\
 4 &   -0.23835  & 0.38603 & 1.4663  & 4.7985 & 0.03476 & 0.74612 & 0.79133 \\
 5 &   -0.33436  & 0.40208 & 1.5302  & 4.8057 & 0.03347 & 0.77812 & 0.79419 \\
 6 &   -0.41963  & 0.41616 & 1.5873  & 4.8142 & 0.03197 & 0.80654 & 0.79747 \\
 7 &   -0.49436  & 0.42836 & 1.6376  & 4.8231 & 0.03038 & 0.83145 & 0.80108 \\
 8 &   -0.55920  & 0.43882 & 1.6816  & 4.8320 & 0.02881 & 0.85307 & 0.80503 \\
 9 &   -0.61506  & 0.44774 & 1.7196  & 4.8406 & 0.02729 & 0.87169 & 0.80935 \\
10 &   -0.66297  & 0.45530 & 1.7524  & 4.8489 & 0.02584 & 0.88766 & 0.81408 \\
50 &   -0.98353  & 0.50113 & 1.9813  & 4.9537 & 0.00779 & 0.99451 & 0.93176 \\
100 &   -0.93643 & 0.49001 & 1.9564  & 4.9926 & 0.00123 & 0.97881 & 0.97201 \\
1000 &  -0.99528 & 0.49933 & 1.9966  & 4.9986 & 0.00023 & 0.99842 & 0.99807  \\
10000 & -0.99952 & 0.49993 & 1.9997  & 4.9999 & 0.00002 & 0.99984 & 0.99979  \\
$\infty$ &  -1   &  0.5    &  2      &  5     &  0      & 1       &  1 \\ \hline
\end{tabular}
\end{table}

\subsection{Time Series}

There exist various time series consisting of sets of data $f_n$ at different
moments of time $t_n$. One of the most important tasks concerning time series is
the possibility of forecasting their future behavior being based on the knowledge
of their past data.

Let a time series be characterized by a data set
\be
\label{3.236}
 \mathbb{D}_k = \{ f_n, \; t_n: ~ n = 0,1,2,\ldots, k \} \;  ,
\ee
in which all $t_n$ are distinct and ordered either in ascending or descending
order. It is straightforward to define the interpolating polynomial
\be
\label{3.237}
f_k(t) = \sum_{n=0}^k c_n t^n
\ee
that passes through all point of the data set,
\be
\label{3.238}
f_k(t_n) = f_n \qquad ( n = 0,1,2,\ldots, k ) \; .
\ee
The latter condition is nothing but a system of linear algebraic equations with
respect to the coefficients $c_m$,
\be
\label{3.239}
  \sum_{m=0}^k c_m t_n^m = f_n \;   .
\ee
The solution to this system of equations involves the Vandermonde determinant
\begin{eqnarray}
\nonumber
V_k = \left | \begin{array}{ccccc}
1 ~ & ~ t_0 ~ & ~ t_0^2 ~ & ~ \ldots ~ & ~ t_0^k \\
1 ~ & ~ t_1 ~ & ~ t_1^2 ~ & ~ \ldots ~ & ~ t_1^k \\
\ldots ~ & ~ \ldots ~ & ~ \ldots ~ & ~ \ldots ~ & ~ \ldots \\
1 ~ & ~ t_k ~ & ~ t_k^2 ~ & ~ \ldots ~ & ~ t_k^k
\end{array} \right | = \prod_{0\leq i < j \leq k} (t_j - t_i) \; .
\end{eqnarray}
For all $t_n$ distinct, the determinant is nonzero, hence the solution is unique.

The other way of defining the interpolating polynomial is by composing a function
\be
\label{3.240}
 f_k(t) = \sum_{n=0}^k f_n \lbd_n^k(t) \;  ,
\ee
which is an expansion over the Lagrange basis polynomials
$$
 \lbd_n^k(t) \equiv \prod_{j(\neq n)}^k \frac{t-t_j}{t_n-t_j} \qquad
( n \leq k ) \;  ,
$$
having the property
$$
 \lbd_n^k(t_m) = \dlt_{mn} \;  .
$$

On the basis of the interpolating polynomial $f_k(t)$, it is possible to construct
a self-similar approximant $f_k^*(t)$ that allows for the extrapolation of the
data set for times $t$ larger than the maximal of $t_n$. Thus we obtain a forecast
$f_k^*(t)$ for future time. Such an extrapolation can be done using exponential
approximants
\cite{Yukalov_168,Yukalov_185,Gluzman_231,Gluzman_232,Gluzman_233,Yukalov_234,Yukalov_235,
Gluzman_236} or factor approximants.

\subsection{Probabilistic Scenarios}

Quite regularly, accomplishing measurements for the same process may result in
different data. This can be due, e.g., to the existence of poorly controlled noise,
because the measurements have been accomplished by different devices or with
differing methodics, or just realized in slightly different times. Then the results
composing different data sets have to be characterized in a probabilistic picture.
The idea of a probabilistic approach comes from the problem of pattern selection in
the case of multiple solutions of nonlinear equations
\cite{Yukalov_185,Yukalov_235,Yukalov_237,Yukalov_238,Yukalov_239}.

Suppose we have several data sets enumerated by an index $\al=1,2,\ldots$,
\be
\label{3.241}
 \mathbb{D}_{k\al} = \{ f_{n\al}, \; t_{n\al}: ~ n = 0,1,2,\ldots, k \} \; .
\ee
For each data set, we can construct, as is explained in the previous section,
a self-similar approximant $f_{k \alpha}^*(t)$ forecasting for times outside the
data set. Following Sec. 3.3, we interpret the sequence of $f_{k \alpha}^*(t)$ as
the points corresponding to a trajectory of a dynamical system for which we can
define the map multipliers
\be
\label{3.242}
 m_{k\al}^*(t) = \frac{\dlt f_{k\al}^*(t)}{\dlt f^*(t)} =
\frac{\prt f_{k\al}^*(t)/\prt t}{ \prt f^*(t)/\prt t} \; ,
\ee
where $f^*(t)$ is a first nontrivial self-similar approximant. Then we introduce
the average multiplier $M_k(t)$ by the relation
\be
\label{3.243}
 \frac{1}{M_k(t)} \equiv \sum_\al  \frac{1}{| m_{k\al}^*(t) |} \;  .
\ee
The probability of a scenario $\alpha$ at time $t$ is
\be
\label{3.244}
 p_\al(t) = \left |  \frac{M_k(t)}{ m_{k\al}^*(t)} \right | \;  .
\ee
By construction, it is normalized,
\be
\label{3.245}
 \sum_\al  p_\al(t) = 1 \;  .
\ee
The optimal scenario corresponds to the maximal probability
\be
\label{3.246}
  p_{opt}(t) \equiv \sup_\al  p_\al(t) \; .
\ee

\subsection{Discrete Scaling}

One calls a function $f(x)$ scale invariant, when it satisfies the scaling relation
\be
\label{3.247}
 f(\lbd x) = \mu f(x) \;  .
\ee
The evident solution to this equation is the power-law behavior
\be
\label{3.248}
  f(x) = C x^\al \; ,
\ee
with the condition
\be
\label{3.249}
\lbd^\al = \mu e^{i2\pi n} \qquad ( n = 0,1,2,\ldots ) \;  .
\ee
Therefore the power $\alpha$ writes as
\be
\label{3.250}
  \al = \frac{\ln\mu}{\ln\lbd} + i \; \frac{2\pi n}{\ln\lbd} \; .
\ee

In the case of continuous scaling, the scale $\lbd$ can be taken as $\lbd=1$,
which requires to set $n=0$. Then
\be
\label{3.251}
   \al = \frac{\ln\mu}{\ln\lbd} \qquad ( n = 0 ) \;  .
\ee
This defines the power $\al$ as a real number, provided $\lbd$ and $\mu$ are
real.

But for discrete scaling, $\lbd$ can pertain to a set not including $\lbd=1$.
In such a case, the power $\al$ becomes complex-valued \cite{Sornette_240}. The
processes with complex-valued exponents, resulting in log-periodic oscillations,
 occur, for instance, in financial time series
\cite{Sornette_241,Johansen_242,Drozdz_299,Johansen_243,Johansen_244,
Feigenbaum_245,Sornette_246,Feigenbaum_247} and materials rupture
\cite{Johansen_248}.

\subsection{Complex Renormalization}

Under discrete scaling, when characteristic exponents become complex-valued,
there appears the necessity of considering renormalization equations on a complex
plane. Below, we give an example of such a procedure.

Let us study a system characterized by a function $I(t)$. This can be a market
price, financial index, or the intensity of acoustic radiation from a loaded
material as a function of time \cite{Moura_249,Yukalov_250}. The function is
considered on the time interval $0<t< t_c$, with $t_c$ being a critical point,
where $I(t)$ experiences a sharp increase or a drastic fall. It is convenient
to introduce the dimensionless variable
\be
\label{3.252}
 x \equiv \frac{t_c-t}{t_c} = x(t) \;  ,
\ee
with the boundary conditions
\be
\label{3.253}
 x(0) = 1 \; , \qquad x(t_c) = 0 \;  .
\ee
Also, we define the dimensionless function
\be
\label{3.254}
f(x) \equiv \frac{I(t_c)-I(t)}{I(t_c)-I(0)}
\ee
satisfying the boundary conditions
$$
f(0) = 0 \qquad ( t = t_c \; , ~ x(t_c) = 0 ) \; ,
$$
\be
\label{3.255}
 f(1) = 1  \qquad ( t = 0 \; , ~ x(0) = 1 ) \;  .
\ee
The function $I(t)$ is expressed through function (\ref{3.254}) as
\be
\label{3.256}
 I(t) = I(t_c) - [ I(t_c) - I(0) ] f(x(t) ) \;  .
\ee

Since we keep in mind a discrete scaling, requiring to work on the complex
plane, we analytically continue the real function $f(x)$ to a complex function
$F(x)$, such that
\be
\label{3.257}
f(x) = {\rm Re} F(x) \;   .
\ee
And $F(x)$ is assumed to obey the same boundary conditions as $f(x)$,
\be
\label{3.258}
 F(0) = 0 \; , \qquad F(1) = 1 \;  .
\ee

The renormalization-group equation for $F(x)$ reads \cite{Sornette_241} as
\be
\label{3.259}
 \frac{d\ln F}{d\ln x} = G \;  ,
\ee
where $G = G(x)$ is a Gell-Mann law function. This equation has to be complemented
either by an equation for the complex-conjugate function $F^*(x)$ or by an equation
for the function
\be
\label{3.260}
 W = | F |^2 = W(x) \;  ,
\ee
with the boundary conditions
\be
\label{3.261}
  W(0) = 0 \; , \qquad W(1) = 1 \; .
\ee
From Eq. (\ref{3.259}), we get
\be
\label{3.262}
\frac{d\ln W}{d\ln x} = 2{\rm Re} G \;   .
\ee

The function $W$ equals the squared amplitude of $F$ describing the main
tendency of the variance of $F$, while $F$ itself also contains a phase producing
oscillations around the main tendency. Therefore it is reasonable to treat $F$ as
a fast function, as compared to the slow function $W$. Then it is admissible to
employ the idea of the averaging techniques and solve the equation for the fast
function $F$, keeping the slow function as an integral of motion. This gives
\be
\label{3.263}
  F = x^G \; .
\ee
Generally, $G$ is a complex function, hence the previous expression can be
represented as
\be
\label{3.264}
 F = x^{{\rm Re} G} \left\{ \cos \left[ ({\rm Im} G)\ln x \right] +
i\sin \left[ ({\rm Im} G)\ln x \right] \right \} \; .
\ee
Note that the necessary and sufficient condition for classifying the function
$F$ as fast, while $W$ as slow is $\rm{Re} G \ll \rm{Im} G$.

We need to find out the dependence of $G$ on $x$. If we assume that the function
$W(x)$ can be expanded in powers of some $x^s$ then from equation (\ref{3.262})
it follows that $G(x)$ is expandable in the same powers, so that its $k$-th order
expansion reads as
\be
\label{3.265}
  G(x) \simeq G_k(x) = \sum_{n=0}^k a_n x^{sn} \qquad ( x \ra 0 ) \; ,
\ee
under the relation
\be
\label{3.266}
 s = 2 {\rm Re} a_0 \equiv 2 \al \;  .
\ee

The extrapolation of the small variable asymptotic expansion (\ref{3.265}) to
the finite values of $x$ can be done by means of the self-similar exponential
approximants (see Sec. 3.10). This yields
\be
\label{3.267}
G_k^*(x) = a_0 \exp \left( C_1 x^{2\al} \exp\left( C_2 x^{2\al} \ldots
\exp \left( C_k x^{2\al} \right) \right) \right) \;   ,
\ee
where $C_j$ are control parameters. To first order, we have
\be
\label{3.268}
 G_1^*(x) = a_0 \exp \left( C_1 x^{2\al} \right) \; .
\ee
Taking into account that the parameters, in general, are complex, we separate
their real and imaginary parts,
\be
\label{3.269}
 a_0 = \al + i\om \; , \qquad C_1 = \bt + i\gm \;  .
\ee
Then we get
$$
G_1^*(x) = ( \al + i\om ) \exp\left \{( \bt + i\gm ) x^{2\al} \right\} =
$$
\be
\label{3.270}
 =
\exp \left( \bt x^{2\al} \right) \left \{ \al\cos\left( \gm x^{2\al} \right) -
\om\sin\left( \gm x^{2\al}\right) + i \left[
\al\sin\left( \gm x^{2\al} \right) + \om\cos\left( \gm x^{2\al} \right) \right]
 \right\} \; .
\ee

Using the approximant (\ref{3.270}), we obtain the sought function
\be
\label{3.271}
 f(x) = {\rm Re} F(x) = \exp [ ({\rm Re} G ) \ln x ]
\cos [ ({\rm Im} G ) \ln x ] \; ,
\ee
in which
$$
{\rm Re} G = \left[ \al\cos\left( \gm x^{2\al} \right) -
\om\sin\left( \gm x^{2\al}\right) \right] \exp \left( \bt x^{2\al} \right) \; ,
$$
\be
\label{3.272}
 {\rm Im} G = \left[ \al\sin\left( \gm x^{2\al} \right) +
\om\cos\left( \gm x^{2\al}\right) \right] \exp \left( \bt x^{2\al} \right) \; .
\ee
The four parameters, $\alpha$, $\beta$, $\gamma$, and $\omega$ should be found from
additional conditions, or from empirical data.

Note that close to the critical point $t_c$, where
$$
x \ra 0 \; , \qquad {\rm Re} G \ra \al \; , \qquad {\rm Im} G \ra \om \;  ,
$$
function (\ref{3.271}) reduces to
$$
 f(x) \simeq x^\al \cos( \om \ln x ) \qquad ( x \ra 0 ) \;  ,
$$
which is the simple case of a power law behavior decorated by log-periodic
oscillations.

Taking higher orders of approximant (\ref{3.267}) makes the function $f(x)$ more
complicated and increases the number of parameters. In the $k$-th order, the number
of control parameters is $2(1+k)$.

\subsection{Quark-Gluon Plasma}

In Quantum Chromodynamics (QCD), perturbation theory with respect to the coupling
parameter is valid at asymptotically large temperatures. One assumes that the
temperature is so high that it is admissible to set the chemical potential zero and
to neglect quark masses. The weak-coupling free energy expansion can be represented
either in powers of the quantum  chromodynamics coupling $\al_s$ or in powers of
the coupling parameter $g$, using the relation
\be
\label{3.273}
\al_s = \frac{g^2}{4\pi} \;   .
\ee

The perturbative expansion for the pressure of high-temperature $SU(N_c)$ gauge
theory, with massless fermions, in four dimensions, with $N_c = 3$ colours and with
$n_f$ quark flavors in the fundamental representation, has been calculated to fourth
order \cite{Arnold_251} and to fifth order \cite{Zhai_252,Braaten_253}. In terms
of $\alpha_s$, the expansion of pressure reads as
\be
\label{3.274}
P(\al_s) \simeq \frac{8\pi^2}{45}\; T^4 \; \left[
\sum_{n=0}^5 a_n \left( \frac{\al_s}{\pi} \right)^{n/2} +
a_4' \left( \frac{\al_s}{\pi} \right)^2 \ln \; \frac{\al_s}{\pi} \right] \; ,
\ee
with the coefficients
$$
a_0 = 1 + \frac{21}{32}\; n_f \; , \qquad a_1 = 0 \; , \qquad
a_2 = -\; \frac{15}{4} \; \left( 1 + \frac{15}{12}\; n_f  \right) \; , \qquad
a_3 = 30 \left( 1 + \frac{1}{6}\; n_f  \right)^{3/2} \; ,
$$
$$
a_4 = 273.2 + 15.97\; n_f - 0.413\; n_f^2 +
\frac{135}{8} \; \left( 1 + \frac{1}{6}\; n_f  \right)
\ln \left( 1 + \frac{1}{6}\; n_f  \right) -
$$
$$
- \;
\frac{165}{8}  \left( 1 + \frac{5}{12}\; n_f  \right)
\left( 1 -\; \frac{2}{32}\; n_f  \right) \ln \; \frac{\mu}{2\pi T} \; ,
$$
$$
a_5 = \left( 1 + \frac{1}{6}\; n_f  \right)^{1/2} \left[ \frac{495}{2}
\left( 1 + \frac{1}{6}\; n_f  \right)\left( 1 -\; \frac{2}{33}\; n_f  \right)
\ln \; \frac{\mu}{2\pi T} - 799.2 - 21.96\; n_f - 1.926\; n_f^2 \right] \; ,
$$
$$
a_4' =  \frac{135}{2}\; \left( 1 + \frac{1}{6}\; n_f  \right) \; .
$$
Here, $\mu$ is the renormalization scale parameter in the $\overline{\rm MS}$
scheme and the dimensional regularization is used.

The weak-coupling expansion for the pressure in terms of $g$ is
\be
\label{3.275}
 P(g) \simeq \frac{8\pi^2}{45} \; T^4 \left( \sum_{n=0}^5 c_n g^n  +
c_4' g^4 \ln g \right) \;  ,
\ee
with the coefficients
$$
c_0 = 1 + \frac{21}{32}\; n_f \; , \qquad c_1 = 0 \; , \qquad
c_2 = -\; \frac{15}{(4\pi)^2} \; \left( 1 + \frac{5}{12}\; n_f \right)
= - 0.09499 \left( 1 + \frac{5}{12}\; n_f  \right) \; ,
$$
$$
c_3 = \frac{240}{(4\pi)^3} \; \left( 1 + \frac{1}{6}\; n_f \right)^{3/2} =
0.12094 \; \left( 1 + \frac{1}{6}\; n_f \right)^{3/2}
$$
$$
c_4 =  0.04331 \left( 1 + \frac{1}{6}\; n_f  \right)
\ln \left( 1 + \frac{1}{6}\; n_f  \right) + 0.01733 - 0.00763\; n_f - 0.00088\; n_f^2 -
$$
$$
- 0.01323 \left( 1 + \frac{5}{12}\; n_f  \right)
\left( 1 -\; \frac{2}{33}\; n_f  \right) \ln \; \frac{\mu}{T} \; ,
$$
$$
c_5 = 0.02527 \left( 1 + \frac{1}{6}\; n_f  \right)^{3/2}
\left( 1 -\; \frac{2}{33}\; n_f  \right) \ln \; \frac{\mu}{T}  - 
$$
$$
-
\left( 0.12806 + 0.00717\; n_f - 0.00027\; n_f^2 \right)
\left( 1 + \frac{1}{6}\; n_f  \right)^{1/2} \; ,
$$
$$
c_4' =  0.08662 \left( 1 + \frac{1}{6}\; n_f  \right) \;  .
$$

It is convenient to introduce the reduced pressure
\be
\label{3.276}
 p_k \equiv \frac{P_k}{P_0} = p_k(g,\mu,T) \;  ,
\ee
that is the ratio of a $k$-th order pressure $P_k$, to the Stefan-Boltzmann pressure
\be
\label{3.277}
 P_0 = \frac{8\pi^2}{45} \; T^4 \left( 1 + \frac{21}{32}\; n_f  \right) \; .
\ee
The expansion for the reduced pressure reads as
\be
\label{3.278}
p_k = 1 + \sum_{n=0}^5 \overline c_n g^n  +  \overline c_4' g^4 \ln g \; ,
\ee
where
$$
 \overline c_n \equiv \frac{c_n}{c_0} \; , \qquad
\overline c_4' \equiv \frac{c_4'}{c_0} \;  .
$$

The parameter $g$ is the running coupling, whose dependence $g(\mu)$ on the
renormalization scale $\mu$ is given by the renormalization group equation
\be
\label{3.279}
 \mu \; \frac{\prt g}{\prt\mu} = \bt(g) \;  .
\ee
The weak-coupling expansion for the Gell-Mann-Low function $\beta(g)$ can be
written as
\be
\label{3.280}
\bt_k(g) = - \sum_{n=0}^k b_n g^{2n+3} \; .
\ee
The coefficients for $N_c = 3$ have been calculated in four-loop order
\cite{Ritbergen_254,Czakon_255} and, recently, in five-loop order
\cite{Luthe_256,Baikov_257}. These coefficients are
$$
b_0 =  \frac{1}{(4\pi)^2} \; \left( 11 -\; \frac{2}{3}\; n_f  \right) \; ,
\qquad
b_1 =  \frac{2}{(4\pi)^4} \; \left( 51 -\; \frac{19}{3}\; n_f  \right) \; ,
$$
$$
b_2 =  \frac{1}{2(4\pi)^6} \;
\left( 2857 -\; \frac{5033}{9}\; n_f + \frac{325}{27}\; n_f^2 \right) \; ,
$$
$$
b_3 =  \frac{1}{(4\pi)^8} \; \left[  \frac{149753}{6}+ 3564\zeta(3) -
\left( \frac{1078361}{162} +  \frac{6508}{27} \;\zeta(3) \right) n_f + 
\right.
$$
$$
+
\left.
\left( \frac{50065}{162} +  \frac{6472}{81} \;\zeta(3) \right) n_f^2 +
\frac{1093}{729} \; n_f^3 \right] \; ,
$$
$$
b_4 = \frac{1}{(4\pi)^{10}} \; \left[  \frac{8157455}{16} +
\frac{621885}{2}\; \zeta(3) - \; \frac{88209}{2}\; \zeta(4) -
288090 \zeta(5) - \right.
$$
$$
-\; \left( \frac{336460813}{1944} + \frac{4811164}{81}\; \zeta(3) -
\; \frac{33935}{6}\; \zeta(4) - \; \frac{1358995}{27}\; \zeta(5) \right) n_f +
$$
$$
+
\left( \frac{25960913}{1944} + \frac{698531}{81}\; \zeta(3) -
\; \frac{10526}{9}\; \zeta(4) - \; \frac{381760}{81}\; \zeta(5) \right) n_f^2 -
$$
$$
- \left.
\left( \frac{630559}{5832} + \frac{48722}{243}\; \zeta(3) -
\; \frac{1618}{27}\; \zeta(4) - \; \frac{460}{9}\; \zeta(5) \right) n_f^3 +
\left( \frac{1205}{2916} -\; \frac{152}{81}\; \zeta(3)\right) n_f^4 \right] \; .
$$
As an initial condition for Eq. (\ref{3.277}), we can take the value of the
running coupling $\al_s(m_Z) = 0.119$ for the $Z^0$ boson mass $m_Z$. In terms of
the coupling $g$, this initial condition is
\be
\label{3.281}
g(m_Z) = 1.222856 \qquad ( m_Z = 91187\; {\rm MeV} ) \;  .
\ee

The equation of state for quantum chromodynamics, employing exponential
self-similar approximants (see Sec. 3.10), has been derived in Ref.
\cite{Yukalov_258}. The procedure has been as follows. The renormalization scale
$\mu$ is treated as a control function defined by the minimal difference condition
$p_4-p_3=0$, which gives $\mu=\mu(g,T)$. By using the expansions $\bt_k(g)$ and
$p_k(g,T)$, exponential self-similar approximants $\beta_k^*(g)$ and $p_k^*(g,T)$
are constructed. The renormalization group equation (\ref{3.276}), with the initial
condition (\ref{3.278}), yields $g = g(\mu)$. From $\mu = \mu(g,T)$ and $g = g(\mu)$,
we find $\mu(T)$ and $g(T)$. Then the self-similar approximant $p_k^*(g,\mu,T)$
defines the pressure as a function of temperature $p_k^*(T)$. When diminishing
temperature, the pressure varies from the Stefan-Boltzmann limit to a sharp drop to
zero at temperature $T_c\approx 150$ Mev. This temperature $T_c$ can be associated
with the confinement-deconfinement phase transition. These results agree with other
models of confinement \cite{Yukalov_259,Yukalov_260,Yukalov_298}, as well as with
lattice data \cite{Borsanyi_261}. Hard-thermal-loop perturbation theory, that is a
variant of optimized perturbation theory, also yields a similar behavior of the QCD
pressure \cite{Andersen_262}.

\subsection{Gell-Mann-Low Function: Quantum Chromodynamics}

The overall behavior of a Gell-Mann-Low function, as a function of a scaling
variable provides important information on the related field theory. Of special
interest is the large variable asymptotic form of the function. In Secs. 3.18
and 3.19, it has been shown that the large-variable behavior of a function can
be predicted by extrapolating its small-variable expansion by means of self-similar
factor approximants. The accuracy of such a prediction is about $10 \%$. Using these
approximants, we also can estimate the large-variable behavior of Gell-Mann-Low
functions.

The Gell-Mann-Low function for QCD can be defined as
\be
\label{3.282}
\bt(a_s) = \mu^2 \; \frac{da_s}{d\mu^2} \qquad
\left ( a_s \equiv \frac{\al_s}{\pi} \right ) \;  .
\ee
In the space-time four dimensions, within the minimal subtraction scheme
$(\overline{\rm MS})$, a weak-coupling expansion of the beta function in powers of $a_s$,
in five-loop order \cite{Baikov_257,Herzog_263}, reads
as
\be
\label{3.283}
 \bt(a_s)  \simeq - \sum_{n=0}^k \bt_n a_s^{n+2} \qquad ( a_s \ra 0 ) \;  ,
\ee
with the coefficients $\beta_n$ expressed through the coefficients $b_n$ from the
previous section by the relation
\be
\label{3.284}
 \bt_n = (2\pi)^{2n+2} b_n \;  .
\ee
In numerical form, these coefficients are
$$
\bt_0 = 2.75 - 0.166667\; n_f \; , \qquad \bt_1 = 6.375 - 0.791667\; n_f \; ,
$$
$$
\bt_2 = 22.3203 - 4.36892\; n_f + 0.0940394 \; n_f^2 \; ,
$$
$$
\bt_3 = 114.23 - 27.1339\; n_f + 1.58238 \; n_f^2 + 0.0058567 \; n_f^3\; ,
$$
$$
 \bt_4 = 524.56 - 181.8\; n_f + 17.16 \; n_f^2 - 0.22586 \; n_f^3 -
0.0017993 \; n_f^4 \;  .
$$

Applying self-similar factor approximants for summing series (\ref{3.283}), we
can find the strong-coupling limit
\be
\label{3.285}
 \bt(a_s) \simeq - B a_s^\gm \qquad ( a_s \ra \infty )
\ee
that is connected with the limit of the function $\beta(g)$, defined by
Eq. (\ref{3.279}), as
$$
\bt(g) = \frac{(2\pi)^2}{g} \; \bt(a_s) \simeq -\;
\frac{B}{(2\pi)^{2\gm-2} } \; g^{2\gm -1} \qquad ( g \ra \infty ) \;  .
$$

It turns out that for the factor approximants with the $n_f$ from zero to five
the limit $g$ tending to infinity does not exist, since the approximants become
complex-valued. For the physically reasonable number of flavors $n_f = 6$, there
exist the real-valued approximants:
$$
\bt_2^*(a_s) = -1.75 a_s^2 ( 1 + 1.1554 a_s)^{0.598} \; , 
$$
\be
\label{3.286}
 \bt_3^*(a_s) = -1.75 a_s^2 (  1 + a_s)^{0.912} ( 1 + 32.53 a_s)^{0.0005} \; ,
\ee
whose strong-coupling asymptotics are
$$
\bt_2^*(a_s) \simeq - 2.277 a_s^{2.598} \; ,
$$
\be
\label{3.287}
\bt_3^*(a_s) \simeq - 1.753 a_s^{2.913} \qquad  ( a_s \ra \infty ) \;  .
\ee

Similarly, applying self-similar factor approximants for the beta function
defined in equation (\ref{3.279}), with expansion (\ref{3.280}), we obtain
$$
 \bt_2^*(g) = - 0.044 g^3 ( 1 + 0.039 g^2)^{0.598} \; , 
$$
$$
 \bt_3^*(g) = - 0.044 g^3 ( 1 + g^2)^{0.000277} ( 1 + 0.028 g^2)^{0.833} \;   .
$$
The strong-coupling limit for these approximants gives
$$
 \bt_2^*(g) \simeq - 0.0064 g^{4.195} \; , \qquad
\bt_3^*(g) \simeq - 0.0022 g^{4.667} \qquad  ( g \ra \infty ) \; .
$$

It seems that there are no reliable estimates of the strong-coupling behavior
of the QCD beta function, which it would be possible to compare with the above
results. There exists an estimate \cite{Suslov_273,Suslov_264} of $\gamma=-13$
for $n_f = 0$ and the negative $g$ tending to minus infinity. The change of the
sign of $g$ is motivated by the fact that the used Borel-Leroy summation can be
applied to alternating series, while the series  (\ref{3.238}) are not alternating,
so that, strictly speaking, the Borel summation is not applicable. Changing the
sign of $g$ produces a kind of alternating series. However, the expansions for
$g\ra\infty$ and $g\ra -\infty$ can be drastically different \cite{Kazakov_290}.

\subsection{Gell-Mann-Low Function: Quantum Electrodynamics}

The Gell-Mann-Low function in Quantum Electrodynamics (QED) is known in the
five-loop approximation \cite{Herzog_263,Kataev_265}. The weak-coupling expansion
of the beta-function depends on the used renormalization scheme.

\vskip 2mm

(i) {\it Minimal subtraction scheme}

\vskip 2mm

In the modified minimal subtraction scheme $(\overline{\rm MS})$, the Gell-Mann-Low
function is defined by the renormalization group equation
\be
\label{3.288}
 \bt_{MS}(\overline\al) = \mu^2 \; \frac{\prt}{\prt\mu^2} \left(
\frac{ \overline\al}{\pi} \right) \; ,
\ee
in which $\overline{\alpha}$ is the renormalized $(\overline{\rm MS})$ scheme QED
coupling parameter and $\mu$ is a scale parameter. In the five-loop approximation
for the beta-function of an electron, neglecting the contributions of leptons with
higher masses, that is, of muons and tau-leptons, the expansion of the beta-function
with respect to $\overline{\alpha}$ \cite{Kataev_265} has the form
\be
\label{3.289}
 \bt_{MS}(\overline\al) \simeq \sum_{n=0}^k b_n
\left( \frac{ \overline\al}{\pi} \right)^{n+2} \;   ,
\ee
with the coefficients
$$
b_0 = \frac{1}{3} \; , \qquad b_1 = \frac{1}{4} \; , \qquad
b_2 = -\; \frac{31}{288} \; , \qquad b_3 = -\; \frac{2785}{31104} \; - \;
\frac{13}{36} \;\zeta(3) \; ,
$$
$$
 b_4 = -\; \frac{195067}{497664} \; - \; \frac{25}{96} \;\zeta(3)
- \; \frac{13}{96} \;\zeta(4) + \frac{215}{96} \;\zeta(5) \; .
$$
In numerical form, the coefficients are
$$
 b_0 = 0.3333\; , \qquad b_1 = 0.25 \; , \qquad b_2 = -0.1076 \; , \qquad
b_3 = -0.5236 \; , \qquad b_4 = 1.471 \;  .
$$

\vskip 2mm

(ii) {\it On-shell scheme}

\vskip 2mm
In the on-shell scheme (OS), one sets $\mu=m$, with $m$ being the electron mass.
The beta-function, is defined by the equation
\be
\label{3.290}
 \bt_{OS}(\al) = m^2 \; \frac{\prt}{\prt m^2} \left(
\frac{\al}{\pi} \right) \;  ,
\ee
where $\alpha=e^2/4 \pi$ is the QED coupling parameter. The weak-coupling
expansion reads as
\be
\label{3.291}
 \bt_{OS}(\al) \simeq \sum_{n=0}^k b_n  \left(\frac{\al}{\pi} \right)^{n+2}
\qquad ( \al \ra 0 ) \;  ,
\ee
with the coefficients
$$
b_0 = 0.3333\; , \qquad b_1 = 0.25 \; , \qquad b_2 = -0.4201 \; , \qquad
b_3 = -0.5712 \; , \qquad b_4 = -0.3462 \; .
$$
The latter are denoted here by the same notation $b_n$, which should not lead
to confusion.

\vskip 2mm

(iii) {\it Momentum subtraction scheme}

\vskip 2mm

In the frame of the momentum subtraction scheme (MOM), the beta-function expansion
can be written as in Eq. (\ref{3.289}), but with different coefficients:
$$
b_0 = 0.3333\; , \qquad b_1 = 0.25 \; , \qquad b_2 = 0.04999 \; , \qquad
b_3 = -0.6010 \; , \qquad b_4 = 1.434 \;   .
$$

\vskip 2mm

Constructing self-similar factor approximants using the beta-function expansions
for each of the scheme, we take those of them that are defined for all coupling
parameters $\alpha \in [0, \infty]$. The highest orders of these approximants give
the strong-coupling limits
$$
\bt_{MS}^*(\overline\al) \simeq 0.476\; \overline\al^{2.096} \; ,
\qquad
\bt_{OS}^*(\al) \simeq 0.433\; \al^{2.189} \; ,
$$
\be
\label{3.292}
\bt_{MOM}^*(\overline\al) \simeq 0.493 \;\overline\al^{2.309} \; .
\ee

As is seen, the strong-coupling asymptotic behavior of the beta function is
practically the same for all schemes. This suggests that self-similar factor
approximants provide scheme independent expressions. The Borel-Leroy summation
\cite{Suslov_264,Suslov_274} in the MOM scheme gives the strong-coupling limit
with the beta function proportional to $\overline{\alpha}$ in first order.

\subsection{Gell-Mann-Low Function: Quartic Field Theory}

The Gell-Mann-Low function for the $O(N)$-symmetric $\vp^4$ field theory,
within the minimal subtraction scheme, is known in the six-loop approximation
\cite{Kompaniets_266}. Defining the beta-function as
\be
\label{3.293}
 \bt(g) = \mu \; \frac{\prt g}{\prt\mu} \;  ,
\ee
where $g = \lbd/(4\pi)^2$, for its weak-coupling expansion one has
\be
\label{3.294}
 \bt(g) \simeq \sum_{n=0}^k b_n g^{n+2} \;  .
\ee
The six coefficients $b_n$ for a different number of components are given in
Table 6.

Schnetz \cite{Schnetz_267} derived the coefficients for $N = 1$ to seven-loop
order:
$$
b_0 = 3 \; , \qquad b_1 = -5.66667 \; , \qquad b_2 = 32.5497 \; , \qquad
b_3 = -271.606 \; ,
$$
$$
 b_4 = 2848.57 \; , \qquad b_5 = -34776.1 \; , \qquad
b_6 = 474651  \;   .
$$

The general six-loop expansion of the beta function for the $N$-component field
theory \cite{Kompaniets_266} has the coefficients
$$
b_0 = \frac{N+8}{3} \; , \qquad b_1 = -\; \frac{3N+14}{3} \; ,
$$
$$
b_2 = \frac{1}{216} \left[ 96(5N+22)\zeta(3) + 33 N^2 + 922 N + 2960 \right] \; ,
$$
$$
b_3 = -\; \frac{1}{3888} \left[ 1920 \left( 2N^2 + 55N + 186 \right) \zeta(5)
- 288(N+8)(5N+22)\zeta(4) + \right.
$$
$$
+ \left.
96\left( 63N^2 + 764 N + 2332 \right)\zeta(3) -
\left( 5N^3 - 6320 N^2 - 80456 N - 196648\right)  \right] \; ,
$$
$$
b_4 =  \frac{1}{62208} \left[  112896 \left( 14 N^2 + 189 N + 526\right)\zeta(7) -
768 \left( 6N^3 + 59N^2 - 446 N - 3264\right) \zeta^2(3) -  \right.
$$
$$
-
9600(N+8) \left( 2N^2 + 55N + 186\right) \zeta(6) +
256\left( 305 N^3 + 7466 N^2 + 66986 N + 165084\right)\zeta(5) -
$$
$$
-
288\left(63 N^3 + 1388 N^2 + 9532 N + 21120\right)\zeta(4) -
$$
$$
-
16\left( 9N^4 - 1248 N^3 - 67640 N^2 - 552280 N - 1314336\right)\zeta(3) +
$$
$$
+ \left.
13 N^4 + 12578 N^3 + 808496 N^2 + 6646336 N + 13177344 \right] \; ,
$$
$$
b_5 =  -\; \frac{1}{41990400} \left[  204800
\left( 1819N^3 + 97823 N^2 + 901051 N + 2150774\right)\zeta(9) +  \right.
$$
$$
+
14745600\left( N^3 + 65 N^2 + 619 N + 1502 \right)\zeta^3(3) +
$$
$$
+
995328 \left( 42N^3 + 2623 N^2 + 25074 N + 59984\right)\zeta(3,5) -
$$
$$
-
20736\left( 28882 N^3 + 820483 N^2 + 6403754 N + 14174864 \right) \zeta(8) -
$$
$$
-
5529600 \left( 8N^3 - 635 N^2 - 9150 N - 25944 \right) \zeta(3)\zeta(5) +
$$
$$
+
11520\left( 440N^4 + 126695 N^3 + 2181660 N^2 + 14313152 N + 29762136
\right)\zeta(7) +
$$
$$
+
207360 (N+8)\left( 6N^3 + 59N^2 - 446N - 3264\right)\zeta(3)\zeta(4) -
$$
$$
-
23040\left( 188 N^4 + 132 N^3 - 93363 N^2 - 862604 N - 2207484\right)\zeta^2(3)
-
$$
$$
-
28800\left( 595 N^4 + 20286 N^3 + 277914 N^2 + 1580792 N + 2998152\right)\zeta(6) +
$$
$$
+
5760\left( 4698 N^4 + 131827 N^3 + 2250906 N^2 + 14657556 N + 29409080\right)\zeta(5)
+
$$
$$
+
2160\left( 9 N^5 - 1176 N^4 - 88964 N^3 - 1283840N^2 - 6794096 N -
12473568\right)\zeta(4) -
$$
$$
- 720\left( 33 N^5 + 2970 N^4 - 477740 N^3 - 10084168 N^2 -
61017200 N - 117867424 \right)\zeta(3) -
$$
$$
- \left.
45\left( 29N^5 + 22644 N^4 - 3225892 N^3 - 88418816 N^2 - 536820560 N -
897712992\right)  \right] \; .
$$
Here $\zeta(3,5)$ implies the double zeta function
$$
\zeta(3,5) = \sum_{1\leq n<m} \frac{1}{n^3m^5} = 0.037707673 \;   .
$$

\begin{table}[h!]
\centering
\caption{Coefficients of weak-coupling expansion for the Gell-Man-Low function
of the $N$-component $\vp^4$ field theory, in the six-loop approximation.}
\vskip 3mm
\label{tab-6}
\renewcommand{\arraystretch}{1.25}
\begin{tabular}{|c|c|c|c|c|c|} \hline
$N$   &     0     &       1   &    2     &    3     &    4     \\ \hline
$b_0$ &    2.667  &     3.000 &    3.333 &    3.667 &    4.000 \\ \hline
$b_1$ & $-$4.667  &  $-$5.667 & $-$6.667 & $-$7.667 & $-$8.667  \\ \hline
$b_2$ &  25.46    &   32.55   &  39.95   &  47.65   &   55.66   \\ \hline
$b_3$ & $-$200.9  &  $-$271.6 & $-$350.5 & $-$437.6 & $-$533.0  \\ \hline
$b_4$ &  2004     &    2849   &    3845  &   4999   &   6318     \\ \hline
$b_5$ &  $-$23315 &  $-$34776 & $-$48999 & $-$66243 & $-$86768 \\ \hline
\end{tabular}
\end{table}

Employing self-similar factor approximants for the six-loop expansions, we find
the beta-functions for a different number of components. We present below the
approximants of even orders that are more reliable then those of odd orders,
although all of them are close to each other. Thus for $N = 0$, we have
$$
\bt_2^*(g) = \frac{2.667 g^2}{( 1 + 9.161 g)^{0.191} } \qquad ( N = 0 ) \; ,
$$
\be
\label{3.295}
 \bt_4^*(g) = \frac{2.667 g^2}{( 1 + 6.083 g)^{0.194}(1+15.544 g)^{0.0366} } \;  ,
\ee
which yields the strong-coupling behavior
\be
\label{3.296}
 \bt_2^*(g) \simeq 1.747 g^{1.809} \; , \qquad  \bt_4^*(g) \simeq 1.699 g^{1.769}
\qquad ( N = 0 , ~ g \ra \infty) \; .
\ee

For $N = 2$, we get
$$
\bt_2^*(g) = \frac{3.333 g^2}{( 1 + 9.984 g)^{0.2003} } \qquad ( N = 2 ) \; ,
$$
\be
\label{3.297}
\bt_4^*(g) = \frac{3.333 g^2}{( 1 + 5.812 g)^{0.206}(1+16.234 g)^{0.0493} } \;   ,
\ee
with the strong-coupling limit
\be
\label{3.298}
 \bt_2^*(g) \simeq 2.102 g^{1.7997} \; , \qquad  \bt_4^*(g) \simeq 2.020 g^{1.7443}
\qquad ( N = 2 , ~ g \ra \infty) \;  .
\ee

In the case of $N = 3$, we find
$$
\bt_2^*(g) = \frac{3.667 g^2}{( 1 + 10.339 g)^{0.202} } \qquad ( N = 3 ) \; ,
$$
\be
\label{3.299}
 \bt_4^*(g) = \frac{3.667 g^2}{( 1 + 5.544 g)^{0.213}(1+16.537 g)^{0.055} } \;    ,
\ee
giving the limit
\be
\label{3.300}
 \bt_2^*(g) \simeq 2.286 g^{1.798} \; , \qquad  \bt_4^*(g) \simeq 2.183 g^{1.732}
\qquad ( N = 3 , ~ g \ra \infty) \;   .
\ee

And the case of $N = 4$ results in
$$
\bt_2^*(g) = \frac{4 g^2}{( 1 + 10.677 g)^{0.203} } \qquad ( N = 4 ) \; ,
$$
\be
\label{3.301}
 \bt_4^*(g) = \frac{4 g^2}{( 1 + 5.274 g)^{0.219}(1+16.866 g)^{0.05988} } \;  ,
\ee
with the strong-coupling limit
\be
\label{3.302}
 \bt_2^*(g) \simeq 2.474 g^{1.797} \; , \qquad  \bt_4^*(g) \simeq 2.345 g^{1.721}
\qquad ( N = 4 , ~ g \ra \infty) \;   .
\ee

Using the seven-loop expansion for $N = 1$ yields the beta functions
$$
\bt_2^*(g) = \frac{3 g^2}{( 1 + 9.599 g)^{0.197} } \qquad ( N = 1 ) \; ,
$$
$$
\bt_4^*(g) = \frac{3 g^2}{( 1 + 6.014 g)^{0.2004}(1+15.9204 g)^{0.0479} } \; ,
$$
\be
\label{3.303}
\bt_6^*(g) = 
\frac{3 g^2}{( 1 + 5.357 g)^{0.187}(1+13.72 g)^{0.0597}(1+22.096)^{0.00316} }\; ,
\ee
whose strong-coupling behavior is
$$
 \bt_2^*(g) \simeq 1.922 g^{1.803} \; , \qquad  
\bt_4^*(g) \simeq 1.859 g^{1.757} \; ,
$$
\be
\label{3.304}
\bt_6^*(g) \simeq 1.857 g^{1.750} \qquad ( N = 1 , ~ g \ra \infty) \;   .
\ee

As is seen, all powers of the strong-coupling asymptotic behavior
$\bt(g)\propto g^\gm$ are close to each other, being between $\gm=1.7$ and $\gm=1.8$.
In literature, it is possible to find the estimates for the single-component field
theory with $N = 1$. For example, Kazakov \cite{Kazakov_268} gives the estimate of
$\gamma = 2$. Borel summation, with conformal mapping \cite{Chetyrkin_269} gives
$\gamma = 1.9$. The use of optimized perturbation theory \cite{Sissakian_270} results
in $\gamma = 1.5$. Borel-Leroy summation \cite{Suslov_264,Suslov_272,Suslov_275} leads
to $\gamma = 1$. So, our result is between that of optimized perturbation theory
\cite{Sissakian_270} and that of Borel summation \cite{Kazakov_268,Chetyrkin_269}.

\subsection{Other Applications}

Self-similar approximation theory has been applied to hundreds of various problems, 
as can be inferred from the literature cited in this review. In addition to the cited 
publications, it is possible to mention here several more applications. For example, 
self-similar approximants were used for describing systems with Bose-Einstein condensate 
\cite{Yukalov_277,Yukalov_278,Yukalov_279,Yukalov_280,Yukalov_281}, barrier crossing 
processes \cite{Drozdov_282,Drozdov_283}, stochastic processes \cite{Chaturverdi_300}, 
seismological phenomena \cite{Gluzman_284,Gluzman_285,Sornette_286}, and time series 
\cite{Andersen_287,Gluzman_288}. Self-similar approximants have been widely used 
for a large class of problems in the theory of structured media 
\cite{Gluzman_324,Gluzman_289}.

\section{Basics of Renormalization Group}

Throughout the previous chapters, it has been mentioned several times that the 
ideas of optimized perturbation theory and especially of self-similar approximation 
theory share many properties with the method of renormalization group. Here we recall 
the basics of the latter approach in order to better explain the similarities in the 
foundations of the different theories.

\subsection{Renormalization-Group Classes}

It is necessary to distinguish two main classes among the available variants 
of renormalization group \cite{Shirkov_291}. One class incorporates the methods 
where the introduction of a renormalization group is based on the existence of 
some properties of internal symmetry of a studied physical system formulated 
in the language of natural variables of this system. Here, as an example, one 
may point to the derivation of the quantum-field renormalization group, as is 
described in the book \cite{Bogolubov_292}. In the models of condensed matter 
physics, one also sometimes tries to discover internal symmetries allowing one, 
at least approximately, involving several simplifications, to formulate 
renormalization group equations 
\cite{Papp_293,Papp_294,Hanckowiak_295,Altenberger_296}.

The other class derives renormalization group equations from the construction of 
a set of models differing from each other by the scale of some variable. It is 
this class of renormalization group that is used in considering critical phenomena
\cite{Wilson_301,Hu_302,Ma_303,Yukalov_304}. This class, actually, corresponds to 
semigroups, since the inverse operation is not defined.

Thus, the first class reflects internal symmetries of the considered systems and 
is formed by groups. While the second class describes scaling transformations and 
represents semigroups. Below we briefly sketch the main ideas of both these classes.

\subsection{Critical Phenomena}

Under critical phenomena one understands the peculiarities of a system behavior in 
the vicinity of a continuous phase transition, that is, near a second order phase 
transition \cite{Yukalov_221,Fisher_305,Stanley_306}. At the critical point, two 
thermodynamic phases become indistinguishable. The point, where several phases are 
indistinguishable, is termed multicritical \cite{Lawrie_307}. A multicritical point 
on a critical line usually separates second- and first-order phase transitions. The 
critical behavior near a multicritical point is more complicated than in the vicinity 
of a critical point. Below we keep in mind critical points.

The vicinity of a system to a critical temperature $T_c$ is described by the 
relative temperature deviation
\be
\label{4.1}
 \tau \equiv \frac{T-T_c}{T_c} \;  .
\ee
The asymptotic behavior of a statistical system near a critical point is 
characterized by critical exponents. Thus the asymptotic behavior of specific heat 
is
\begin{eqnarray}
\label{4.2}
C_V \propto \left\{ \begin{array}{ll}
(-\tau)^{\al'}\; , ~ & ~ T < T_c \\
\tau^\al \; , ~ & ~ T > T_c
\end{array} \right. .
\end{eqnarray}
The order parameter, such as magnetization, behaves as
\begin{eqnarray}
\label{4.3}
M \propto \left\{ \begin{array}{ll}
(-\tau)^{\bt'}\; , ~ & ~ T < T_c \\
\tau^\bt \; , ~ & ~ T > T_c
\end{array} \right. .
\end{eqnarray}
For the dependence of a magnetic field $B$ on magnetization, at the critical 
temperature, one has
\be
\label{4.4}
 B \propto M^\dlt \qquad ( T = T_c ) \;  .
\ee
Susceptibility diverges as
\begin{eqnarray}
\label{4.5}
\chi \propto \left\{ \begin{array}{ll}
(-\tau)^{-\gm'}\; , ~ & ~ T < T_c \\
\tau^{-\gm} \; , ~ & ~ T > T_c
\end{array} \right. .
\end{eqnarray}
Correlation length also diverges as
\begin{eqnarray}
\label{4.6}
\xi \propto \left\{ \begin{array}{ll}
(-\tau)^{\nu'}\; , ~ & ~ T < T_c \\
\tau^{-\nu} \; , ~ & ~ T > T_c
\end{array} \right. .
\end{eqnarray}
The spatial dependence of correlation function at the critical temperature,
\be
\label{4.7}
  C(r) \propto r^{-(d-2+\eta)} \qquad ( T = T_c ) \; ,
\ee
where $d$ is the space dimensionality, is characterized by the exponent $\eta$.
The derivative of the Gell-Mann-Low function at a fixed-point coupling defines 
the exponent
\be
\label{4.8}
\om \equiv \lim_{g\ra g^*}\; \frac{d\bt(g)}{dg}
\ee
specifying the correction to the critical scaling behavior.

Assuming that the critical exponents are the same above as well as below the 
critical temperature, one gets the symmetry relations
\be
\label{4.9}
 \al = \al' \; , \qquad \bt = \bt' \; , \qquad \gm = \gm' \; , 
\qquad \nu = \nu' \;  .
\ee
And one also assumes the validity of the scaling relations
$$
\al + 2\bt + \gm = 2 \; , \qquad \al + \bt(1 + \dlt) = 2 \; , \qquad
\gm + \bt( 1 - \dlt) = 0 \; ,
$$
\be
\label{4.10}
\gm + \nu( \eta - 2) = 0 \; , \qquad   \al + \nu d = 2 
\ee
motivated by thermodynamic equalities.

In the case of a multicritical point, the symmetry relations do not hold. 
Approaching a multicritical point from the side of a second-order transition, it 
is possible to introduce the critical exponents as above. But at the point itself, 
the exponents change by a jump.

\subsection{Universality Classes}

When all critical exponents for two systems coincide, one tells that these systems 
are equivalent in the sense of their critical behavior. And the collection of the 
critical exponents is termed the set of universal quantities.

Historically, the assumption on the universality of critical behavior has been based 
on the results of mean-field theory, where all critical exponents are the same for 
all physical systems \cite{Brout_308}. A particular case of mean-field theory is the 
Landau method \cite{Landau_309} of expanding the free energy of an $N$-particle 
system
\be
\label{4.11}
f \equiv \frac{F}{N} = -\; \frac{T}{N} \; \ln {\rm Tr} e^{-\hat H/T}
\ee
in powers of an order parameter, say magnetization,
\be
\label{4.12}
 f \simeq f_0 + f_2 M^2 + f_4 M^4 \qquad ( M \ll 1 \; , ~ T < T_c ) \; .
\ee
If $(f_0 - f)/f_2 \propto \tau$, then
$$
 M \propto (-\tau)^\bt \qquad \left( \bt = \frac{1}{2} \right) \; ,
$$
irrespectively of the system nature.

The Landau expansion is applicable in the temperature region, where the order 
parameter is already small, while the fluctuations, increasing in the vicinity 
of the critical temperature, are not yet too large, so that the Ginzburg criterion 
be valid:
$$
Gi \propto | \tau|^{(d-4)/2} \ll 1 \;   .
$$
This criterion shows that in the space of dimensionality $d>4$ critical exponents 
can be defined by mean-field theory. While for dimensionality $d<4$ the Landau 
expansion is not appropriate. The dimensionality $d=4$ is marginal, requiring a 
separate investigation.

Experiments have demonstrated that critical exponents of the majority of systems 
are not of the mean-field type, being quite different for different systems. At 
the same time, there are many systems demonstrating very similar critical behavior, 
with the critical exponents being practically the same, within the accuracy of 
experiments. Thus the notion of universality classes has arisen.

{\it A universality class is a family of systems, equivalent in the sense of 
their critical behavior, hence possessing the same set of critical exponents}.

Among the characteristics of physical systems there are the main that define 
a class of universality of a system. The systems having at least one of such 
characteristics different, pertain to different universality classes. These main 
characteristics are:

\begin{enumerate}[label=(\Roman{*})]
\item
The system dimensionality.

\item
The number of the order parameter components.

\item
The order parameter symmetry.

\item
The range of particle interactions.

\item
The interaction potential anisotropy.

\end{enumerate}

Particle interactions can be of the following types.

\vskip 2mm

{\it Long-range potentials}. For a discrete system, such potentials
$$
 J_{ij} \equiv J(\br_{ij}) \qquad ( \br_{ij} \equiv \br_i - \br_j ) \;  ,
$$
in thermodynamic limit, satisfy the limiting property
\be
\label{4.13}
 \lim_{N\ra\infty} J_{ij} = 0 \; , \qquad 
\lim_{N\ra\infty} \; \frac{1}{N} \sum_{i\neq j}^N J_{ij} = const \neq 0 \;  .
\ee
The equivalent property for a continuous system is
\be
\label{4.14}
 \lim_{V\ra\infty} \Phi(\br) = 0 \; , \qquad 
\lim_{V\ra\infty} \; \int_V \Phi(\br) \; d\br = const \neq 0 \; .
\ee
The models with such long-range potentials allow for asymptotically exact 
solutions in thermodynamic limit \cite{Bogolubov_310} and enjoy the critical 
exponents of mean-field type.

\vskip 2mm

{\it Short-range potentials}. These are the potentials with a finite interaction 
radius. For instance, the exponential potentials
\be
\label{4.15}
 J_{ij} \propto \exp\left( - \; \frac{|\br_{ij}|}{b} \right) \; ,
\ee
or a sufficiently quickly decaying power-law potentials
\be
\label{4.16}
 J_{ij} \propto  \frac{1}{|\br_{ij}|^{d+n}} \qquad ( n > 2 ) \; .
\ee
Short-range potentials can be well classified into different universality 
classes.

\vskip 2mm

{\it Mid-range potentials}. These potentials are intermediate between long-range 
and short-range potentials, e.g.,
\be
\label{4.17}
 J_{ij} \propto  \frac{1}{|\br_{ij}|^{d+n}} \qquad ( 0 < n < 2 ) \;  .
\ee
Such potentials do not allow for classifying systems into universality classes, 
since critical exponents depend on the power $n$.

\vskip 2mm

{\it Anisotropic potentials}. This is a type of potentials depending on the 
angles and requiring a special consideration. Thus the dipolar potential
\be
\label{4.18}
 J_{ij} \propto  \frac{1}{|\br_{ij}|^d} \; \left( \dlt_{\al\bt} -\;
\frac{r_{ij}^\al r_{ij}^\bt}{|\br_{ij}|^2} \right) \; ,
\ee
although it looks like a long-range potential, but has the critical exponents 
coinciding, up to logarithmic corrections, with those of short-range systems. 
This is because of the property
$$
\frac{1}{N} \sum_{i\neq j} J_{ij} = 0 \;  ,
$$
valid for sufficiently large systems. Dipolar potentials also lead to some 
complications in describing the systems with such forces. Thus the Fourier 
transform of a dipolar potential is ill-defined, so that to overcome this problem 
one needs to take into account screening effects
\cite{Yukalov_311,Yukalov_312,Yukalov_313}.

\vskip 2mm

The reason why the universality classes do exist is as follows. In the vicinity 
of critical points, large-scale fluctuations become predominant and the correlation 
length tends to infinity. This makes small-scale peculiarities less important, while 
the system behavior is governed by long-range fluctuations. But let us recall that 
the concept of universality classes is not applicable to mid-range interactions and 
to some anisotropic models \cite{Baxter_314}.

In practice, the universality can be employed by describing critical phenomena 
resorting to the simplest system in the given universality class.

\subsection{Kadanoff Transformations}

Let a statistical system be characterized by a Hamiltonian $\hat{H}$ being a 
functional of a set of variables $s$. It is convenient to define the dimensionless 
Hamiltonian
\be
\label{4.19}
 H[ s ] \equiv \frac{\hat H}{T} \;  ,
\ee
where $T$ is temperature. Thermodynamic characteristics are described by the 
free energy
\be
\label{4.20}
 f = -\; \frac{T}{N} \; \ln Z \;  ,
\ee
in which $Z$ is the partition function
\be
\label{4.21}
Z = {\rm Tr}_s \exp ( - H[s] ) \;   .
\ee

The Kadanoff idea is to define a set of transformations moving from small to 
large scales and allowing for a step-by-step simplification of the Hamiltonian 
\cite{Kadanoff_315}. At the first step, one splits the set of variables $s$ into 
two subsets, $s_1$ and $s_1'$, and integrates out (or sums out) the variables 
$s_1'$ according to the rule
\be
\label{4.22}
\exp ( - H_1[s_1] ) \equiv {\rm Tr}_{s_1'} \exp ( - H[s_1,s_1'] ) 
\ee
defining the effective Hamiltonian
\be
\label{4.23}
 H_1[s_1] = - \ln {\rm Tr}_{s_1'} \exp ( - H[s_1,s_1'] ) \; .
\ee
Hence the partition function $Z$ becomes
\be
\label{4.24}
 Z = {\rm Tr}_{s_1} \exp ( - H_1[s_1] ) \;  .
\ee
Splitting again the set of variables $s_1$ into two subsets, $s_2$ and $s_2'$, 
and summing out the variables $s_2'$,
\be
\label{4.25}
 \exp ( - H_2[s_2] ) \equiv {\rm Tr}_{s_2'} \exp ( - H_1[s_2,s_2'] ) \; ,
\ee
one introduces the effective Hamiltonian
\be
\label{4.26}
 H_2[s_2] = - \ln {\rm Tr}_{s_2'} \exp ( - H_1[s_2,s_2'] ) \;  .
\ee
This yields the partition function
\be
\label{4.27}
 Z = {\rm Tr}_{s_2} \exp ( - H_2[s_2] ) \;  .
\ee
Repeating this procedure $n$ times results in the recurrence relation
\be
\label{4.28}
H_n[s_n] = - \ln {\rm Tr}_{s_n'} \exp ( - H_{n-1}[s_n,s_n'] ) 
\ee
and in the partition function
\be
\label{4.29}
  Z = {\rm Tr}_{s_n} \exp ( - H_n[s_n] ) \;  ,
\ee
where $H_0[s_0] \equiv H[s]$.

Introducing the renormalization transformation
\be
\label{4.30}
\hat R(s_n') H_{n-1}[s_n,s_n'] \equiv - 
\ln {\rm Tr}_{s_n'} \exp ( - H_{n-1}[s_n,s_n'] ) 
\ee
makes it straightforward to rewrite the recurrence relation (\ref{4.28}) as
\be
\label{4.31}
 H_n[s_n] = \hat R(s_n') H_{n-1}[s_n,s_n'] \qquad ( n \geq 1 ) \; .
\ee
Defining the double transformation
\be
\label{4.32}
\hat R(s_m', s_n') H_k \equiv  - \ln {\rm Tr}_{s_m'}{\rm Tr}_{s_n'} 
\exp ( - H_k )
\ee
and the identity transformation
\be
\label{4.33}
\hat R(0) \equiv 1
\ee
gives us the property
\be
\label{4.34}
 \hat R(s_m', s_n') = \hat R(s_m') \hat R(s_n') = 
\hat R(s_n') \hat R(s_m') \; .
\ee
This tells us that the family of the renormalization operators $\hat R(s_j')$ forms 
a commutative semigroup. This is a semigroup, since the inverse operation is not 
defined.

The free energy (\ref{4.20}) acquires the form
$$
f = \frac{T}{N} \; \hat R(s_n) H_n[s_n] = \frac{T}{N} \; \hat R(s_n) \hat R(s_n')
\hat R(s_{n-1}') \ldots \hat R(s_1') H[s] =
$$
\be
\label{4.35}
 = -\; \frac{T}{N}\; \ln {\rm Tr}_{s_n} \exp ( - H_n[s_n]) \;  .
\ee

The above scheme lies in the core of renormalization semigroups used in 
statistical physics. Practical realizations can be different, but the principal 
idea is as described above.

\subsection{Momentum Scaling}

A widely employed approach of introducing renormalization operations is based 
on different variants of scaling in the momentum space
\cite{Wilson_301,Ma_303,Polchinski_316,Boer_317,Heemskerk_318}. The typical 
procedure can be characterized as follows.

Let the Hamiltonian $H[\varphi]$ be a functional of a field depending on momentum,
\be
\label{4.36}
\vp = \vp(\bk) \qquad ( 0 \leq k \leq \Lbd \; , ~ k \equiv | \bk| ) \;  ,
\ee
with $\Lambda$ being an ultraviolet cutoff. The partition function, or generating 
functional, is given by the functional integral
\be
\label{4.37}
 Z = \int \exp ( - H[\vp] ) \; {\cal D}\vp \;  .
\ee

One separates the momenta into two regions, one is the region of low energy, where
\be
\label{4.38}
 0 \leq k \leq \frac{\Lbd}{b} \; ,
\ee
and the other is the region of high energy, where
\be
\label{4.39}
 \frac{\Lbd}{b} < k \leq \Lbd \;  ,
\ee
with $b \geq 1$ being an arbitrary real number. Then the field splits into two 
parts:
\begin{eqnarray}
\label{4.40}
\vp(k) = \left\{ \begin{array}{ll}
\vp_b(\bk) \; , ~ & ~ k \leq \Lbd/b \\
\vp_b'(\bk) \; , ~ & ~ k > \Lbd/b 
\end{array} \right. .
\end{eqnarray}
Hence partition function (\ref{4.37}) takes the form
\be
\label{4.41}
 Z = \int \exp( - H[\vp_b,\vp_b'] )\; 
{\cal D}\vp_b  {\cal D}\vp_b' \; .
\ee

One defines the effective Hamiltonian $H_b[\varphi_b]$ by integrating out 
high-energy momenta in the equality
\be
\label{4.42}
\exp (- H_b[\vp_b] ) \equiv \int \exp( - H[\vp_b,\vp_b'] )\; {\cal D}\vp_b' \; ,
\ee
which gives
\be
\label{4.43}
 H_b[\vp_b]  = - \ln \int \exp( - H[\vp_b,\vp_b'] )\; {\cal D}\vp_b' \; .
\ee
So that the partition function reads as
\be
\label{4.44}
  Z = \int \exp( - H_b[\vp_b] )\; {\cal D}\vp_b \; .
\ee

The domain of the momenta in the field $\varphi_b(k)$ can be extended to the 
whole momentum space by the change
\be
\label{4.45}
\bk \ra b\bk \; , \qquad \vp_b(\bk) \ra \zeta \vp(\bk) \;   ,
\ee
in which $\zeta$ is a scale factor to be specified later. Then partition function 
(\ref{4.44}) transforms into
\be
\label{4.46}
  Z = \int \exp( - H_b[\vp] ) J(\zeta)\; {\cal D}\vp \;  ,
\ee
where $J(\zeta)$ is the Jacobian of transformation (\ref{4.45}) and the effective 
Hamiltonian is
\be
\label{4.47}
 H_b[\vp]  \equiv  -\ln \int \exp( - H[\zeta\vp,\vp_b'] )\; {\cal D}\vp_b' \; .
\ee

The procedure of obtaining the effective Hamiltonian (\ref{4.47}) can be 
symbolically denoted as the action of a renormalization transformation $\hat{R}(b)$,
\be
\label{4.48}
 \hat R(b) H[\vp] = H_b[\vp ] \;  .
\ee
The unitary element is defined by the equality
\be
\label{4.49}
 H_1 [\vp ] = H[\vp ] \; , \qquad \hat R(1) = 1 \; .
\ee
The transformation $\hat{R}(b)$ enjoys the property
\be
\label{4.50}
 \hat R(b_1b_2) =  \hat R(b_1) \hat R(b_2) = \hat R(b_2) \hat R(b_1) \; .
\ee
Thus the family of transformations $\hat{R}(b)$ forms a commutative renormalization 
semigroup.

\subsection{Fixed Points}

The renormalization transformations $\hat{R}(b)$ generate a continuous sequence 
of effective Hamiltonians (\ref{4.48}) starting from the initial Hamiltonian 
$H[\vp]$. One says that the Hamiltonian moves along a trajectory parameterized 
by the scaling factor $b\in [1,\infty)$. However, in that motion the form of the 
Hamiltonian varies. Thus, if the initial Hamiltonian contains a set of coupling 
parameters $g$, in the effective renormalized Hamiltonian, not merely the coupling 
parameters become renormalized, but also there appear new terms with additional
coupling parameters. This can be formulated as the motion in the space of 
coupling parameters:
\be
\label{4.51}
g \ra \overline g(b,g) = \{ \overline g_1(b,g), \overline g_2(b,g) ,
\ldots \} 
\ee
with the initial condition
\be
\label{4.52}
 \overline g(1,g) = g \; .
\ee
During this motion, one confronts with the problem of proliferation of coupling 
parameters. Even if initially the Hamiltonian $H[\varphi]$ had a finite number of 
such parameters, with increasing $b$, their number tends to infinity.

But the hope is that in the vicinity of a critical point the Hamiltonian becomes 
invariant with respect to the renormalization transformations, since at a critical 
point the correlation length diverges, making irrelevant the fine structure of the 
Hamiltonian. The Hamiltonian invariance implies the existence of a fixed point 
$H^*[\varphi]$, where
\be
\label{4.53}
 \hat R(b) H^*[\vp] = H^*[\vp ] \; .
\ee
In the space of coupling parameters, this is equivalent to the equation
\be
\label{4.54}
 \overline g(b,g^*) = g^* \; .
\ee

As far as at the fixed point the Hamiltonian is invariant, conversely, assuming 
the Hamiltonian invariance, one could approximately find the fixed point.

\subsection{Wilson Method}

To be more specific, let us illustrate how the Wilson method works by considering 
the scalar $\vp^4$ field theory. We shall follow the scheme of Refs. 
\cite{Wilson_301,Ma_303,Yukalov_319}.

The Hamiltonian of the scalar $\vp^4$ field theory can be represented in the form
$$
H [\vp] = \frac{1}{2} \int_0^\Lbd \left( m^2 + k^2 \right) | \vp(\bk) |^2 \; d\bk
+
$$
\be
\label{4.55}
+ g \int_0^\Lbd \vp(\bk_1)\vp(\bk_2)\vp(\bk_3)\vp(\bk_4) 
\dlt_d(\bk_1+\bk_2+\bk_3+\bk_4) \; d\bk_1 d\bk_2 d\bk_3 d\bk_4 \; ,
\ee
where the momentum is a vector in a $d$-dimensional space,
$$
 \bk = \{ k_\al : ~ \al = 1, 2, \ldots, d\} \; ,
$$
and the notations are used:
$$
\int_0^\Lbd d\bk \equiv \left( \frac{a}{2\pi}\right)^d 
\prod_{\al=1}^d \int_0^\Lbd dk_\al \; , 
\qquad 
\dlt_d(\bk) \equiv \left( \frac{2\pi}{a}\right)^d 
\prod_{\al=1}^d \dlt(k_\al) \; .
$$
Here $1/a^d \sim N/V$ plays the role of an effective density.

Integrating out the high-energy momenta, as is explained in Sec. 4.5, the integrals 
are calculated involving perturbation theory with respect to the coupling parameter 
$g$. And, assuming that we are close to a fixed point, hence the Hamiltonian has to 
be invariant under the renormalization transformation, the terms spoiling the 
Hamiltonian structure are neglected. Then the term $m^2+k^2$ transforms as
\be
\label{4.56}
 m^2 + k^2 \ra m^2(b) + f(b) k^2 \;  ,
\ee
where
\be
\label{4.57}
 m^2(b) = \frac{\zeta^2}{b^d} \; \left[ m^2 + 12g \int_{\Lbd/b}^\Lbd
\frac{d\bk}{m^2 + k^2} \right] + {\cal O}(g^2) \; , \qquad
f(b) = \frac{\zeta^2}{b^{d+2}} + {\cal O}(g^2) \;  .
\ee
The coupling parameter experiences the transformation
\be
\label{4.58}
  g \ra g(b) \; ,
\ee
in which
\be
\label{4.59}
  g(b) = \frac{\zeta^4}{b^{3d}} \; \left[ g - 36g^2 \int_{\Lbd/b}^\Lbd
\frac{d\bk}{(m^2 + k^2)^2} \right] + {\cal O}(g^3) \; .
\ee

The scaling parameter, introduced in scaling (\ref{4.45}), is arbitrary and can 
be chosen such that the factor at $k^2$ be unchanged, that is, setting it to one, 
which means
\be
\label{4.60}
 f(b) = 1 \; , \qquad \zeta^2 = b^{d+2} \;  .
\ee
Thus we get the recurrence relations
$$
m^2(b) = b^2\; \left[ m^2 + 12g \int_{\Lbd/b}^\Lbd
\frac{d\bk}{m^2 + k^2} \right] + {\cal O}(g^2) \; , 
$$
\be
\label{4.61}
g(b) = b^{4-d} g \left[ 1 - 36g \int_{\Lbd/b}^\Lbd
\frac{d\bk}{(m^2 + k^2)^2} \right] + {\cal O}(g^2) \;   .
\ee

The fixed point in the parameter space is defined by the equations
$$
(m^*)^2 = b^2\; \left[ (m^*)^2 + 12g^* \int_{\Lbd/b}^\Lbd
\frac{d\bk}{(m^*)^2 + k^2} \right] \; , 
$$
\be
\label{4.62}
g^* = b^{4-d} g^* \left[ 1 - 36g^* \int_{\Lbd/b}^\Lbd
\frac{d\bk}{((m^*)^2 + k^2)^2} \right] \;   .
\ee
These equations have a trivial fixed point, where both $m^*$ and $g^*$ are zero. 
But we wish to find a nontrivial fixed point. By noticing that $(m^*)^2$ is of 
order $g^*$, the integrals in Eqs (\ref{4.62}) can be simplified:
$$
 \int_{\Lbd/b}^\Lbd \frac{d\bk}{(m^*)^2 + k^2} \simeq \int_{\Lbd/b}^\Lbd
\frac{d\bk}{k^2} = \frac{C_d}{d-2} \left[ \Lbd^{d-2}  -
\left( \frac{\Lbd}{b}\right)^{d-2}\right] \; ,
$$
$$
 \int_{\Lbd/b}^\Lbd \frac{d\bk}{((m^*)^2 + k^2)^2} \simeq \int_{\Lbd/b}^\Lbd
\frac{d\bk}{k^4} = \frac{C_d}{d-4} \left[ \Lbd^{d-4} -
\left(\frac{\Lbd}{b}\right)^{d-4}\right] \; ,
$$
where
\be
\label{4.63}
  C_d = \left( \frac{a}{2\pi}\right)^d \int d^d \Om =
\frac{a^d}{2^{d-1}\pi^{d/2}\Gm(d/2)} \; .
\ee
Thus we come to the fixed point characterized by the approximate values
$$
(m^*)^2 = -\; \frac{12g^*b^2C_d}{(b^2-1)(d-2)} \; \left[ \Lbd^{d-2} -
\left( \frac{\Lbd}{b}\right)^{d-2} \right] \; , 
$$
\be
\label{4.64} 
g^* =  \frac{(1-b^{d-4})(d-4)}{36C_d} \; \left[ \Lbd^{d-4} -
\left( \frac{\Lbd}{b}\right)^{d-4} \right]^{-1} \; .
\ee

This answer, however, is not yet satisfactory, since it contains an arbitrary 
scaling factor $b$ that makes the obtained values unphysical.

\subsection{Fractional Dimension}

We may notice that the dependence on $b$ disappears for dimension $d=4$, although 
then the fixed point reduces to trivial zero. But it would be possible to get rid 
of $b$ dependence supposing that the deviation from the dimension $d=4$ is small, 
however not zero. This suggests to introduce the variable
\be
\label{4.65}
\ep \equiv 4 - d
\ee
assumed to be small. Then it is admissible to expand the quantities containing 
$\ep$ in powers of $\varepsilon$ employing the expansion
$$
x^\ep = \exp(\ep \ln x) \simeq 1 + \ep \ln x \; .
$$
Such an $\varepsilon$- expansion for the fixed point (\ref{4.64}), to first order in
$\varepsilon$, gives
\be
\label{4.66}
 (m^*)^2 = -\; \frac{\Lbd^2}{6} \; \ep \; , \qquad 
g^* = \frac{\Lbd}{36C_d}\; \ep \;  .
\ee
At the end, one has to remember that the real space is not $d = 4$, but it is 
$d=3$, or in the limiting cases it can be $d=1$ or $d=2$. Respectively, one should 
set $\varepsilon = 1$ or even $\varepsilon = 2$ or $\varepsilon = 3$. For instance, 
quantity (\ref{4.63}) becomes
$$
C_1 = \frac{a}{\pi} \qquad ( d = 1 \; , ~ \ep = 3 ) \; ,
$$
$$
C_2 = \frac{a^2}{2\pi} \qquad ( d = 2 \; , ~ \ep = 2 ) \; ,
$$
$$
C_3 = \frac{a^3}{2\pi^2} \qquad ( d = 3 \; , ~ \ep = 1 )  \; .
$$
And for the fixed point (\ref{4.66}), we find
\begin{eqnarray}
\label{4.67}
\left( \frac{m^*}{\Lbd}\right)^2 = \left\{ \begin{array}{ll}
-0.5 \; , ~ & ~ d = 1 \\
-0.333 \; , ~ & ~ d = 2 \\
-0.167 \; , ~ & ~ d = 3 \end{array} \right. \; , 
\qquad
g^*\;  \frac{a^d}{\Lbd} = \left\{ \begin{array}{ll}
0.262 \; , ~ & ~ d = 1 \\
0.349 \; , ~ & ~ d = 2 \\
0.548 \; , ~ & ~ d = 3 \end{array} \right. \;  .
\end{eqnarray}

The described $\varepsilon$ - expansion results in asymptotic series. Substituting 
there finite values of $\ep$, strictly speaking, is not correct. Before such a 
substitution, the series have to be extrapolated to finite values of $\ep$, e.g., 
as is done in Sec. 3.25.

\subsection{Differential Formulation}

It is possible to derive differential equations for the renormalized parameters
\cite{Ma_303,Chang_320} by considering infinitesimally small variations of the 
scaling factor
\be
\label{4.68}
 b = 1 + \dlt x \qquad ( \dlt x \ra 0 ) \;  .
\ee
Then the integrals in the recurrence relations (\ref{4.61}) can be represented as
$$
 \int_{\Lbd/b}^\Lbd \frac{d\bk}{m^2 + k^2} \simeq 
\frac{C_d\Lbd^d}{m^2+\Lbd^2}\; \dlt x \; , 
\qquad
 \int_{\Lbd/b}^\Lbd \frac{d\bk}{(m^2 + k^2)^2} \simeq 
\frac{C_d\Lbd^d}{(m^2+\Lbd^2)^2}\; \dlt x \; .
$$
Hence for the quantities
\be
\label{4.69}
m^2(b) = m^2(1 + \dlt x) \; , \qquad g(b) = g(1 + \dlt x)
\ee
we obtain the relations
$$
\frac{m^2(1+\dlt x)-m^2}{\dlt x} = 2m^2 + 
\frac{12C_d\Lbd^d}{m^2+\Lbd^2}\; g \; ,
$$
\be
\label{4.70}
\frac{g(1+\dlt x)-g}{\dlt x} = ( 4 - d )g -\; 
\frac{36C_d\Lbd^d}{(m^2+\Lbd^2)^2}\; g^2 \;   .
\ee
Thus we come to the differential equations
$$
\frac{dm^2(x)}{dx} = 2m^2(x) + 
\frac{12C_d\Lbd^d}{m^2(x)+\Lbd^2}\; g(x) \; ,
$$
\be
\label{4.71}
 \frac{dg(x)}{dx} = ( 4 - d )g(x) -\; 
\frac{36C_d\Lbd^d}{(m^2(x)+\Lbd^2)^2}\; g^2(x) \; ,
\ee
with the initial conditions
\be
\label{4.72}
  m(0) = m \; , \qquad g(0) = g \; .
\ee

The fixed points of the dynamical system are defined by the zero derivatives
\be
\label{4.73}
 \left. \frac{dm^2(x)}{dx} \right \vert_{m^*,g^*} = 0 \; , 
\qquad  
\left. \frac{g(x)}{dx} \right \vert_{m^*,g^*} = 0 \; ,
\ee
which leads to the equations
$$
6g^* C_d \Lbd^d + (m^*)^2\left( (m^*)^2 + \Lbd^2 \right) = 0 \; ,
$$
\be
\label{4.74}
 36g^* C_d \Lbd^d - (4 - d)\left( (m^*)^2 + \Lbd^2 \right)^2 = 0 \;  .
\ee
Taking into account that the ultraviolet cutoff is sufficiently large, such that
$\Lambda \gg |m|$, we find
\be
\label{4.75}
 (m^*)^2 = -\; \frac{4-d}{6}\; \Lbd^2 \; , \qquad
 g^* =  \frac{4-d}{36C_d}\; \Lbd^{4-d} \;  .
\ee

Let us stress that in this way, we have not involved the $\varepsilon$-expansion. 
Nevertheless, if we set $4-d=\ep$, we return to the fixed point (\ref{4.66}). That 
is, the differential formulation is equivalent to the $\varepsilon$-expansion, 
resulting in the same form of fixed points.

\subsection{Decimation Procedure}

Renormalization group in real space can be accomplished employing the Kadanoff 
transformations described in Sec. 4.4. Summing out a part of variables leads to 
the coarsening of the real-space scale \cite{Kadanoff_315,Efrati_321}. Such a 
procedure is often called {\it decimation}. This term is borrowed from the language 
of the ancient Rome, where the decimation meant the killing of one in every ten of 
soldiers from a mutinous or retreated Roman legion.

Let us consider the two-dimensional Ising model with the Hamiltonian
\be
\label{4.76}
  \hat H = N \; \frac{U}{2} \; - \; 
\frac{1}{2}\sum_{\lgl ij\rgl} J \sgm_i\sgm_j \; ,
\ee
where the spin variables take the values $\sigma_j = \pm 1$ and the summation is 
over the nearest neighbors. We keep here a constant term to show that it also needs 
renormalization \cite{Yukalov_319}. The set of all spin variables is denoted as
\be
\label{4.77}
 s = \{ \sgm_j = \pm 1\; : ~ j = 1,2,\ldots, N\} \;  .
\ee
Introducing the dimensionless parameters
\be
\label{4.78}
u \equiv \frac{U}{2T} \; , \qquad g \equiv \frac{J}{2T} \; ,
\ee
gives us the dimensionless Hamiltonian (\ref{4.19}) in the form
\be
\label{4.79}
 H[s] = N u - g \sum_{\lgl ij\rgl}  \sgm_i\sgm_j \;  .
\ee

Considering a square lattice with the side $a$, we sum out the nearest neighbors, 
which coarsens the lattice side into $\sqrt{2} a$. The effective Hamiltonian 
(\ref{4.23}) becomes
\be
\label{4.80}
H_1[s_1] = N u_2 - g_2 \sum_{\lgl ij\rgl}  \sgm_i\sgm_j -
g_2' \sum_{(ij)}  \sgm_i\sgm_j - 
g_2'' \sum_{[ijkl]}  \sgm_i\sgm_j \sgm_k\sgm_l\; ,
\ee
where the initial parameters $u$ and $g$ are renormalized into
\be
\label{4.81}
u_2 = 2u - \ln 2 -\; \frac{1}{8}\; [ \ln\cosh(4g) + 4\ln\cosh(2g) ] \; , 
\qquad
g_2 = \frac{1}{4} \; \ln\cosh(4g) \; ,
\ee
and there arise two additional terms, with the couplings
\be
\label{4.82}
 g_2' = \frac{1}{2}\; g_2 \; , \qquad 
g_2'' =  \frac{1}{8}\; [ \ln\cosh(4g) - 4\ln\cosh(2g) ] \; .
\ee
In these additional terms, the notation $(ij)$ implies the next to nearest neighbors, 
and $[ijkl]$ denotes a placket, that is a square of sites containing no sites inside.

As is seen, the number of the coupling parameters has been doubled,
$$
 \{ u, g\} \ra \{ u_2,g_2,g_2',g_2'' \} \;  .
$$
This is an example of the coupling-parameter proliferation, discussed in Sec. 4.4.

Assuming that near a fixed point the Hamiltonian is invariant under the 
renormalization transformation, one neglects the appearance of the additional terms. 
It is straightforward to continue the procedure, as is explained in Sec. 4.4. If we 
limit ourselves by one transformation, we get the fixed point equations
\be
\label{4.83}
 u^* = \ln 2 + \frac{1}{2} \; [ g^* + \ln\cosh(2g^*) ] \; , \qquad
 g^* = \frac{1}{4}\; \ln\cosh(4g^*) \;  .
\ee
However, these equations have only a trivial solution $u^*=\ln 2$ and $g^*=0$ 
yielding incompatible values for the critical temperature. Thus the first equality 
gives $T_c = U/2 \ln 2$, while the second equality means that $T_c \ra \infty$. 
This signifies that just one step of the renormalization procedure is not sufficient 
for defining a nontrivial fixed point. But the procedure has to be continued.

\subsection{Exact Semigroup}

There exist rather rare cases, when the proliferation of parameters does not occur. 
In such exceptional cases, the renormalization procedure can be realized exactly. 
As an illustration, let us consider the one-dimensional Ising model with the 
Hamiltonian
\be
\label{4.84}
 H[s] = \sum_{j=1}^N ( u - g \sgm_j\sgm_{j+1} ) \;  ,
\ee
in which $\sigma_j = \pm 1$ and the periodic boundary condition
\be
\label{4.85}
\sgm_{N+1} = \sgm_1
\ee
is imposed.

Following the scheme of Sec. 4.4, we split the set of spin variables into two 
subsets, $s_1$ and $s_1'$, the first subset comprising the spins with odd indices, 
while the second, containing even spins,
$$
s_1 = \left\{ \sgm_{2j+1}: ~ j = 0,1,2,\ldots,\frac{N}{2}\right\} \; ,
\qquad
s_1' = \left\{ \sgm_{2j}: ~ j = 1,2,\ldots,\frac{N}{2}\right\} \; .
$$
Summing out the even spins at the first renormalization step, we pass from a chain 
with the interspin distance $a$ to that with the distance $2a$. For the effective 
Hamiltonian (\ref{4.23}), we get
\be
\label{4.86}
 H_1[s_1] = \sum_{j=0}^{N/2} ( u_2 - g_2 \sgm_{2j+1}\sgm_{2j+3} ) \;  ,
\ee
with the renormalized parameters
$$
u_2 = 2u - \ln 2 - 2\ln\cosh g + \ln\cosh g_2 \; ,
$$
\be
\label{4.87}
 g_2 = {\rm arctanh}\left( \tanh^2 g \right) \; .
\ee
To simplify the recurrence relations, it is useful to introduce the notations
$$
c \equiv e^{-u}\cosh g \; , \qquad t \equiv \tanh g \; ,
$$
\be
\label{4.88}
 c_n \equiv e^{-u_n}\cosh g_n \; , \qquad t_n \equiv \tanh g_n \; .
\ee
Then relations (\ref{4.87}) take the form
\be
\label{4.89}
 c_2 = 2u^2 \; , \qquad t_2 = t^2 \;  .
\ee

At the second renormalization step, we pass to the chain with the interspin 
distance $4a$. The related twice renormalized parameters read as
\be
\label{4.90}
 c_4 = 2c_2^2 = 2^3 c^4 \; , \qquad t_4 = t_2^2 = t^4 \;   .
\ee
After the third renormalization, we get a chain with the interspin distance $8a$, 
and with the parameters
\be
\label{4.91}
c_8 =  2^7 c^8 \; , \qquad t_8 =  t^8 \;   .
\ee

Accomplishing $k$ renormalizations results in the parameters
\be
\label{4.92}
 c_n =  2^{n-1} c^n \; , \qquad t_n =  t^n \qquad ( n = 2^k ) \; ,
\ee
which can be rewritten as
$$
u_n = nu - ( n - 1 )\ln 2 - n \ln \cosh g + \ln\cosh g_n \; ,
$$
\be
\label{4.93}
 g_n = {\rm arctanh} \left( \tanh^n g \right) \qquad ( n = 2^k ) \;  .
\ee
At all renormalization steps, no additional parameters arise.

Fixed points are given by the equations
\be
\label{4.94}
c^* =  2^{n+1} (c^*)^n \; , \qquad t^* =  (t^*)^n \; ,
\ee
having the solutions $c^* = 0, 1/2$ and $t^* = 0, 1$. The sole solution, yielding 
compatible values in terms of the parameters $u$ and $g$, is
\be
\label{4.95}
 u^* = \infty \; , \qquad g^* = \infty \;  .
\ee
This defines the critical point $T_c = 0$, which is the exact answer for the 
one-dimensional Ising model.

The $k$ times renormalized effective Hamiltonian is
\be
\label{4.96}
H_k[ s_k] = H_{1,n+1} + H_{n+1,2n+1} + H_{2n+1,3n+1} + \ldots
\ee
containing $N/n$ terms
$$
  H_{i,j} \equiv u_n - g_n \sgm_i \sgm_j \qquad ( n = 2^k ) \; .
$$
Increasing the number $k$ of renormalizations, we get the parameters
\be
\label{4.97}
  u_n \simeq n [ u - \ln(2\cosh g) ] \; , \qquad g_n \simeq 0 
\qquad ( n \ra\infty ) \;  .
\ee
Then the effective Hamiltonian (\ref{4.96}) tends to
\be
\label{4.98}
 H_k [ s_k ] \simeq u_n \qquad ( n = 2^k \ra \infty ) \; .
\ee

The free energy (\ref{4.35}) becomes
\be
\label{4.99}
  f \simeq \frac{T}{N} \; u_n \qquad ( n \ra \infty ) \; .
\ee
Since the maximal number of the described renormalizations is connected with the 
number of spins by the relations
$$
 N = 2^k \; , \qquad \max\;k = \frac{\ln N}{\ln 2} \;  ,
$$
we finally obtain the free energy
\be
\label{4.100}
 f \simeq T [ u - \ln(2\cosh g) ] \;  .
\ee
This result is asymptotically exact in thermodynamic limit.

Recall that the situation when there is no proliferation of parameters is quite 
exceptional. In general, the process of renormalization generates additional terms
with additional coupling parameters. To overcome this problem, it is possible to 
use the idea of optimized perturbation theory and to introduce control functions 
canceling the appearing additional terms \cite{Swendsen_322}.

\subsection{Field-Theory Group}

Statistical renormalization groups, considered above, are actually semigroups, 
since they do not include inverse transformations. They are based on some kind of 
scaling transformations resulting in the consecutive coarse graining of the system 
and constructing a sequence of effective models. Thus in the Wilson method, it is 
the scaling of momentum variables, while in the decimation procedure, it is the 
real-space coarsening of a lattice.

Field-theory renormalization groups are groups, containing inverse transformations. 
It is based on discovering the natural symmetries of the system. This approach can 
be used in field theory as well as in statistical physics. As an application of 
the latter kind, the field-theory approach is illustrated below for the Ising 
model \cite{Pasquale_323}.

The spin variables in the Ising model can be scaled by the transformation
$\sigma_j \ra \nu \sigma_j$, which results in the Hamiltonian
\be
\label{4.101}
  \hat H_\nu = -\; \frac{\nu^2}{2} \sum_{\lgl ij\rgl} J \sgm_i \sgm_j - 
\nu \sum_j B_j \sgm_j \; ,
\ee
where $B_j$ is an external magnetic field. With the dimensionless parameters
\be
\label{4.102}
 g \equiv \frac{J}{2T} \; , \qquad h_j \equiv \frac{B_j}{T} \;  ,
\ee
we have the dimensionless Hamiltonian
\be
\label{4.103}
H_\nu [s] = -\nu^2 g \sum_{\lgl ij\rgl} \sgm_i \sgm_j - \nu \sum_j h_j \sgm_j \; .
\ee
The partition function reads as
\be
\label{4.104}
 Z(\nu,g,h) = \sum_s \exp(-H_\nu[s] ) \;  ,
\ee
where $h$ is the set of the fields $h_j$.

We wish to find such a transformation of the parameters
\be
\label{4.105}
\nu \ra z_1 \nu \; , \qquad g \ra z_2 g \; , \qquad h_j \ra z_3 h_j
\ee
that would leave invariant the partition function,
\be
\label{4.106}
 Z(z_1\nu,z_2g,z_3h) = Z(\nu,g,h) \;  .
\ee
From Eq. (\ref{4.104}), it follows that the condition of invariance is satisfied 
if
$$
 z_1^2 z_2 = 1 \; , \qquad z_1 z_3 = 1 \;  .
$$
This tells us that of the three quantities $z_1$, $z_2$, and $z_3$ only one is 
independent. Denoting one of them, say $z_2 \equiv z$, we have
\be
\label{4.107}
 z_1 = z^{-1/2} \; , \qquad z_2 = z \; , \qquad z_3 = z^{1/2} \;  .
\ee
The property of invariance (\ref{4.106}) takes the form
\be
\label{4.108}
 Z(z^{-1/2}\nu,zg,z^{1/2}h) = Z(\nu,g,h) \;  .
\ee

Because of invariance (\ref{4.108}), the correlation function
\be
\label{4.109}
C_{ij}(\nu,g) \equiv \left. \frac{\prt^2\ln Z(\nu,g,h)}{\prt h_i \prt h_j} 
                     \right \vert_{h=0}
\ee
enjoys the property
\be
\label{4.110}
 C_{ij}(z^{-1/2}\nu,zg) = z^{-1} C_{ij}(\nu,g) \; .
\ee
The Fourier transform
\be
\label{4.111}
 C(\bk,\nu,g) \equiv 
\frac{1}{N} \sum_{ij} C_{ij}(\nu,g) e^{-i\bk\cdot\br_{ij}} \;  ,
\ee
in which $\br_{ij}\equiv \br_i-\br_j$, possesses the same property
\be
\label{4.112}
 C(\bk,z^{-1/2}\nu,zg) = z^{-1}C(\bk,\nu,g) \;  .
\ee

Keeping in mind a spatially isotropic situation, we can define the function
\be
\label{4.113}
K(k^2,\nu,g) \equiv g C(\bk,\nu,g)
\ee
that is invariant under the transformation
\be
\label{4.114}
 K(k^2,z^{-1/2}\nu,zg) = K(k^2,\nu,g) \;  .
\ee
The latter means that this function actually depends on two combinations of the 
variables,
\be
\label{4.115}
  K(k^2,\nu,g) = \overline K(k^2,\nu^2g) \;  .
\ee

In the vicinity of a critical point, where the correlation length diverges, it 
is possible \cite{Pasquale_323} to introduce the dimensionless function
\be
\label{4.116}
D\left( \frac{k^2}{\lbd^2}\; , g \right) \equiv 
\frac{\overline K(k^2,g)}{\overline K(\lbd^2,g)}
\ee
having the boundary condition
\be
\label{4.117}
 D(1,g) = 1 \;  .
\ee
Representing the correlation functions as series in powers of the coupling 
parameter $g$, it is possible \cite{Pasquale_323} to explicitly demonstrate that 
function (\ref{4.116}) is invariant under the transformation
\be
\label{4.118}
 \lbd \ra \mu \; , \qquad g \ra zg \; , \qquad D \ra zD \;  ,
\ee
satisfying the relation
\be
\label{4.119}
 D\left( \frac{k^2}{\lbd^2}\; , g \right) = 
z  D\left( \frac{k^2}{\mu^2}\; , zg \right)\;  ,
\ee
with the property
\be
\label{4.120}
D\left( \frac{\mu^2}{\lbd^2}\; , g \right) = z \;   .
\ee

From here, with the use of the dimensionless variables,
\be
\label{4.121}
 x \equiv \frac{k^2}{\lbd^2} \; , \qquad t \equiv \frac{\mu^2}{\lbd^2}\;  ,
\ee
we come to the renormalization group equation
\be
\label{4.122}
D(x,g) = D(t,g) D\left( \frac{x}{t}\; , gD(t,g)\right)
\ee
for which condition (\ref{4.117}) is valid.

\subsection{Invariant Charge}

The renormalization group equations of type (\ref{4.122}) are often met in 
quantum field theory. The other form of such equations can be given by introducing 
the invariant charge
\be
\label{4.123}
G(x,g) \equiv g D(x,g)
\ee
with the boundary condition
\be
\label{4.124}
 G(1,g) = g \;  .
\ee

Scaling the variable $x$ as $x\ra tx$, we get the renormalization group equation
\be
\label{4.125}
 gG(tx,g) = G(t,g) G(x,G(t,g) ) \;  .
\ee
Differentiating with respect to $x$ and setting $x\ra 1$ and $t\ra x$, we obtain 
the renormalization group equation in differential form
\be
\label{4.126}
  \frac{\prt G(x,g)}{\prt\ln x} = \bt ( G(x,g),g) \; ,
\ee
where the Gell-Mann-Low function is
\be
\label{4.127}
 \bt ( G(x,g),g) = \frac{G(x,g)}{g} \;  \left.
\left[ \frac{\prt}{\prt y} \; G(y,G(x,g) ) \right] \right \vert_{y=1} \;  .
\ee
Integrating this equation yields
\be
\label{4.128}
 \int_g^G \frac{df}{\bt(f,g)} = \ln x \;  .
\ee

Fixed points correspond to the condition
\be
\label{4.129}
G(x,g^*) = g^* \;   ,
\ee
hence to the zero of the Gell-Mann-Low function,
\be
\label{4.130}
 \bt(g^*,g^*) = 0 \;  .
\ee

Thus the field-theory group is a Lie group of continuous transformations, 
possessing the inverse transformation and allowing for the presentation in the 
form of a differential equation.

\section{Discussion}

The aim of this review has been twofold: First, it presents two approaches in 
approximation theory, developed by the author, Optimized Perturbation Theory and 
Self-Similar Approximation Theory. Secondly, it emphasizes that these approaches 
are based on the ideas having several similarities with renormalization group 
techniques.

The main idea of {\it Optimized Perturbation Theory} is to reorganize a divergent 
sequence into a convergent sequence by introducing control functions. Such a 
reorganization of the approximants for the sought quantity is analogous to the 
renormalization of these approximants. The optimization conditions for control 
functions follow from the Cauchy criterion, which for sequences plays the role of 
a fixed-point condition. The derivation of the optimization conditions from the 
Cauchy criterion makes the theory logical and self-consistent. This makes the theory 
principally different from ad hoc prescriptions used by some authors. The developed
theory allows for the derivation of different optimization conditions, justifying 
them on the basis of the Cauchy criterion. Since the latter plays the role of a 
fixed-point condition, different particular forms of optimization conditions, such 
as the minimal-difference and minimal-derivative conditions, have the same meaning 
of fastest-convergence conditions. Control functions can be incorporated either 
into an initial approximation or into the perturbative series with respect to an 
asymptotically small variable by means of a change of the variable. The theory makes 
it possible to obtain good approximation accuracy with just a few perturbative
terms.

In {\it Self-Similar-Approximation Theory}, the passage between subsequent 
approximations is treated as a motion with respect to the approximation order. The 
sequence limit is equivalent to a fixed point. In the vicinity of a fixed point, 
there exists a self-similar relation analogous to a renormalization group equation. 
The transformation between subsequent approximants forms a dynamical system in 
discrete time, called an approximation cascade, whose trajectory is bijective to 
the sequence of the approximants. By embedding the approximation cascade into a 
flow results in an approximation flow, where the approximation order is considered 
as a continuous variable. Using a continuous approximation order reminds us the use 
of a continuous dimensionality in the Wilson renormalization group. Since both the 
equations of renormalization group and self-similar approximation theory can be 
written either in a functional form or as Lie differential equations, they allow 
for the analysis of stability by considering the related map multipliers, as for 
dynamical systems. Control functions can be introduced either into an initial 
approximation or by a transformation of the treated sequence. A very convenient 
is the fractal transform effectively rising the approximation order. By employing 
the fractal transform it is possible to derive different forms of approximants, 
such as self-similar root approximants, self-similar nested approximants, 
self-similar exponential approximants, self-similar additive approximants, 
self-similar factor approximants, and self-similar combined approximants. The wide 
applicability of these approximants is illustrated by several examples. The main 
features of the self-similar approximants is their generality, simplicity, and good 
accuracy. In the cases, where a problem enjoys an exact solution, these approximants 
reconstruct this exact solution.

The main equation of self-similar approximation theory is based on the assumption 
of the existence of the sequence limit, similarly to the assumption of the existence 
of a critical point in statistical renormalization group. The field-theory 
renormalization group equations are derived by employing the natural symmetry of 
the considered system. The efficiency of both optimized perturbation theory and 
self-similar approximation theory can be understood as being due to their ability 
of discovering the hidden symmetry in asymptotic sequences. By their formulation, 
they make it possible to discover the hidden transformation law when moving from 
one to another terms of a perturbative sequence. The discovered transformation law
allows us to represent it as an equation of motion analogous to renormalization 
group equations. The difference is that in the renormalizaion group equations the 
motion is with respect to some kind of a scaling parameter. While in self-similar 
approximation theory, the motion is considered with respect to the approximation 
order.

\section*{Acknowledgement}

I am grateful for many useful discussions and collaboration to S. Gluzman and 
E.P. Yukalova.

\end{document}